\documentclass[final,12pt,a4paper,oneside,titlepage]{book}

\usepackage{microtype}
\usepackage{emptypage}
\usepackage{pdfpages}
\usepackage{textcomp} 
\usepackage{amssymb}
\usepackage{amsmath}
\usepackage{amsthm}
\usepackage{amscd}
\usepackage{amsfonts}
\usepackage{mathrsfs}
\usepackage{graphicx}
\usepackage[latin1]{inputenc}
\usepackage[T1]{fontenc}
\usepackage[italian,english]{babel}
\usepackage{varioref}
\usepackage{color}
\usepackage[pagebackref]{hyperref}

\renewcommand{\epsilon}{\varepsilon}
\renewcommand{\theta}{\vartheta}
\renewcommand{\rho}{\varrho}
\renewcommand{\phi}{\varphi}

\newcommand{\re}[1]{(\ref{#1})}

\newcommand{\hubbleconstant}{H_0\approx73.8\,km\,s^{-1}\,Mpc^{-1}}
\newcommand{\hubbletime}{t_H\doteq\frac{1}{H_0}\approx13.6\,Gyr}

\newcommand{\m}{\mathcal{M}}
\newcommand{\tm}{\mathcal{T}\mathcal{M}}
\newcommand{\tpm}{\mathcal{T}_p\mathcal{M}}
\newcommand{\cotm}{\mathcal{T}^{\ast}\mathcal{M}}
\newcommand{\cotpm}{\mathcal{T}_p^{\ast}\mathcal{M}}
\newcommand{\w}{W}
\newcommand{\pw}{W_p}
\newcommand{\s}{S}

\newcommand{\av}[1]{\langle #1\rangle}
\newcommand{\fdt}{\left(t\right)}
\newcommand{\q}{\mathcal{Q}}
\newcommand{\nq}{\mathcal{Q}}
\newcommand{\tq}{\mathcal{Q}}
\newcommand{\dr}{{ ^{\left(2\right)}R}}
\newcommand{\tr}{{ ^{\left(3\right)}R}}
\newcommand{\nmur}{{ ^{\left(N\right)}R}}
\newcommand{\avdr}{\av{ ^{\left(2\right)}R}}
\newcommand{\avtr}{\av{ ^{\left(3\right)}R}}
\newcommand{\avnmur}{\av{ ^{\left(N\right)}R}}
\newcommand{\ecm}{\chi\left(\mathcal{M}\right)}

\newcommand{\g}{\textrm{g}}  %define the genus

\newcommand{\ev}[1]{\textsl{#1}}  %ora un comando perevidneziare in modo diverso dal comando \emph, che uso per temrini fisici appena introdotti...\ev servirà invece per termini da mettere in risalto, non per la loro importanza ma per distinguerli visivamente.

\newenvironment{citazione}
{\begin{quotation}\small}
{\end{quotation}}

\theoremstyle{plain}
\newtheorem{theorem}{Theorem}

\theoremstyle{definition}
\newtheorem{definition}{Definition}

\begin{document}

\frontmatter

\includepdf{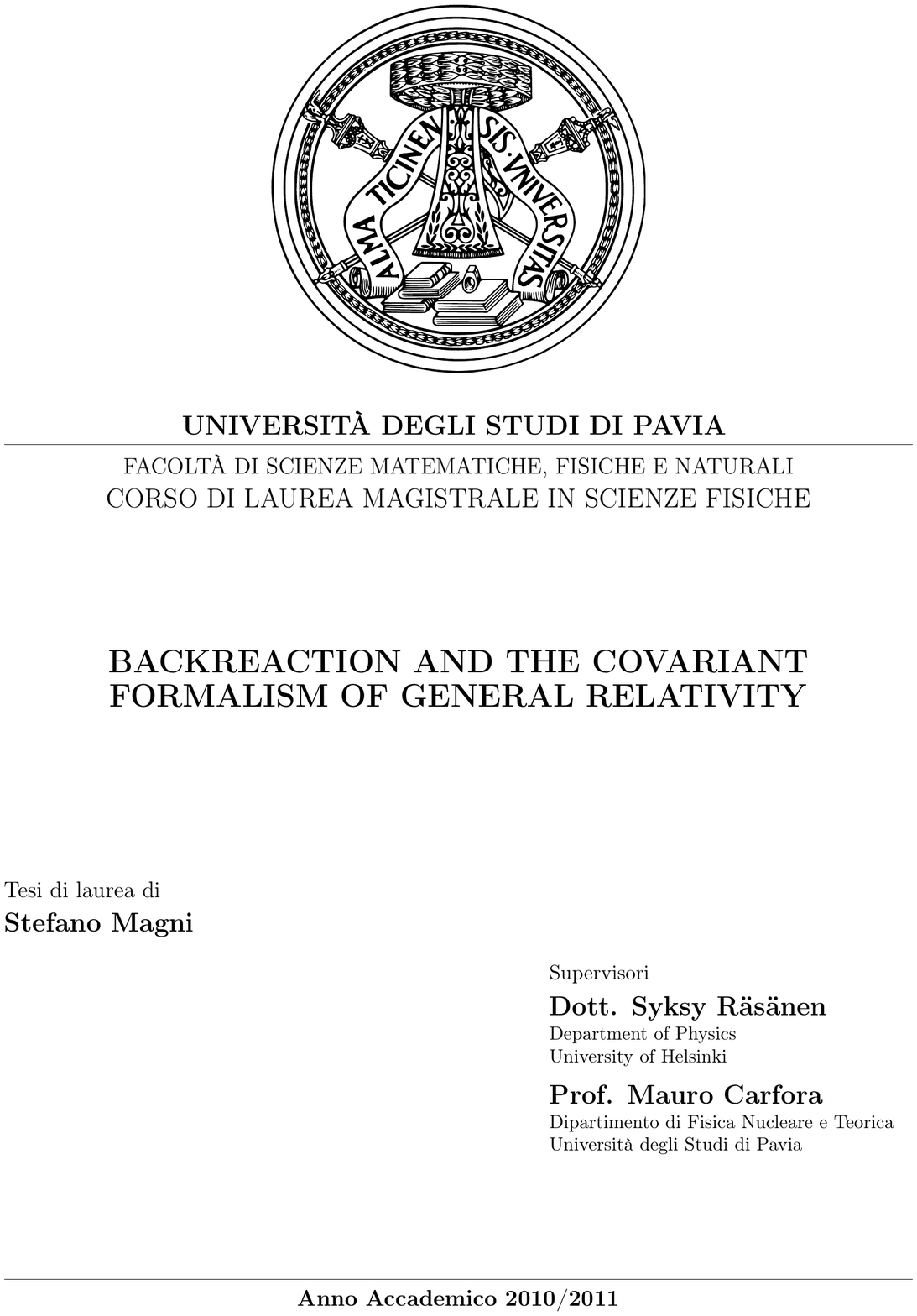}

\paragraph{ABSTRACT.}Backreaction represents a possible alternative to dark energy and modified gravity.
The simplest modification of gravity, equivalent to the existence of vacuum energy, the main candidate for dark energy, is the cosmological constant $\Lambda$, which is inserted in the Einstein equation in the framework of the $\Lambda CDM$ model in order to account for the observations.
This model is based on the FLRW metric, which is the exact solution of the Einstein equation directly obtained from the hypothesis of exact homogeneity and isotropy of the spacetime.
The real universe is not exactly homogeneous and isotropic, so it is interesting to consider the effect on the evolution of the universe of the inhomogeneities, analysed trough the study of averaged physical quantities (instead of local ones), which is precisely backreaction.

In the beginning of this thesis we present the ideas underlying the study of backreaction, and we point out its role as a possible alternative to dark energy and modified gravity.
We present the FLRW model, we describe in detail the 3+1 covariant formalism and we recall all the results of general relativity we use.
Then we analyse Frobenius' theorem and we explain what it tells us about the splitting of spacetime into space and time.
We present the averaging procedure developed by Buchert, the Buchert equations and their generalization to the case of general matter. 
We then generalize the procedure and the equations to an arbitrary number of spatial dimensions. We focus on the case of 2+1 dimensions, where the relation between the topology and the geometry of a surface imposes a global constraint on backreaction.

\cleardoublepage

\tableofcontents

\mainmatter

\chapter{Introduction}

\enlargethispage{-1\baselineskip}

\paragraph{Cosmology, general relativity and the cosmological constant.}Cosmology is the branch of physics that studies the universe as a whole. In this framework we work on scales such that gravity is the only fundamental interaction that matters.\footnote{Even though gravity is the weakest of the four fundamental interactions, it possesses an infinite range, unlike strong and weak interactions. Electromagnetism too has an infinite range, but celestial bodies are electrically neutral.}
Einstein's theory of general relativity provides a widely accepted description of it.

This theory describes \emph{spacetime} as a \emph{manifold} equipped with a \emph{Lorentzian metric}. The geometry of spacetime governs the motion of matter, and matter in turn tells spacetime how to curve. This mutual influence is encoded in the \emph{Einstein equation}, where matter is described by the \emph{energy-momentum tensor}, while curvature is represented by the \emph{Einstein tensor}, constructed from the \emph{Riemann tensor}.

The Einstein equation can be generalized by adding a term containing a constant $\Lambda$, known as the \emph{cosmological constant}. This modification was first introduced by Einstein in order to allow a stationary solution of his equation (that he believed to better describe the real universe), which is otherwise possible only if we have matter with sufficiently negative pressure.

The idea of a static universe (and then the cosmological constant) was abandoned in favor of the idea of a dynamic universe when, in the late twenties, observations showed that the universe is expanding. Measurements of galaxies' distances, combined with redshift observations, showed that there is a proportionality between the two; the explanation is that self-gravitating objects, such as isolated galaxies or clusters of galaxies, are receding further away from each other and from us with a speed roughly proportional to their distances.\footnote{The authorship of these ideas is an intricate issue, and a detailed discussion on the topic can be found in \cite{beginningoftheworld}. E. Hubble published in 1929 the results of his observations of galaxies' distances, and using V. Slipher's determination of their velocities (obtained through redshift measurements) he suggested the linear distance-velocity relation that bears his name. He derived that equation empirically, and he (wrongly) interpreted the galaxy redshift as a pure Doppler effect. Two years before, in 1927, G. Lema\^itre published a paper containing the same relation, and an exact solution of the Einstein equation. In the light of his results, he gave the first interpretation of cosmological redshifts in terms of space expansion, instead of a real motion of galaxies. This idea represents a cornerstone for cosmology.} Many subsequent observations have confirmed the expansion of the universe at the present time.

\paragraph{The FLRW solution.}An exact solution of the Einstein equation is the \emph{Friedmann-Lema\^itre-Robertson-Walker} (FLRW) \emph{metric}, that is based on the hypothesis of \emph{exact homogeneity and isotropy} of the universe, i.e. all points and all directions in the universe are equivalent.\footnote{Exact definitions are given in section \re{twoassumptions}.} From this assumption we can determine the most general form of the energy-momentum tensor allowed (i.e. the one that describes an \emph{ideal fluid}), and the form of the metric up to one free function (the \emph{scale factor}) and one constant (that determines the geometry of the three-dimensional space, which can be only \emph{spherical}, \emph{hyperbolic} or \emph{Euclidean}).

Applying the Einstein equation to the FLRW metric we obtain the \emph{Friedmann equations}, that govern the evolution of the scale factor. Formulated in the twenties and thirties, this solution was the simplest one able to predict expansion.

The early universe seems to be well described by a homogeneous and isotropic FLRW model (i.e. a cosmological model based on the FLRW metric) plus linear perturbations, but for the real universe at late times it is not clear whether the metric remains close to the FLRW one.
Moreover, from the nineties independent observations of \emph{type Ia supernovae}, \emph{cosmic microwave background} (CMB) \emph{anisotropies} and  \emph{large scale structure} provide data sets that disagree with the predictions obtained for the late time universe using the matter dominated spatially-flat FLRW model.

The FLRW models are able to account for these observations only either considering modified gravity (of which insert in the Einstein equation the cosmological constant is the simplest possibility) or allowing the existence of a form of energy with negative pressure, \emph{dark energy}.\footnote{These solutions are widely discussed in the section titled: ``dark energy and modified gravity''.}

\paragraph{The $\Lambda$CDM model.}In general relativity, the only rigorous way of describing the matter that fills the universe on a cosmic scale is through fluid dynamics. Einstein's theory does not determine what kind of matter the universe contains, i.e. it does not tell us what kind of fluid better describes the various forms of \emph{ordinary matter} (i.e. particles of the \emph{Standard Model of particle physics} and \emph{cold dark matter}) that fill the cosmos. It does not even tell which is the metric that better describes our universe, except for the fact that it must be a solution of the Einstein equation. These information are instead provided by the particular \emph{cosmological model}.

The simplest model that is in general agreement with the observed phenomena is the $\Lambda$CDM (\emph{Lambda-Cold Dark Matter}). It describes \emph{dark matter} as being \emph{cold} (i.e. constituted of non-relativistic particles), and the cosmological constant is present. It is frequently referred to as the ``\emph{standard model of cosmology}'', and it is based on a remarkably small number of parameters.\footnote{The most important of these parameters are defined in section \ref{friedmannsection}.} The best fit values of these parameters are obtained from the above mentioned observations.
After the \emph{Big Bang} a very short period of rapid expansion, which is described by the theory of \emph{inflation}, follows. The universe after the inflationary epoch is described in $\Lambda$CDM using the FLRW metric and the Friedmann equations.

From the end of this period to the time when the universe was approximately $50000$ years old (corresponding to a redshift $z\approx3500$) the density of \emph{radiation} (i.e. relativistic particles) exceeded the density of \emph{matter} (i.e. non-relativistic particles). This period is known as the \emph{radiation dominated era}. After that the density of matter became dominant, and the following epoch is called the \emph{matter dominated era}.\footnote{The above change is due to the fact that the scale factor grows with time as a consequence of the expansion of the universe, and the densities of matter and radiation depend on the scale factor in a different way.}

When the universe was around $380000$ years old ($z\approx1090$) \emph{recombination} (i.e. the formation of hydrogen atoms from electrons and protons) occurred, and soon after the photons {decoupled} from baryonic matter (i.e. they were no more in thermal equilibrium with baryonic matter, mainly constituted at that point of hydrogen atoms) because the rate of Compton scattering dropped. The universe became transparent to photons, which started propagating freely through space, and constitute what we observe today as the cosmic microwave background. After that the growth of \emph{baryonic structures} (previously inhibited by photon pressure) could start.\footnote{If dark matter exists, it almost certainly decoupled during the radiation-dominated era, and before recombination. \emph{Dark matter structures} started grow earlier than baryonic structures.}

The epoch that started when the universe was about 5 billion years old and lasts until today is sometimes called the \emph{dark energy dominated era}. This is due to the fact that according to $\Lambda$CDM in this epoch the density of dark energy exceeded the density of matter. In the framework of $\Lambda$CDM, dark energy is currently estimated to constitute about $74\%$ of the energy density of the present universe, dark matter is considered to account for $22\%$, while the remaining $4\%$ is due to matter and radiation. Then the \emph{coincidence problem} arises: why does the repulsive component (i.e. dark energy or the cosmological constant) dominate today when the past era was matter dominated? A possible answer is provided further in this chapter.

\paragraph{Dark energy and modified gravity.}The model that describes the early time universe is based on three hypothesis: homogeneity and isotropy (it uses the FLRW metric), ordinary matter (i.e. matter with non-negative pressure), standard gravity (i.e. general relativity based on the Einstein-Hilbert action). Because set in this way the model gives us predictions that are in conflict with late time observations, and these observations are beyond doubt, at least one of the three assumptions must be wrong.

Usually in the framework of the $\Lambda$CDM model the agreement with data is restored by introducing a term in the Einstein equation. We can insert a term on the geometric side (as we do when we insert the term containing the cosmological constant), and consider it a modification of the law of gravity; or we can insert it into the matter side and write it in the form of an energy-momentum tensor.\footnote{See equation \re{efe} and the related footnote.} This means to discard the second or the third hypothesis, i.e. state that standard gravity must be modified, or that in addition to ordinary matter we have also exotic matter (dark energy, with its negative pressure that must satisfy the condition $p<-\frac{1}{3}\mu$, which defines it).\footnote{In this introduction we give an overview of the problem; a wider explanation can be found in \cite{deandmg}.}

As the cosmological constant and dark energy enter the Einstein equation in the same way, and their only signature is their effect on spacetime, they cannot be distinguished by observations.

The aforementioned observations, together with the Friedmann equations (containing the cosmological constant or \emph{vacuum energy}, which is the principal candidate for dark energy, as explained below), lead to the conclusion that the expansion of the universe at late times is accelerating, i.e. the scale factor grows at an increasing rate (see for instance \cite{deandmg}).\footnote{This conclusion holds if we describe the universe with a FLRW model, i.e. we assume homogeneity and isotropy, while for different solutions of the Einstein equation we may not obtain the same result. For instance, if we use a \emph{Lema\^itre-Tolman-Bondi} (LTB) solution (that is spherically symmetric) with our galaxy cluster at the center, either acceleration is not necessarily implied, or it is present but without requiring dark energy; which possibility occurs is not yet clear. The LTB solution describes a system that is isotropic (w.r.t. us) but not homogeneous. By construction this use of the LTB solution puts us in the center of the universe. This violates the \emph{Copernican Principle} that we do not occupy a somehow privileged position in the universe.}

It is important to underline that, without the above mentioned modifications, from the Friedmann equations it follows that the expansion of the universe decelerates. This is expected, on the ground that (unmodified) gravity is attractive. Introducing dark energy with its negative pressure or modifying the gravity in order to allow acceleration means that we explain the observations with repulsive gravity.

The use of the cosmological constant represents the simplest way to modify gravity, but there are other ways of doing it. Some examples are \emph{scalar-tensor models} and \emph{brane-world models}. However, it turns out to be extremely difficult to modify general relativity without violating observational constraints or introducing instabilities in the theory, and for these reasons the modified gravity models proposed until now (apart from $\Lambda$CDM) seem to be ruled out (see \cite{deandmg}).

There are also many candidates for dark energy. The principal is the vacuum energy, that unlike most other proposal, is theoretically on very solid ground, because quantum field theory predicts that there is an energy density associated to the vacuum state. Its main problem is that it is not yet clear if it can provide a contribution to the Einstein equation with the right order of magnitude to represent the dark energy term, see for instance \cite{syksy2011}.\footnote{Nevertheless, note for instance that the \emph{Casimir effect} (usually invoked as a proof of the existence of vacuum energy) gives no more (or less) support for the reality of vacuum energy than any other one-loop effect in quantum electrodynamics; see \cite{casimireffect}, where it is shown how Casimir force can be calculated without reference to the vacuum.}

The situation is quite different from that of dark matter. Even if its nature is still an open question, its existence is needed in order to explain many different physical phenomena related to physically distinct situations (the motion of stars in spiral galaxies, the structure of the cosmic microwave background,\dots), so providing an alternative to its existence is very difficult, if not impossible.

On the other hand, the existence of dark energy is needed if, without allowing modifications of gravity, we require the universe to be described by a FLRW model, i.e. we assume the universe to be homogeneous and isotropic. In fact, even if also in the case of dark energy data come from different physical phenomena, all of them give us the behavior of the same quantity: the scale factor. However, any model or theory able to give the same behavior of the scale factor represents a valid alternative to dark energy.

\paragraph{Inhomogeneous cosmologies and backreaction.}While for such a modification of gravity or for the introduction of such an energy there is no other evidence, we know that the real universe is far from being exactly homogeneous and isotropic due to the formation, at late times, of non-linear structures. The hypothesis of homogeneity and isotropy describes a universe where such structures, i.e. galaxies, clusters of galaxies, voids, etc. do not exist. 

Then emerges the idea that forsaking this hypothesis, i.e. constructing models where inhomogeneities and/or anisotropies are present (an interesting pursuit in its own right), it is possible to make predictions that fit the observations without modifying gravity or introducing dark energy. The effect of these inhomogeneities/anisotropies on the expansion of the universe is called \emph{Backreaction}.\footnote{If we abandon the hypothesis of homogeneity and/or isotropy, then the universe must be described with a different solution of the Einstein equation than FLRW. For instance we can consider the LTB solution with ourselves at the center, but as already stated it violates the Copernican principle.}

\paragraph{The averages.}If we deal with inhomogeneities and/or anisotropies the idea also emerges of considering quantities averaged over a certain portion of space, instead of local quantities. This can always be done, but the averaged quantities are useful only if the conditions of \emph{statistical homogeneity and isotropy} hold.\footnote{These concepts are explained in the beginning of chapter 4, and are analysed in detail, both from the theoretical and from the observational point of view, in \cite{sylosbook}.}
Roughly speaking, if we consider a large box, we can put it in different places in the universe and evaluate the change of the mean quantities considered inside the box from one location to another, and which is (if any) the length scale over which these differences are small. If this scale exists, we refer to it as \emph{homogeneity scale}.

Considering averaged quantities gives rise to a couple of problems: first, defining a procedure that allow us to average quantities; second, determining whether the averaged quantities satisfy the same equations satisfied by the local ones. Usually, when following the evolution of inhomogeneities into the non-linear regime, the mean quantities are assumed, rather than demonstrated, to obey Friedmann equations. But if we carefully consider what happens, we discover that the averaged quantities in general do not obey the Friedmann equations.

Averaging in general relativity is a very involved problem, one of the reasons being that there are no preferred time-slices one could average over, and in the case of non vanishing vorticity not even hypersurfaces orthogonal to the velocity vector field (which describes the velocity of a set of observers spread out in spacetime, and give rise to a family of preferred world lines representing their motion).
Another problem is the non-linearity of the Einstein equation. The problem of averaging inhomogeneous cosmologies has been studied by many authors with different approaches. In this work we follow the approach of T. Buchert, that has been used by several authors in the study of backreaction.
An averaging procedure is developed in Newtonian gravity by Buchert and J. Ehlers in \cite{buchert1996}. In a general relativistic framework the problem is analysed by Buchert in \cite{buchert1999} and \cite{buchert2001}, for the case of dust and for a general perfect fluid, respectively. The generalization of the procedure to the case of general matter and non-vanishing vorticity is worked out, in the covariant formalism, in \cite{syksy2009b} by S. Räsänen.

\paragraph{Backreaction as an alternative to dark energy and modified gravity.}The equations that hold in the inhomogeneous case differ from their local counterpart by the presence of some additional terms due to the existence of inhomogeneities and anisotropies. When we consider the averaged versions of these equations, the effects of inhomogeneities and anisotropies can be collected in the \emph{backreaction variable $\q$}. The backreaction is also an expression of the fact that time evolution and averaging do not commute, i.e. evolving through the Einstein equation the local quantities and taking the average, or evolving the averaged quantities, gives different results.

If by taking into account the effect of inhomogeneities and anisotropies (i.e. the formation/presence of structures) on the expansion rate we are able to make predictions that agree with the observations, we can avoid the introduction of dark energy or the modification of gravity.
Two questions arise: \ev{does backreaction provide acceleration, and, if so, is its effect sufficient to explain observations?}

Backreaction can provide acceleration. This has been proven with a model made of two regions in \cite{syksy2006a} and \cite{syksy2006b} by S. Räsänen. Acceleration has been demonstrated also in the exact spherically symmetric dust solution, the Lema\^itre-Tolman-Bondi model, in \cite{acceleration34}, \cite{acceleration35} and \cite{acceleration36}. In \cite{acceleration34} acceleration arises in some examples where a fluid consistent of three regions is analysed. In \cite{acceleration35} acceleration is demonstrated for an unbound LTB model constituted of a single region, in which the contribution of the matter density is negligible compared to the contribution of the curvature. In \cite{acceleration36} acceleration is shown in some swiss cheese models, with LTB spherical regions inserted in a Einstein-De Sitter background.
In conclusion, accelerated average expansion due to inhomogeneities is possible.

While the average expansion rate is given by the equations containing backreaction, the local expansion rate is governed by the local equations which without the cosmological constant and vorticity force it to decelerate. So it seems to be paradoxical that the average expansion rate accelerates even if the local one slows down everywhere.
This can be intuitively understood through the following argument, as simple as it is amazing.
If the space is inhomogeneous, different regions expand at different rates. Regions with faster expansion rate increase their volume more rapidly than regions with slower expansion rate, by definition. As a consequence, the fraction of the total volume of the universe that is expanding faster rises. So also the average expansion rate can rise. If all the local expansion rates are decelerating,\footnote{This happens for instance when we describe matter as dust (i.e. when the energy density is the only non-negligible term in the energy-momentum tensor, see section \re{dustsect}) and we have vanishing vorticity (irrotational dust case).} then whether the average expansion rate actually rises depends on how rapidly the fraction of fast expanding regions grows relative to the rate at which their expansion rate is decreasing.

In \cite{syksy2008b} a semi-realistic model, where a more realistic distribution of matter is considered, is analysed. That work employs a spatially flat FLRW model with a Gaussian field of density fluctuations, and studies what happens while structures form. In this case it turns out that there is less deceleration, and the era when the structures' formation becomes important for the expansion of the universe comes out roughly correctly, showing how backreaction might solve the coincidence problem, i.e. why the acceleration has started in the recent past.

Whether backreaction provides the right amount of acceleration (or less deceleration) for the real universe is still an open question; the state of the art is analysed in \cite{syksy2010} and \cite{syksy2011}. It should also be noted that the backreaction idea has been criticized; some of these critiques, and useful references, can be found in \cite[sections 5.4.1, 5.5.2]{ilcerchiosichiude}.
In conclusion, backreaction represents a possible alternative to dark energy and modified gravity, but there is still work to be done.

\paragraph{Covariance.}Tensor fields are objects abstract enough so that the vast majority of the quantities that one considers in physics can be viewed as tensor fields. The laws of physics governing these quantities can then be expressed as tensor equations, i.e. equations between tensor fields.

An important principle that applies to the form of the laws of physics is \emph{general covariance}.\footnote{This principle applies both in special relativity as well as general relativity. The adjective ``general'' in the name of Einstein's theory of general relativity came from this principle, see \cite[chapter 4]{wald}.} Roughly speaking, it can be stated as follows: the metric of spacetime is the only quantity pertaining to spacetime that can appear in the laws of physics, i.e. there is no preferred basis of vector fields pertaining only to the structure of spacetime which can appear in any law of physics.\footnote{In the relativistic context, also the time orientation and space orientation of spacetime can enter the laws of physics. A detailed discussion can be found in \cite[chapter 4]{wald}.}

Often a coordinate system is chosen, and equations have been written out in component form using the coordinate basis. If the principle of general covariance were violated, it would be possible to find a preferred basis, and it can be shown that in this case the form of the equations is not preserved under general coordinates transformations, and they lose the tensorial character, because they do not transform in the right way.
\\

\paragraph{The 3+1 splitting.}The goal of cosmology is to find a model that best describes the universe, and we have to bear in mind that from any such model we have to extract observational predictions.
Einstein's equations are not particularly intuitive, so it is desirable to break them into a more intuitive set of equations retaining their covariant character, and at the same time introduce quantities that are directly measurable.

For a complete cosmological model one must specify not only a metric defined on a manifold, but also a family of observers spread out in spacetime, whose velocity is described by a velocity vector field that
gives rise to a family of preferred world lines representing their motion. This velocity vector field can be used to look at tensors along these world lines, and orthogonal to them; this represents a 3+1 splitting of quantities. It is also covariant because the velocity vector field can be defined uniquely and without any coordinates. In this work we employ this 3+1 covariant formalism

The derivative of the fundamental congruence can then be split into irreducible quantities:\footnote{In the 3+1 covariant formalism all irreducible quantities are either scalars, projected vectors or projected, symmetric, trace-free tensors.} the \emph{expansion rate} of fluid elements, the \emph{acceleration}, the \emph{shear}, which describes how the congruence
will distort in time, and the \emph{vorticity}, that is the rate of rotation of the congruence.

This procedure can be seen as a 3+1 splitting of spacetime into space plus time. This interpretation is correct only when the vorticity vanishes. In fact as a consequence of Frobenius' theorem we have that a family of three-dimensional spaces orthogonal to the four-velocity vector field exists if and only if the vorticity vanishes. Otherwise we can still project quantities orthogonal to the velocity vector field, but we have to bear in mind that they are not projections into a three-dimensional space.
This result is of fundamental importance for our discussion on average quantities, because we want to average over three-dimensional hypersurfaces.

However, even if dealing with averaging we are mainly interested in the above consequence of Frobenius' theorem, it must be pointed out that this theorem represents a more general result, interesting on its own, and as a powerful mathematical tool. Moreover, it provides another very useful consequence for general relativity: for every smooth vector field a family of integral curves (which represent the world lines associated to the velocity vector field of the above observers) can be found.

Eventually, the outlined approach allows to split the Einstein equation into a set of evolution and constraint equations that hallow a more intuitive interpretation, and contain quantities whose physical meaning is more transparent.

\paragraph{The 2+1 dimensions.}The $\left(2+1\right)$-dimensional gravity is interesting in its own right. In this case, no gravity outside matter is allowed. Matter curves the spacetime only locally, and then there are no gravitational waves.\footnote{This is described by the fact that the \emph{Weyl tensor} (which describes the propagation of \emph{gravitational waves}, see section \ref{curvtenssect}) vanishes in 2+1 dimensions.} We cannot obtain Newtonian gravity as a limit of Einstein theory.

Anyway, we could ask why one should investigate physical phenomena in the ``physically unrealistic'' $\left(2+1\right)$-dimensional case, while the observed universe possesses (at least) four dimensions. The answer is not that in this case it is simpler to work out calculations (although this is usually true), but that in this case sometimes the situation is simpler (for instance we have vanishing Weyl tensor, see above), and this allows to better understand the relation between different physical phenomena.

Moreover, some of the results obtained in the $\left(2+1\right)$-dimensional case may be universal and independent of spacetime dimensions. On the other hand, comparing the results belonging to the low-dimensional case with those obtained for the high-dimensional case, we can enlighten the latter.

In particular, also for our purposes the $\left(2+1\right)$-dimensional case is interesting. In fact, due to Gauss-Bonnet theorem, the geometry and the topology of a two-dimensional manifold (i.e. a surface, the ``spatial'' part in the 2+1 decomposition) are related. Using this result, we obtain that the average curvature of the above surface is inversely proportional (through a constant that can also vanish) to the square of the scale factor, so it is constrained. As a consequence, also the backreaction variable is constrained to be inversely proportional (through a constant that can also vanish) to the fourth power of the scale factor.

\paragraph{Structure of this thesis.}The first chapter of this work is represented by this introduction, and its aim is to outline the topic and to provide the setting for the following material.

In the second chapter we present the \emph{3+1 covariant formalism}, that constitutes the formal framework of this work. This consists in a fluid dynamical description and allows us to operate with tensor quantities. We explain in detail the formalism, and also summarize the results of general relativity that are important for the sequel. Projecting the Einstein equation and separating out the trace, the antisymmetric parts and the symmetric trace-free parts we derive the \emph{evolution} and the \emph{constraint equations of general relativity}, that drive the dynamics of the universe, and give us the equations that we want to average.

The third chapter is dedicated to Frobenius' theorem, that leads us to clarify some aspects of the central concept of space and time. In relativity we assume the existence of \emph{spacetime}, considered as one, while \emph{space} and \emph{time} are just derived concepts, and we must be careful when we think about them separately.
The theorem is presented in an abstract way, and after analysing it from the geometrical point of view we clarify its role in general relativity.

The fourth chapter is focused on backreaction. We analyse the problem of averaging, and we describe the averaging procedure used. We consider the scalar parts of the Einstein equation and we show how to average them. In the case of irrotational dust we derive the Buchert equations, that contain the backreaction term. We analyse some particular solutions, and we derive the integrability condition that relates the evolution of the spatial curvature with the backreaction. Finally we derive the generalization of the Buchert equations to the case of general matter with non-zero vorticity.

In the fifth chapter we generalize the Buchert equations and the integrability condition to the case of $N+1$ dimensions. We analyse some differences between the case of $N+1$ dimensions, $3+1$ dimensions and  $2+1$ dimensions, and some particular solutions. In the $\left(2+1\right)$-dimensional case we relate the average of the spatial Ricci curvature to the Euler characteristic.

The appendix is dedicated to the FLRW model, that has been included because we often refer to it and to its equations in the text. We give precise definitions of exact homogeneity and isotropy. We present the metric, and we obtain Friedmann equations as a particular case of the Raychaudhuri equation and the Hamiltonian constraint. We give some useful definitions. We point out the role of the cosmological constant, and we show how it provides acceleration.

\clearpage

\paragraph{Notation and conventions.}
\begin{itemize}
  \item Spacetime indices are indicated by Greek letters and run from 0 to 3, 
        i.e. $\alpha,\beta,\gamma,\ldots=0,1,2,3$.
  \item Spatial indices are indicated by Latin letters and run from 1 to 3, i.e. $a,b,c,\ldots=1,2,3$.
  \item The metric tensor is indicated by $g_{\alpha\beta}$.
  \item The signature of the metric is $\left(-+++\right)$.
  \item Einstein summation convention is used.
  \item We employ units such that the speed of light and 
        the Newton gravitational constant satisfy $c=1={8\pi G_N}/{c^2}$.
  \item Equality by definition is indicated by $\doteq$ \ .
  \item Symmetrization is indicated by round brackets, for example we have:
        $T_{\left(\alpha\beta\right)}\doteq\frac{1}{2}\left(T_{\alpha\beta}+T_{\beta\alpha}\right)$.
  \item Antisymmetrization is indicated by square brackets, for example we have:        
        $T_{\left[\alpha\beta\right]}\doteq\frac{1}{2}\left(T_{\alpha\beta}-T_{\beta\alpha}\right)$.
  \item The partial derivative w.r.t. $x^\alpha$ is indicated with $\partial_\alpha$ or with a comma, 
        for example we have: $u_{\beta,\alpha}=\partial_\alpha u_\beta$.
  \item The covariant derivative w.r.t. $x^\alpha$ is indicated with $\nabla_\alpha$ or with a semicolon, 
        for example we have: $u_{\beta;\alpha}=\nabla_\alpha u_\beta$.
\end{itemize}

\chapter{The 3+1 covariant formalism}
\label{3+1covform}

One of the aims of cosmology is to describe the large scale structure of the universe. On distances greater than the scale of the solar system, and in particular on the scale of clusters of galaxies, gravity is the dominant long-range force (we consider the heavenly bodies to be electrically neutral, according to observations). General relativity describes gravitation on these scales, and because we do not believe (as discussed in the introduction) that there are sufficient reasons for modifications of this theory, we assume its validity.

The core of general relativity may be summarized as follow: spacetime can be described as a manifold $\left(\mathcal{M},g\right)$ on which there is defined a Lorentzian metric $g$; the curvature of the metric is related to the matter distribution by the \emph{Einstein equation} (\cite[p. 73]{wald}).

In general relativity matter can be described either with a model of \emph{point masses}, or as a \emph{continuous medium}. The latter is the only one that can be carried out in a rigorous mathematical way, because the definition of a singularity of the metric field (that describes a particle) has not yet been obtained. So we use the latter approach, i.e. a \emph{fluid dynamical description} (see \cite{covform1}, \cite{covform2} for a review).

We know from observations that the \emph{peculiar velocities} of stars or galaxies w.r.t. surrounding objects are small when compared to the general motion of the clusters (that is an overall expansion), so we are able to determine a \emph{local velocity} which represents to a good approximation the over-all motion of matter. Then we assume the existence at every point of spacetime of a vector field representing this local velocity.

We can describe spacetime via \emph{3+1 covariantly defined variables}, via the metric $g$ described in a particular set of \emph{local coordinates} by $g_{\alpha\beta}\left(x^{\mu}\right)$, or via the metric described by means of particular \emph{tetrads}. Because in general relativity we have complete coordinate freedom, it is better when possible to describe physics and geometry by tensor relations and quantities, that remain valid whatever coordinate system is chosen. In this work we use the \emph{3+1 covariant approach} (see \cite{ilcerchiosichiude,covform1,covform2,covform3}), that is summarized in this chapter together with some important relations in general relativity, and the evolution and constraint equations that arise from the Einstein equation.

Note that all the relations obtained in this chapter are independent of any particular cosmological model. For the sake of generality, we have also included the cosmological constant.

\section{Kinematical variables}
\label{kinvarsec}

\subsection{Velocity vector field}
\label{velvecfield}

We consider a set of observers spread out in spacetime. We assume that their velocity is described by a unique vector field of components $u_{\mu}$ that exists at each point of spacetime. As a consequence there is a family of preferred world lines representing their motion.\footnote{This family of preferred world lines is often referred to as a \emph{congruence}, i.e., given an open subset $O\subset\mathcal{M}$, a family of curves such that through each $p\in O$ there passes precisely one curve in this family \cite[sec. 9.2]{wald}. The existence of this family of curves is proved in chapter 3, section \ref{intcurvecase}.}

In \emph{general coordinates $x^{\mu}$} this \emph{four-velocity} is
\begin{equation}
u^{\mu}\doteq\frac{dx^{\mu}}{d\tau} \ ,
\label{four-velocity}
\end{equation}
where $\tau$ is the \emph{proper time} measured along the fundamental world lines, and the normalization reads
\begin{equation}
u_{\mu}u^{\mu}=-1 \ .
\label{norm}
\end{equation}

Instead of writing the four-velocity in a general way, one can choose to use \emph{comoving coordinates $(x^{a},t)$} defined as follows. Choose arbitrarily a space section of the spacetime and label the fluid particles by coordinates $x^{a}$; at all later times label the same particles by the same coordinate values, so that the fluid flow lines in spacetime are the curves $x^{a}=const$. The time coordinate is then determined by measuring proper time, from the initial space section, along the flow lines. Expressed in these coordinates the four-velocity takes the form
\begin{equation}
u^{\mu}=\delta^{0\mu} \ .
\label{normcoord}
\end{equation}
Note that this represents a particular choice of coordinates, so we avoid using it unless necessary.\footnote{When useful, after the definition of a quantity we give also its expression in comoving coordinates.}

\subsection{The projection tensors}
\label{prosection}

A $(3+1)$ split of the spacetime is determined, given $u^{\mu}$, by the \emph{projection tensors}:
\begin{equation}
U_{\mu\nu}\doteq-u_{\mu}u_{\nu}
\label{prou}
\end{equation}
and
\begin{equation}
h_{\mu\nu}\doteq g_{\mu\nu}+u_{\mu}u_{\nu} \ .
\label{proh}
\end{equation}

It is easy to show the projector character of these quantities, in fact by construction they satisfy the sets of relations:
\begin{gather}
U_\alpha^{\phantom{\alpha}\beta}U_\beta^{\phantom{\beta}\gamma}=U_\alpha^{\phantom{\alpha}\gamma} \ ,\qquad
U_\alpha^{\phantom{\alpha}\alpha}=1 \ ,\qquad
U_{\alpha\beta}u^\beta=u_\alpha \ , 
\label{propU} \\
h_\alpha^{\phantom{\alpha}\beta}h_\beta^{\phantom{\beta}\gamma}=h_\alpha^{\phantom{\alpha}\gamma} \ ,\qquad
h_\alpha^{\phantom{\alpha}\alpha}=3 \ ,\qquad
h_{\alpha\beta}u^\beta=0 \ .
\label{proph}
\end{gather}

The tensor \re{prou} projects parallel to the four-velocity vector $u_\mu$, while \re{proh} projects into the instantaneous rest space of an observer moving with four-velocity $u_\mu$. In the sequel we often project quantities using the above tensors. The meaning of quantities projected using $h_{\alpha\beta}$ is that they represent what observers moving with $u^\mu$ measure.\footnote{If we consider another velocity vector field $n^\nu$ different from $u^\mu$, we can construct for it projectors analogous to \re{prou} and \re{proh}. So two observers, one moving with $u^\mu$, and the other with $n^\nu$, in general measure different values of the same physical quantity, because their projectors are different.}

Because of \re{proh} the expression for $ds^2$ can be written as
\begin{equation}
ds^2=g_{\mu \nu}dx^\mu dx^\nu=h_{\mu\nu}dx^{\mu}dx^{\nu}-\left( u_\mu dx^{\mu} \right)^2 \ .
\label{ds2}
\end{equation}

\subsection{Acceleration vector}
  \label{accefreefallsec}

The effective time derivative of a tensor $T$ measured by an observer moving with the velocity $u_\mu$ is denoted by $\dot{T}$. So
\begin{equation}
\dot{T}^{\alpha \ldots \beta} _{\phantom{\alpha \ldots \beta} \gamma \ldots \delta} \doteq u^\mu \nabla_\mu T^{\alpha \ldots \beta} _{\phantom{\alpha \ldots \beta}\gamma \ldots \delta} 
\label{timederiv}
\end{equation}
is the \emph{covariant time derivative along the fundamental world lines}.

The \emph{acceleration vector $\dot{u}_\mu$} is then defined as
\begin{equation}
\dot{u}_\mu \doteq u^{\nu} \nabla_\nu u_\mu \ ,
\label{acceleration}
\end{equation}
and it represents the degree to which the matter moves under the influence of any forces (remember that in general relativity gravity and inertia, which cannot be covariantly separated from each other, are not forces). The acceleration identically vanishes if and only if matter is moving under gravity plus inertia alone (geodesic flow), i.e. it is in \emph{free fall}.

Note that the normalization \re{norm} and the above definition imply that
\begin{equation}
\dot{u}_\mu u^\mu = 0 \ ,
\label{normacc}
\end{equation}
and so the acceleration can be considered, in this sense, spacelike.

In comoving coordinates we can express the acceleration vector, in terms of the \emph{Christoffel symbols $\Gamma_{\phantom{\beta}\mu \alpha} ^{ \beta }$}, as\footnote{We have used the expression of the covariant derivative of a vector field in terms of the Christoffel symbols $\nabla_\beta v^\alpha= \partial_\beta v^\alpha-\Gamma^{\alpha}_{\phantom{\alpha}\beta\gamma}u^\gamma$ (see \cite[p. 34]{wald}).}
\begin{equation}
\dot{u}^\alpha = \Gamma_{\phantom{\alpha}00} ^{ \alpha} \ .
\end{equation}

\subsection{Volume elements}
  \label{volelsec}

Because $\left( \mathcal{M},g \right) $ is an \emph{oriented pseudo-Riemannian manifold} by construction, it possesses a natural \emph{volume form} that, in local coordinates, can be expressed as
\begin{multline}
\epsilon\doteq\frac{1}{4!} \epsilon_{\alpha _1 \alpha _2 \alpha _3 \alpha _4} \sqrt{ \lvert g \rvert}dx^{\alpha_1}\wedge dx^{\alpha_2}\wedge dx^{\alpha_3}\wedge dx^{\alpha_4}=\\
=\sqrt{ \lvert g \rvert}dx^{1}\wedge dx^{2}\wedge dx^{3}\wedge dx^{4}=\ast\left( 1 \right) \ ,
\label{volume form}
\end{multline}
where
\begin{equation}
g\doteq det \left( g_{\mu\nu} \right)<0 \ ,
\end{equation}
and the quantity $\epsilon_{\alpha _1 \alpha _2 \alpha _3 \alpha _4}$ is the four-dimensional \emph{Levi-Civita symbol}. We have emphasized that we can also write the volume form as $\ast\left( 1 \right)$, i.e. the \emph{Hodge dual} of the constant map on the manifold.

So $\eta_{\alpha\beta\gamma\delta}$ is the \emph{four-dimensional volume element} that arises from the above form, and we have
\begin{equation}
\eta_{\alpha \beta \gamma \delta}=\frac{1}{4!} \sqrt{\lvert g \rvert} \epsilon_{\alpha \beta \gamma \delta}, \qquad
\text{and} \qquad
\eta_{\alpha \beta \gamma \delta}=\eta_{[\alpha \beta \gamma \delta]} \ .
\label{volelemxxx}
\end{equation}
Using it we can also define the \emph{three-dimensional volume element} of the rest-space of an observer moving with four-velocity $ u^{\mu} $ as
\begin{equation}
\eta_{\alpha \beta \gamma} \doteq \eta_{\alpha \beta \gamma \delta} u^{\delta} \ ,
\label{3volel}
\end{equation}
for which
\begin{equation}
\eta_{\alpha \beta \gamma} = \eta_{[\alpha\beta\gamma]} \qquad \text{and} \qquad \eta_{\alpha \beta \gamma} u^{\gamma} = 0 \ .
\label{antiseta}
\end{equation}

\subsection[Velocity's covariant derivative decomposition]{Decomposition of the covariant derivative of the four-velocity}
\label{veldec}

We can also define a \emph{fully orthogonally projected covariant derivative $\hat{\nabla}$} that, for any tensor $T^{\alpha \ldots \beta} _{\phantom{\alpha \ldots \beta} \gamma \ldots \delta}$, is:
\begin{equation}
\hat{\nabla}_{\mu}T^{\alpha \ldots \beta} _{\phantom{\alpha \ldots \beta} \gamma \ldots \delta}=h^{\alpha}_{\phantom{\alpha} \lambda }  h^{\rho}_{\phantom{\rho} \gamma } \ldots h^{\beta}_{\phantom{\beta} \nu } h^{\sigma}_{\phantom{\sigma} \delta } h^{\phi}_{\phantom{\phi} \mu}\nabla_{\phi} T^{\lambda \ldots \nu} _{\phantom{\lambda \ldots \nu} \rho \ldots \sigma} \ ,
\label{spatderiv}
\end{equation}
with total projection on all free indices. The <<hat>> over the symbol of the covariant derivative is used as a reminder of the fact that if $u_{\mu}$ has non-zero vorticity, then $\hat{\nabla}$ is not a proper three-dimensional covariant derivative.\footnote{See chapter \ref{fbthchapter} for an exhaustive discussion about the role of vorticity.}

It is also useful to denote the \emph{orthogonal projections of vectors} and the \emph{orthogonally projected symmetric trace-free part of tensors} with angle brackets, so we have
\begin{equation}
v^{\left\langle \alpha \right\rangle}=h^\alpha _{\phantom{\alpha} \beta} v^\beta \qquad
\text{and} \qquad
T^{\left\langle \alpha \beta \right\rangle}=\left[h^{(\alpha} _{\phantom{(\alpha }\gamma} h^{ \beta)} _{\phantom{ \beta)} \delta} -\frac{1}{3}h^{\alpha \beta} h_{\gamma \delta} \right]T^{\gamma \delta} \ .
\label{ortpro}
\end{equation}

Now we can split the covariant derivative of the four-velocity $u_\mu$ into its irreducible parts, defined by their symmetry properties:\footnote{Note that some authors define $\omega_{\alpha\beta}$ in a different way, and this can lead to a certain confusion. For instance, while \cite{syksy2009c} and \cite{buchert1996} use the same convention we use, i.e. $\omega_{\alpha\beta}\doteq\hat{\nabla}_{[\beta}u_{\alpha]}$, \cite{covform3} uses $\omega_{\alpha\beta}\doteq\hat{\nabla}_{[\alpha}u_{\beta]}$. In that case instead of equation \re{split} you obtain $\nabla_{\alpha} u_{\beta}=-u_\alpha \dot{u}_\beta+\hat{\nabla}_{\alpha}u_{\beta}=-u_\alpha \dot{u}_\beta+\frac{1}{3}\Theta h_{\alpha\beta} + \omega_{\alpha \beta} + \sigma_{\alpha\beta} $, and also some other equations differ in some signs from ours.}
\begin{equation}
u_{\alpha;\beta}=\nabla_{\beta} u_{\alpha}=-u_\beta \dot{u}_\alpha+\hat{\nabla}_{\beta}u_{\alpha}=-u_\beta \dot{u}_\alpha+\frac{1}{3}\Theta h_{\alpha\beta} + \omega_{\alpha \beta} + \sigma_{\alpha\beta} \ ,
\label{split}
\end{equation}
where $\Theta$ is the \emph{expansion rate}, $\omega_{\alpha\beta}$ the \emph{vorticity tensor} and $\sigma_{\alpha\beta}$ the \emph{shear tensor}.

\subsubsection{Generalized Hubble law}
Before stating the properties and the definitions of the quantities introduced by \re{split}, let us obtain two equations that are useful in clarifying their physical meaning.

Given a \emph{deviation vector} $\eta^{\alpha}$ for the family of fundamental world lines, which is defined as\footnote{See \cite[p. 46]{wald}.}
\begin{equation}
u^\alpha \nabla_\alpha \eta^\beta = \eta^\alpha \nabla_\alpha u^\beta \ ,
\end{equation}
a \emph{relative position vector} is obtained by using the projector \re{proh}
\begin{equation}
\eta^\alpha _{\bot}=h^\alpha _{\phantom{\alpha}\beta}\eta^\beta \ .
\end{equation}
Introducing a \emph{relative distance} $\delta l $ and a \emph{relative direction vector} $e^\alpha$ for which $e_\alpha e^\alpha = 1 $ and $e_\alpha u^\alpha = 0$, we can write the relative position vector in the form
\begin{equation}
\eta^\alpha _{\bot}= \delta le^\alpha \ .
\end{equation}
We can obtain the two propagation equations
\begin{equation}
   \frac{\dot{\delta l}}{\delta l}=\frac{1}{3}\Theta +\sigma_{\alpha\beta}e^\alpha e^\beta
   \label{genhubble}
\end{equation}
and
\begin{equation}
   \dot{e}^{\langle \alpha \rangle}= \left[ \sigma^\alpha _{\phantom{\alpha}\beta} - \left( \sigma_{\gamma\delta} e^\gamma e^\delta \right) h^\alpha _{\phantom{\alpha}\beta} - \omega^{\alpha} _{\phantom{\alpha} \beta} \right] e^\beta \ ,
   \label{prop}
\end{equation}
that give respectively the \emph{rate of change of relative distance} and the \emph{rate of change of direction}.

Considered in a cosmological model, equation \re{genhubble} is a \emph{generalized Hubble law}, allowing for possible anisotropic expansions.
It is valid for distances large enough to ensure that random velocities are small if compared with velocities associated with the general motion of matter, but small enough for the Hubble relation to be linear, and also for the change in distance of the galaxies to be relatively small during the time of light travel between the galaxies and the observer. Thus we might expect its range of validity to be roughly from 50 to 500 Mpc.
Equation \re{prop}, that we expect to be valid roughly on the same length scale, gives us the \emph{ rate of change of position in the sky} of neighboring clusters of galaxies, with respect to an observer at rest in a \emph{local inertial frame} (\emph{L.I.F.}).

Now we can come back to the kinematic quantities previously introduced, and clarify their meaning in light of these equations.

\subsubsection{Expansion rate}
\label{exprate}
The expansion rate $\Theta$ is a scalar quantity and it is defined as the trace of the velocity gradient, i.e.
\begin{equation}
\Theta \doteq \nabla_\mu u^\mu = \hat{\nabla}_\mu u^\mu \ .
\label{deftheta}
\end{equation}

Thinking of a sphere of fluid particles that changes according to \re{genhubble} during a small increment of proper time, it is easy to understand that $\Theta$ describes the isotropic volume expansion of that sphere. We may then define a representative length $l$ by the equation
\begin{equation}
\frac{\dot{l}}{l}=\frac{1}{3}\Theta \ ,
\end{equation}
which is nothing other than what we obtain from equation \re{genhubble} if the shear tensor vanishes. The quantity $l$, which represents completely the volume behavior of the fluid, in a Friedmann-Lema\^itre-Robertson-Walker model (where isotropy holds by construction) corresponds to the \emph{scale factor} $a(t)$.\footnote{For a brief review on FLRW models, see appendix A.}
From it we can define the \emph{Hubble parameter} to be 
\begin{equation}
H \left( t \right) \doteq \frac{\dot{l}}{l}=\frac{1}{3}\Theta \ ,
  \label{hubbleparam}
\end{equation}
that is the slope of the curve $ l \left( t \right) $. Let us call \emph{Hubble Constant} the value $H_0\doteq H\left( t_0 \right)$ assumed today by this parameter ($t_0$ being the \emph{present age of the universe}).\footnote{Note that the value of $t_0$ is strongly model-dependent, i.e. it can be obtained using the measured value of $H_0$, but in order to do that we must assume the validity of a particular cosmological model. In a \emph{$\Lambda$CDM} model we also need to know the value of the parameters $\Omega_M$, $\Omega_\Lambda$,\dots (defined in the appendix), but roughly we can obtain a value for $t_0$ that is $t_0\approx\hubbletime$, where we have used the value $\hubbleconstant$; see also section \ref{evouniv}.}

In comoving coordinates, the expansion rate expressed in terms of the Christoffel symbols turns out to be\footnote{Equation \re{thetacomcoord} is obtained using the expression of the covariant derivative in terms of the Christoffel symbols and the property $\Gamma^{\alpha}_{\phantom{\alpha}\beta\alpha}=\partial_\beta g / 2g = \partial_\beta \left( \ln \sqrt{\left|g\right|} \right)$. The same result can be obtained by the use of the expression for the divergence of a four-vector $\nabla_{\alpha} v^\alpha = \frac{1}{\sqrt{-g}} \partial_{\alpha} \left( \sqrt{-g} v^\alpha \right)$.}
  \begin{equation}
    \Theta = \partial_0 \left( \ln \sqrt{ \left| g \right| } \right)\ .
      \label{thetacomcoord}
  \end{equation}

\subsubsection{Vorticity}
\label{vorticitysection}
The vorticity tensor $\omega_{\alpha\beta}$ is defined as the antisymmetric part of the orthogonally projected covariant derivative of the velocity field, i.e.\footnote{Sometimes a different definition is used, see footnote 6 on the page \pageref{split}.}
\begin{equation}
\omega_{\alpha\beta}\doteq \hat{\nabla}_{[\beta} u_{\alpha]} \ ,
\label{defomega}
\end{equation}
so it is obvious that
\begin{equation}
\omega_{\alpha\beta}=\omega_{[ \alpha \beta ]} \ , \qquad
\omega_{\alpha \beta} u^\beta = 0 \ , \qquad
\omega^{\alpha}_{\phantom{\alpha}\alpha}=0 \ .
\label{propomega}
\end{equation}
From \re{defomega} it follows that the tensor $\omega_{\alpha\beta}$ has only three independent components, so instead of it we can use without loss of information the \emph{vorticity vector}
\begin{equation}
\omega^\alpha \doteq \frac{1}{2}\eta^{\alpha\beta\gamma}\omega_{\beta\gamma} \ ,
\label{vorticityvector}
\end{equation}
from which the vorticity tensor can be obtained by
\begin{equation}
\omega_{\alpha\beta}=\eta_{\alpha\beta\gamma}\omega^{\gamma} \ ,
\end{equation}
and we have
\begin{equation}
\omega_\alpha u^\alpha = 0 \ .
\end{equation}
We can also define a scalar quantity that is used in the sequel: the \emph{vorticity scalar}\footnote{Note that $\omega$ is not the trace of $\omega_{\alpha\beta}$, which is anyway traceless.}
\begin{equation}
\omega \doteq \frac{1}{2} \left( \omega_{\alpha\beta}\omega^{\alpha\beta} \right)^{\frac{1}{2}}= \left( \omega_\alpha \omega^\alpha \right)^{\frac{1}{2}} \ .
\label{omegascal}
\end{equation}
Note that the conditions of \ev{vanishing vorticity tensor ($\omega_{\alpha\beta}=0$)}, \ev{vanishing vorticity vector ($\omega^{\alpha}=0$)} and \ev{vanishing vorticity scalar ($\omega=0$)} are equivalent, as can be seen from their definitions, i.e.:
\begin{equation}
\omega_{\alpha\beta}=0  \Leftrightarrow  \omega_\alpha=0  \Leftrightarrow  \omega=0 \ .
\end{equation}

The vorticity tensor $\omega_{\alpha\beta}$ determines a rigid rotation of our sphere of fluid with respect to a local inertial frame. The vorticity vector $\omega_\alpha$ makes it more clear: its direction is the axis of rotation of the matter, because it is the only one left unchanged by the action of vorticity alone.\footnote{For a wide discussion about the role of vorticity see chapter 3.}

In comoving coordinates the vorticity turn out to be
  \begin{equation}
    \omega_{0\alpha}=0 \ ,  \qquad \omega_{ij}=\partial_{[j} u_{i]} \ ,
  \end{equation}
where $u_i=g_{0i}$.

\subsubsection{Shear}
The shear tensor is defined as the trace-free symmetric part of the spatial projection of the covariant derivative of the velocity vector field, i.e.
\begin{equation}
\sigma_{\alpha\beta}\doteq\hat{\nabla}_{ ( \beta } u_{ \alpha ) } \ .
\end{equation}
From this equation we can easily obtain the properties
\begin{equation}
\sigma_{(\alpha\beta)}=\sigma_{\alpha\beta} \ , \qquad
\sigma_{\alpha\beta}u^\beta=0 \ , \qquad
\sigma^\alpha _{\phantom{\alpha}\alpha}=0 \ ,
\end{equation}
and we can define the scalar quantity $\sigma$, the \emph{shear scalar}, as 
\begin{equation}
\sigma \doteq \left( \frac{1}{2} \sigma_{\alpha\beta} \sigma^{\alpha\beta} \right)^{\frac{1}{2}} \ .
\end{equation}
Note that
\begin{equation}
\sigma=0 \Leftrightarrow \sigma_{\alpha\beta}=0 \ .
\end{equation}

The action of the tensor $\sigma_{\alpha\beta}$ determines the distortion of a sphere of fluid particles, leaving the volume and the principal axis of shear (i.e. the eigenvectors of the shear tensor) unchanged, while all other directions change.

In comoving coordinates we can express the shear as
  \begin{equation}
    \sigma_{\alpha\beta}=-\left( \Gamma^0 _{\phantom{0}\alpha\beta} + \delta_{\alpha}^{\phantom{\alpha} 0} \Gamma^0_{\phantom{0}\beta 0 } 
    + \Gamma^0_{\phantom{0} \alpha 0} \delta_{\beta} ^{\phantom{\beta}0 } 
    + \delta_{\alpha }^{\phantom{\alpha}0}\delta_{\beta}^{\phantom{\beta} 0} \Gamma^0_{\phantom{0}00}\right) \ .
  \end{equation}

\section{The curvature tensor}
\label{curvtenssect}

In general relativity the curvature of spacetime is described by the \emph{Riemann curvature tensor}, whose components satisfy the properties:\footnote{This relations are proved in \cite[p. 142]{jost}, where can also be found the definition of the Riemann tensor as the \emph{curvature tensor} of the \emph{Levy-Civita connection}. Note that equation \re{ric3} is usually called the \emph{first Bianchi identity}.}
\begin{subequations}
  \begin{align}
    R_{ [ \alpha\beta ] [ \gamma\delta ] } &= R_{\alpha\beta\gamma\delta} \ ,
      \label {ric1} \\
    R_{\alpha\beta\gamma\delta}&=R_{\gamma\delta\alpha\beta} \ ,
      \label{ric2} \\
    R_{\alpha [ \beta\gamma\delta] } &= 0 \ .
      \label{ric3}
  \end{align}
\end{subequations}

This tensor, which possesses 20 independent components, can be algebraically separated into the \emph{Ricci tensor} $R_{\alpha\beta}$, defined as
\begin{equation}
  R_{\alpha\beta} \doteq R^\gamma _{\phantom{\gamma} \alpha \gamma \beta} = R_{\alpha\phantom{\gamma}\beta\gamma} ^{\phantom{\alpha} \gamma} \ ,
   \label{riccitensor}
\end{equation}
and the Weyl tensor (often called the \emph{conformal curvature tensor}), whose components are defined by\footnote{Here we are working in four dimensions, for a $N$-dimensional version of this definition see \cite[p. 40]{wald}.}
\begin{equation}
  C^{\alpha\beta}_{\phantom{\alpha\beta}\gamma\delta} \doteq R^{\alpha \beta} _{\phantom{\alpha\beta}\gamma \delta} -2g^{[\alpha}_{\phantom{[\alpha}\gamma} R^{\beta ]}_{\phantom{\beta [ }\delta}
  +\frac{R}{3}g^{[\alpha}_{\phantom{[\alpha}\gamma} g^{\beta ]}_{\phantom{\beta [ }\delta} \ ,
\end{equation}
where $R$ is the \emph{Ricci scalar} defined as
\begin{equation}
  R \doteq R^\alpha_{\phantom{\alpha}\alpha} \ .
    \label{ricciscalar}
\end{equation}

The Ricci tensor $R_{\alpha\beta}$ possesses 10 independent components, and from \re{ric2} and its definition it follows that
\begin{equation}
  R_{\alpha\beta}=R_{\beta\alpha} \ ,
\end{equation}
i.e. the Ricci tensor is symmetric.
It describes the amount by which the volume element of a geodesic ball on our Riemannian manifold $\left( \mathcal{M},g\right)$ deviates from that of the standard ball in Euclidean space. We can think of it as the trace part of $R_{\alpha\beta\gamma\delta}$.

The Weyl tensor $C_{\alpha\beta\gamma\delta}$ also has 10 independent components, and it conveys the part of information contained in the Riemann tensor that describes how the shape of a body is distorted by tidal forces when moving along geodesics, but not information about changes in the volume.
It possesses all the symmetries of the Riemann tensor, plus the additional property
\begin{equation}
  C^{\alpha\beta}_{\phantom{\alpha\beta}\alpha\gamma}=0 \ ,
\end{equation}
and we can think of it as the trace-free part of the Riemann tensor.

The Weyl tensor is the only part of the curvature tensor that exists in free space, a solution of the vacuum Einstein equation,\footnote{More generally, the Weyl tensor is the only part of the curvature tensor that exists for Ricci-flat manifolds. It vanishes identically in dimensions 2 and 3, while in dimensions $\geq4$ it is in general non-zero.} so it represents the \emph{free gravitational field}, enabling gravitational action at a distance, and describing tidal forces and gravitational waves, 
while the Ricci tensor $R_{\alpha\beta}$ is determined locally at each point by the energy-momentum tensor through the Einstein equation, and it vanishes identically in the vacuum case.\footnote{This is true only for vanishing cosmological constant, otherwise the Ricci tensor in the vacuum case is proportional to the metric tensor, see \re{vacuumricci}.}
The tensors $C_{\alpha\beta\gamma\delta}$ and $R_{\alpha\beta}$ together completely represent the Riemann curvature tensor $R_{\alpha\beta\gamma\delta}$, which can be decomposed as\footnote{See for instance \cite{hawking}.}
  \begin{equation}
    R_{\alpha\beta\gamma\delta}=C_{\alpha\beta\gamma\delta}-g_{\alpha [ \delta}R_{\gamma ] \beta} 
    - g_{\beta [ \gamma} R_{\delta ] \alpha} - \frac{R}{3} g_{\alpha [ \gamma} g_{\delta] \beta}\ .
      \label{splitriemann2}
  \end{equation}

The Weyl tensor can be split relative to $u^{\alpha}$ into the \emph{electric Weyl curvature part}:
\begin{equation}
  E_{\alpha\beta} \doteq C_{\alpha\beta\gamma\delta} u^\gamma u^\delta \ ,
\end{equation}
for which
\begin{equation}
  E^{\alpha}_{\phantom{\alpha}\alpha} = 0 \ , \qquad
  E_{\alpha\beta}=E_{\left( \alpha \beta \right)} \ , \qquad
  E_{\alpha\beta} u^{\beta} = 0 \ ,
\end{equation}
and the \emph{magnetic Weyl curvature part}:
\begin{equation}
  H_{\alpha\beta} \doteq \frac{1}{2} \eta_{\alpha\gamma\delta} C^{\gamma\delta}_{\phantom{\gamma\delta}\beta\nu} u^{\nu} \ ,
\end{equation}
for which
\begin{equation}
  H^{\alpha}_{\phantom{\alpha}\alpha} = 0 \ , \qquad
  H_{\alpha\beta}=H_{\left( \alpha \beta \right)} \ , \qquad
  H_{\alpha\beta} u^{\beta} = 0 \ .
\end{equation}
The quantities $E_{\alpha\beta}$ and $H_{\alpha\beta}$ are projected orthogonal to $u^\mu$ by construction, so they are what an observer moving with four-velocity $u^\mu$ measures.\footnote{If we consider an observer that is moving with a different four-velocity, the tensor $C_{\alpha\beta\gamma\delta}$ is the same, but he will measure a different value of $E_{\alpha\beta}$ and $H_{\alpha\beta}$. See also section \ref{prosection}.}

Expressed in terms of $E_{\alpha\beta}$ and $H_{\alpha\beta}$, the Weyl tensor $C_{\alpha\beta\gamma\delta}$ reads:
  \begin{equation}
    C_{\alpha\beta}^{\phantom{\alpha\beta}\gamma\delta}=
    4 \left( u_{ [ \alpha}u^{ [ \gamma}+h_{ [ \alpha}^{\phantom{ [ \alpha} [ \gamma} \right)
    E_{\beta ] }^{\phantom{\beta ] }\delta ] } + 2 \eta_{\alpha\beta\epsilon}u^{ [ \gamma} 
    H^{ \delta ] \epsilon } + 2u_{ [ \alpha} H_{\beta ] \epsilon} \eta^{\gamma\delta\epsilon}\ ,
      \label{splitweyl}
  \end{equation}
and an equivalent expression can be found in \cite[p. 11]{ilcerchiosichiude}.

\section{Dynamics}

\subsection{The Einstein field equation}
\label{efesection}

\begin{citazione}
\emph{Space tells matter how to move, matter tells space how to curve.}\footnote{Quoted from \cite[p. 5]{gravitation}.}
\end{citazione}
Throughout this work we assume the validity of Einstein's general relativity as the theory that describes the geometry of spacetime.
In this theory the geometry of spacetime is specified once given a metric tensor $g_{\alpha\beta} \left( x^\mu \right) $. The behavior of matter is described by the \emph{energy-momentum tensor} which contains information about each matter component and their non-gravitational interactions.
On the other hand, the interaction between geometry and matter, i.e. how matter determines the geometry, which in turn determines the motion of matter, is encoded in the Einstein equation.

The Einstein field equation\footnote{The wording ``field equation'' was particularly dear to Einstein, and this is the reason we use it in this section. In the rest of this work we refer to equation \re{efe} simply as the ``Einstein equation''.} (E.F.E.) is:\footnote{According to our choice of inserting in the equations the cosmological constant, this version of the Einstein equation contains the additional term $\Lambda g_{\alpha\beta}$. For vanishing cosmological constant we obtain the original version of the Einstein equation, $G_{\alpha\beta}=T_{\alpha\beta}$. As explained in chapter 1, in order to account for acceleration we can modify the geometric side of this last equation (modified gravity approach), obtaining $G_{\alpha\beta}+ T_{\alpha\beta}^{'dark}=T_{\alpha\beta}$, or the matter side (dark energy approach), that gives $G_{\alpha\beta}=T_{\alpha\beta}+T_{\alpha\beta}^{''dark}$ (see \cite{deandmg}). The case of the cosmological constant corresponds to the choice $T_{\alpha\beta}^{'dark}=\Lambda g_{\alpha\beta}$ that gives equation \re{efe}.}
\begin{equation}
  G_{\alpha\beta} \doteq R_{\alpha\beta}-\frac{1}{2}g_{\alpha\beta}R=T_{\alpha\beta}-\Lambda g_{\alpha\beta} \ ,
    \label{efe}
\end{equation}
where $G_{\alpha\beta}$ is called the \emph{Einstein tensor}, $R_{\alpha\beta}$ is the Ricci tensor, $R$ is the Ricci scalar, $T_{\alpha\beta}$ the \emph{energy-momentum tensor} and we have included $\Lambda$, the cosmological constant, for the sake of generality.
This equation is equivalent to a set of coupled non-linear second order partial differential equations for the components $g_{\alpha\beta}$ of the metric. Each of the tensor quantities that it relates possesses 10 independent components, so we have 10 equations, but because of the 4 differential \emph{twice contracted Bianchi identities}
\begin{equation}
  \nabla^\alpha G_{\alpha\beta}=0
    \label{twicebianchi}
\end{equation}
we can reduce to 6 the number of independent equations by the choice of the gauge (i.e. of coordinates).\footnote{For more details see \cite[pp. 259--260]{wald}.}

By taking the covariant derivative of the Einstein equation, substituting the twice contracted Bianchi identities \re{twicebianchi}, and remembering that 
\begin{equation}
  \nabla^\alpha g_{\alpha\beta}=0
\end{equation}
by definition and that $\Lambda$ is defined as a constant in space and time, i.e.
\begin{equation}
  \nabla^{\alpha} \Lambda = 0 \ ,
\end{equation}
we easily obtain the equation\footnote{In a local inertial frame (where there is no gravity) and in absence of external forces, the conservation equations $\partial^{\alpha}T_{\alpha\beta}=0$ hold \cite[p. 155]{rindler}. If we write the equations \re{evot} using the expression of the covariant derivative in terms of the Christoffel symbols $\nabla_\gamma T_{\alpha\beta} = \partial_\gamma T_{\alpha\beta} - \Gamma^{\delta}_{\alpha\gamma}T_{\delta\beta}-\Gamma^{\delta}_{\beta\gamma}T_{\alpha\delta}$ we obtain $\partial^{\alpha}T_{\alpha\beta}=\Gamma^{\mu}_{\beta\nu} T^{\nu}_{\phantom{\nu}\mu}+g^{\alpha\nu}\Gamma^\mu_{\alpha\nu}T_{\mu\beta}$, so we can think of \re{evot} as an evolution equation, rather than a conservation equation.}
\begin{equation}
  \nabla^\alpha T_{\alpha\beta} = 0 \ .
    \label{evot}
\end{equation}

In the case of vanishing energy-momentum tensor the Einstein equation reduces to the \emph{vacuum field equation}
\begin{equation}
  R_{\alpha\beta}-\frac{1}{2}g_{\alpha\beta}R+g_{\alpha\beta}\Lambda=0 \ .
    \label{vacuumefe}
\end{equation}
Contracting these equations by the metric we obtain that\footnote{Manifolds with vanishing Ricci tensor are called \emph{Ricci-flat manifolds} while manifolds with a Ricci tensor proportional to the metric are known as \emph{Einstein manifolds} (see \cite[p. 145]{jost}).}
\begin{equation}
  R=4\Lambda,\qquad\text{or equivalently}\qquad R_{\alpha\beta}=g_{\alpha\beta}\Lambda \ .
    \label{vacuumricci}
\end{equation}
Examples of \emph{vacuum solutions} (i.e. solutions of the vacuum field equations) are the \emph{Minkowski spacetime} used in special relativity, the  \emph{Schwarzschild solution} that describes static black holes and the \emph{Kerr solution} for rotating black holes.

\subsection{The 3+1 decomposition of the energy-momentum tensor}
\label{tdec}

The energy-momentum tensor $T_{\alpha\beta}$ (sometimes called \emph{stress-energy tensor}, \emph{stress-energy-momentum tensor} or \emph{matter tensor}) describes the density and flux of energy and momentum in spacetime, and enters the Einstein equation as the source term.

The tensor $ T_{\alpha\beta} $ can be decomposed with respect to $ u^\alpha $ as
\begin{equation}
  T_{\alpha\beta} = \mu u_\alpha u_\beta + q_\alpha u_\beta + u_\alpha q_\beta + p h_{\alpha\beta} + \pi_{\alpha\beta} \ .
    \label{ttensor}
\end{equation}
In the above decomposition, $\mu$ is the \emph{total energy density} of matter relative to $u^\alpha$, defined as
\begin{equation}
  \mu \doteq T_{\alpha\beta}u^\alpha u^\beta \ .
\end{equation}
The \emph{isotropic pressure} $p$ is defined as
\begin{equation}
  p\doteq \frac{1}{3}T_{\alpha\beta}h^{\alpha\beta} \ ,
\end{equation}
while the \emph{anisotropic pressure} (or \emph{anisotropic stress}) $\pi_{\alpha\beta}$ as
\begin{equation}
  \pi_{\alpha\beta}\doteq T_{\gamma\delta} h^\gamma _{\phantom{\gamma} \langle \alpha} h^{\delta} _{\phantom{\delta} \beta \rangle} \ ,
\end{equation}
from which we obtain the following properties:
\begin{equation}
  \pi_{\alpha\beta}=\pi_{\left( \alpha \beta \right)} \ , \qquad
  \pi^\alpha_{\phantom{\alpha}\alpha} = 0 \ , \qquad
  \pi_{\alpha\beta}u^\alpha = 0 \ .
\end{equation}
The \emph{relativistic momentum density} $q^\alpha$ is defined as
\begin{equation}
  q^\alpha \doteq - T_{\beta\gamma}u^\beta h^{\gamma \alpha} \ ,
\end{equation}
it is spacelike, in the sense that
\begin{equation}
  q_\alpha u^\alpha = 0 \ ,
\end{equation}
and it represents also the energy flux relative to $u^\alpha$.

\subsection{The 3+1 decomposition of the Riemann tensor}

Now the Riemann tensor $R_{\alpha\beta\gamma\delta}$ can be put into a fully $\left(3+1\right)$-decomposed form.
In order to do this, we consider its decomposition \re{splitriemann2} in terms of the Weyl tensor $C_{\alpha\beta\gamma\delta}$ and the Ricci tensor $R_{\alpha\beta}$. Using \re{splitweyl} we can write $C_{\alpha\beta\gamma\delta}$ in terms of the projected quantities $E_{\alpha\beta}$ and $H_{\alpha\beta}$. With the Einstein equation \re{efe} we express $R_{\alpha\beta}$ in terms of the energy-momentum tensor. Finally we insert the $3+1$ decomposition \re{ttensor} of $T_{\alpha\beta}$.
We obtain:\footnote{Here $P$ stands for the perfect fluid part (i.e. ideal fluid part), $I$ for the imperfect fluid part, $E$ marks the part due to the electric Weyl curvature and $H$ that due to the magnetic Weyl curvature.}
\begin{equation}
  R^{\alpha\beta}_{\phantom{\alpha\beta}\gamma\delta} = R^{\alpha\beta}_{P\phantom{\beta}\gamma\delta} + R^{\alpha\beta}_{I\phantom{\beta}\gamma\delta}
  + R^{\alpha\beta}_{E\phantom{\beta}\gamma\delta} + R^{\alpha\beta}_{H\phantom{\beta}\gamma\delta} \ ,
    \label{riemanndec}
\end{equation}
where we have defined
\begin{subequations}
    \label{splitriemann}
  \begin{align}
    R^{\alpha\beta}_{P\phantom{\beta}\gamma\delta} &\doteq \frac{2}{3} \left( \mu+3p-2\Lambda \right)
    u^{[\alpha}u_{[\gamma}h^{\beta]}_{\phantom{\beta]}\delta]}+\frac{2}{3}\left(\mu+\Lambda\right)
    h^\alpha_{\phantom{\alpha}[\gamma}h^{\beta}_{\phantom{\beta}\delta]} \ ,
      \label{perf} \\ 
    R^{\alpha\beta}_{I\phantom{\beta}\gamma\delta} &\doteq -2u^{[\alpha}h^{\beta]}_{\phantom{\beta]}[\gamma}
    q_{\delta]}-2u_{[\gamma}h^{[\alpha}_{\phantom{[\alpha}\delta]}q^{\beta]}-2u^{[\alpha}u_{[\gamma}
    \pi^{\beta]}_{\phantom{\beta]}\delta]}+2h^{[\alpha}_{\phantom{[\alpha}[\gamma}\pi^{\beta]}_{\phantom{\beta]}\delta]} \ ,
      \label{imp} \\
    R^{\alpha\beta}_{E\phantom{\beta}\gamma\delta} &\doteq
    4u^{[\alpha}u_{[\gamma}E^{\beta]}_{\phantom{\beta]}\delta]}+4h^{[\alpha}_{\phantom{[\alpha}[\gamma}E^{\beta]}_{\phantom{\beta]}\delta]} \ ,
      \label{elect} \\
    R^{\alpha\beta}_{H\phantom{\beta}\gamma\delta} &\doteq 2\eta^{\alpha\beta\epsilon}u_{[\gamma}
    H_{\delta]\epsilon}+2\eta_{\gamma\delta\epsilon}u^{[\alpha}H^{\beta]\epsilon} \ .
      \label{magn}
  \end{align}
\end{subequations}
The former two terms arise from the decomposition of the Ricci tensor, and the latter two from the decomposition of the Weyl tensor.

\subsection{Energy conditions and equations of state}
\label{eqstate}

The Einstein equation does not give information on the form of the energy-momentum tensor, i.e. it does not specify which kinds of states of matter (or of non-gravitational fields) that it describes are admissible in a description of the universe. This allows more generality, because in this way we can describe how gravity works for arbitrary forms of matter, but on the other hand it means that without any further criterion the Einstein equation admits solutions with properties that don't seem to resemble anything in the real universe, even approximately.

For this reason usually one or more of the following conditions are imposed (see \cite[p. 218]{wald}):
\begin{itemize}
  \item the \emph{weak energy condition}:
    \begin{equation}
      \mu=T_{\alpha\beta}u^\alpha u^\beta \geq 0 \ ,
    \end{equation}
    i.e. the requirement that the total energy density not be negative;
  \item the \emph{strong energy condition}:
    \begin{equation}
     \mu + 3p = T_{\alpha\beta} u^\alpha u^\beta + \frac{1}{2}T \geq 0 \ ;
       \label{strong}
    \end{equation}
  \item the \emph{dominant energy condition}:\footnote{A vector $\textbf{v}$ is \emph{causal} when $||\textbf{v}||=g\left( \textbf{v},\textbf{v}\right)=g_{\mu\nu}v^\mu v^\nu\leq0$, i.e. when it is either \emph{timelike} ($||\textbf{v}||<0$) or \emph{lightlike} ($||\textbf{v}||=0$). A causal vector $\textbf{v}$ at a point $p$ of the manifold $\left(\mathcal{M},g\right)$ is future pointing when it lies in the conventionally named <<future half>> of the \emph{\mbox{light cone of $p$}}. Note that the light cone of $p$ is defined as the light cone passing through the origin of the tangent space $T_p\mathcal{M}$ to the manifold $\mathcal{M}$ at the point $p$, that is isomorphic to Minkowsky spacetime (see \cite[p. 189]{wald}).}
    \begin{equation}
      \begin{split}
        &-T_{\alpha\beta}u^\beta \,
        \text{is a future-pointing causal vector} \\
        &\forall \, u^\alpha \,
        \text{that is a future-pointing causal vector},
      \end{split}
    \end{equation}
\end{itemize}
i.e. mass-energy can never flow faster than light.
The physical (thermodynamical) description of the fluid lies also in the \emph{equation of state} that relates these quantities, which depends on the type of matter we consider.

\subsection{Particular fluids}
  \label{partfluids}

Often a particular physical situation may be well described by using, instead of the \emph{general matter} form of the energy-momentum tensor given in equation \re{ttensor}, a particular kind of fluid represented by a simpler form of $T_{\alpha\beta}$.
An interesting case is an \emph{ideal fluid},\footnote{While in thermodynamics it is common to use the name \emph{perfect fluid}, in cosmology this kind of fluid is usually called an ideal fluid, and we follow this custom.} and especially the \emph{dust} sub-case.

\subsubsection{Ideal fluid}

An ideal fluid is characterized by the matter tensor
\begin{equation}
  T_{\alpha\beta}=\mu u_\alpha u_\beta + p h_{\alpha \beta} \ ,
  \label{perfectfluid}
\end{equation}
which is obtained from equation \re{ttensor} by choosing $q_\alpha=0=\pi_{\alpha\beta}$.

After this statement a question may arise: what is the meaning of the description carried by this simpler form of the energy-momentum tensor? A fluid is in general described through the form \re{ttensor} of the energy-momentum tensor $T_{\alpha\beta}$. The pressure $p$, the energy flux $q_\alpha$ and anisotropic pressure $\pi_{\alpha\beta}$ describe the way the <<fluid's elements>> interact with each other. If $q_\alpha$ and $\pi_{\alpha\beta}$ are much smaller than $\mu$ and $p$, then we can approximately consider them to vanish, so we can use the form \re{perfectfluid} of the energy-momentum tensor as an approximation of the correct form \re{ttensor}.

An \emph{equation of state} is a relation that expresses $p$ as a function of other dynamical quantities, such as $\mu$. Let us recall, as an example, that the main component of the \emph{radiation content} of the universe (the cosmic microwave background) can be represented as a ideal fluid with the equation of state\footnote{See \cite[p. 599]{covform1}; other examples are given by equation \re{eqstateflrw}.}
\begin{equation}
  p= \frac{1}{3} \mu \ .
\end{equation}

\subsubsection{Dust}
\label{dustsect}

The dust case, also referred to as \emph{pressure-free matter} or \emph{cold dark matter}, is characterized by
\begin{equation}
  T_{\alpha\beta}=\mu u_\alpha u_\beta \ .
    \label{tdust}
\end{equation}
In this case we are considering a fluid for which also $p$ is much smaller than $\mu$, so we can neglect it too and describe the fluid with the energy-momentum tensor given by equation \re{tdust}.
In this case we have\footnote{These two equations arise by inserting the above form of the energy-momentum tensor in the equations \re{enconseq} and \re{momconseq} respectively.}
\begin{equation}
  \mu \propto a^{-3} \qquad
  \text{and} \qquad
  \dot{u}_{\alpha}=0 \ .
    \label{dotuzero}
\end{equation}

\section{Evolution and constraint equations}
\label{evoeconsteqsect}

There are three sets of equations that arise from the Einstein equation, one set using the \emph{Ricci identities}, the others using the \emph{Bianchi identities} respectively contracted once and twice.\footnote{These equations are briefly summarized in \cite[pp. 9-12]{covform3}, and \cite{syksy2009c}, while here we also show how we obtain them.}

\subsection{Ricci identities}
\label{fromricci}
From the definition of the Riemann tensor (and sometimes <<as>> the definition of the Riemann tensor, see \cite[p. 37]{wald}) for the vector field $u^\alpha$ we have
\begin{equation}
  2\nabla_{[\alpha} \nabla_{\beta ]} u^\gamma = R^{\phantom{\alpha\beta}\gamma}_{\alpha\beta\phantom{\gamma}\delta}u^\delta \ ,
  \label{ricciid}
\end{equation}
known as Ricci identities.
To extract the physical information stored in these identities, we project them along the world lines originated by $u_\alpha$ by use of the projector $U_{\alpha\beta}$, and into the hypersurfaces orthogonal to $u_\alpha$ using $h_{\alpha\beta}$. Separating the two equations that we obtain into \emph{trace}, \emph{antisymmetric part} and \emph{symmetric trace-free part} we have the following six equations.

\subsubsection{Propagation equations}
Projecting \re{ricciid} parallel to the vector field $u_\alpha$ by the projector \re{prou} we obtain three evolution equations for the three quantities $\Theta$, $\omega^\alpha$ and $\sigma^{\alpha\beta}$.
\begin{itemize}
	\item
	  By taking the trace and using in turn the expression \re{split} for the gradient of the four-velocity, the Einstein equation \re{efe},
	  the decomposition \re{ttensor} of the matter tensor and the property of the metric tensor 
	  \mbox{$g_{\alpha\beta}g^{\alpha\gamma}= \delta_{\beta}^{\phantom{\beta}\gamma}$}, we obtain the \emph{Raychaudhuri equation}:
	    \begin{equation}
	      \dot{\Theta}+\frac{1}{3}\Theta^2 = - \frac{1}{2} \left( \mu +3p\right)-2\sigma^2 +2\omega^2 
	      + \hat{\nabla}_\alpha \dot{u}^\alpha + \dot{u}_\alpha \dot{u}^\alpha + \Lambda \ .
	        \label{raycheq}
	    \end{equation}
	    This is the basic equation of gravitational attraction; it also shows the repulsive nature of a positive cosmological constant
	    and leads to the identification of $ \left( \mu +3p \right) $ as the active gravitational mass density 
	    (see \cite{raych} for the original derivation).
	\item
	  By taking the antisymmetric part and using the decomposition of the velocity gradient \re{split} we get the \emph{vorticity propagation equation}:
	  \begin{equation}
	    \dot{\omega}^{ \langle \alpha \rangle} = - \frac{2}{3} \Theta \omega^\alpha + \sigma^\alpha_{\phantom{\alpha}\beta}\omega^{\beta}
	    - \frac{1}{2} \eta^{\alpha\beta\gamma} \hat{\nabla}_{\beta}\dot{u}_{\gamma} \ ,
	      \label{vortpropeq}
	  \end{equation}
	  which is independent of the matter content and of the Einstein equation. In particular this equation shows that, if vorticity vanishes initially for a perfect fluid with \emph{barotropic} equation of state (i.e. \mbox{$p=p\left(\mu\right)$}), it will remain zero throughout the whole time evolution.\footnote{This can be see from equation \re{momconseq}.}
	\item
	  By taking the spatially projected symmetric trace-free part through the use of \re{ortpro}, inserting again the 
	  Einstein equation \re{efe} and the decomposition \re{ttensor} of the matter tensor, 
	  and splitting the Riemann tensor as in \re{splitriemann}, we obtain the \emph{shear propagation equation}:
	  \begin{equation}
	   \begin{split}
	    \dot{\sigma}^{\langle \alpha \beta \rangle} =& \hat{\nabla}^{\langle \alpha} \dot{u} ^{\beta \rangle} -\frac{2}{3}\Theta\sigma^{\alpha\beta}
	    +\dot{u}^{\langle\alpha} \dot{u}^{\beta\rangle} -\sigma^{\langle\alpha}_{\phantom{\langle\alpha}\gamma}\sigma^{\beta\rangle\gamma} \\
	    &-\omega^{\langle\alpha}\omega^{\beta\rangle}-E^{\alpha\beta}+\frac{1}{2}\pi^{\alpha\beta} \ ,
	      \label{shearpropeq}
	   \end{split}
	  \end{equation}
	  where we can see that the Weyl tensor (which represents tidal gravitational forces) induces shear through $E_{\alpha\beta}$,
	  that then feeds into the Raychaudhuri and vorticity propagation equations, influencing the nature of the fluid flow.
\end{itemize}

\subsubsection{Constraint equations}

Now we project the Ricci identities \re{ricciid} orthogonally to the velocity vector field $u_\alpha$ with the projector $h_{\alpha\beta}$ (projecting over every free index). Separating the trace, the symmetric trace-free part and the antisymmetric part we have three constraint equations.
\begin{itemize}
  \item
    Taking the trace, using again \re{split}, \re{efe}, \re{ttensor} and the split \re{splitriemann}, 
    we obtain the \emph{shear divergence constraint}:
    \begin{equation}
      \hat{\nabla}_{\beta}\sigma^{\alpha\beta}=\frac{2}{3}\hat{\nabla}^{\alpha}\Theta -\eta^{\alpha\beta\gamma}\left(\hat{\nabla}_{\beta}\omega_{\gamma}
      +2\dot{u}_{\beta}\omega_{\gamma}\right) - q^\alpha \ ,
    \end{equation}
    which shows how the momentum flux $q^\alpha$ relates to the spatial inhomogeneity of the expansion.
  \item
    Taking the antisymmetric part and inserting the expression \re{split} for the covariant derivative of the four-velocity 
    we have the following \emph{vorticity divergence constraint}:
      \begin{equation}
        \hat{\nabla}_\alpha\omega^\alpha = \dot{u}_\alpha\omega^\alpha \ ,
          \label{vortdiveq}
      \end{equation}
    which is independent of the matter tensor and the Einstein equation.
  \item
    Taking the symmetric trace-free part as we did for \re{shearpropeq}, inserting \re{split} 
    and the split of the Riemann tensor \re{splitriemann} we obtain the \emph{magnetic constraint}:
      \begin{equation}
        H_{\alpha\beta} = 2 \dot{u}_{\langle\alpha}\omega_{\beta\rangle}+\hat{\nabla}_{\langle\alpha}\omega_{\beta\rangle}
        +\eta_{\gamma\delta\langle\alpha}\hat{\nabla}^\gamma\sigma^\delta_{\phantom{\delta}\beta\rangle} \ ,
      \end{equation}
      which characterizes the magnetic Weyl curvature, and like \re{vortpropeq} and \re{vortdiveq} is independent of the matter content and of the Einstein equation.
\end{itemize}

\subsection{Bianchi identities}
\label{frombianchi}

The Riemann tensor satisfies the Bianchi identities as follows:\footnote{Sometimes these identities are called \emph{second Bianchi identity}, in order to distinguish them from the first Bianchi identity \re{ric3}. They are proved in \cite[p. 143]{jost}.}
  \begin{equation}
    \nabla_{[\epsilon}R_{\gamma\delta ] \alpha\beta}=0 \ .
      \label{bianchi}
  \end{equation}
By inserting the splitting \re{splitriemann2} of the Riemann tensor, using the Einstein equation \re{efe} and contracting (over the indices $\epsilon$ and $\delta$), the once contracted Bianchi identities are found to be:
  \begin{equation}
    \nabla^{\delta}C_{\alpha\beta\gamma\delta}+\nabla_{[ \alpha}R_{\beta ] \gamma} + \frac{1}{6} \delta_{\gamma [ \alpha} \nabla_{\beta ] }R =0 \ .
  \end{equation}
  
In a similar manner to the Ricci identities, we can extract the information stored in the above equation by projecting it along the world lines originated by $u_{\alpha}$, and on the orthogonal hypersurfaces. These projections yield, respectively, two propagation and two constraint equations:
  \begin{align}
    h^\alpha_{\phantom{\alpha}\langle \mu} h^\gamma_{\phantom{\gamma}\nu\rangle}u^\beta\left( \nabla^{\delta}C_{\alpha\beta\gamma\delta}
    +\nabla_{[ \alpha}R_{\beta ] \gamma} + \frac{1}{6} \delta_{\gamma [ \alpha} \nabla_{\beta ] }R \right) &=0 \ ,
      \label{aaa} \\
    \eta_{\lambda\sigma\langle\mu}h^{\gamma}_{\phantom{\gamma}\nu\rangle}h^{\lambda\alpha}h^{\sigma\beta}
     \left( \nabla^{\delta}C_{\alpha\beta\gamma\delta}
    +\nabla_{[ \alpha}R_{\beta ] \gamma} + \frac{1}{6} \delta_{\gamma [ \alpha} \nabla_{\beta ] }R \right) &=0 \ ,
      \label{bbb} \\
    h^{\alpha}_{\phantom{\alpha}\mu}h^{\gamma\beta} \left( \nabla^{\delta}C_{\alpha\beta\gamma\delta}
    +\nabla_{[ \alpha}R_{\beta ] \gamma}+ \frac{1}{6} \delta_{\gamma [ \alpha} \nabla_{\beta ] }R \right) &=0 \ ,
      \label{ccc} \\
    \eta_{\lambda\sigma\mu}u^\gamma h^{\lambda \alpha}h^{\sigma\beta}\left( \nabla^{\delta}C_{\alpha\beta\gamma\delta}
    +\nabla_{[ \alpha}R_{\beta ] \gamma} + \frac{1}{6} \delta_{\gamma [ \alpha} \nabla_{\beta ] }R \right) &=0 \ .
      \label{ddd}
  \end{align}

The below sets of equations are explicitly determined by inserting in equations \re{aaa}--\re{ddd} the splitting of the Weyl tensor that arises from \re{splitriemann}, and the expression of the Ricci tensor that we obtain substituting the form \re{ttensor} of the energy-momentum tensor into the Einstein equation \re{efe}.

\subsubsection{Evolution equations}
  These two equations together show how gravitational radiation, i.e. \emph{gravitational waves}, arises \cite[p. 11]{covform3}.
  \begin{itemize}
    \item
      The \emph{electric propagation equation} is:
      \begin{equation}
      \begin{split}
        \dot{E}_{\langle\alpha\beta\rangle}=&-\Theta E_{\alpha\beta} +3\sigma_{\langle\alpha}^{\phantom{\langle\alpha}\gamma}E_{\beta\rangle\gamma}
        -\frac{1}{2}\left( \mu+p\right)\sigma_{\alpha\beta}-\frac{1}{2}\sigma_{\langle\alpha}^{\phantom{\langle\alpha}\gamma}\pi_{\beta\rangle\gamma} \\
        &-\frac{1}{2}\dot{\pi}_{\langle\alpha\beta\rangle}-\frac{1}{6}\Theta\pi_{\alpha\beta}-\frac{1}{2}\hat{\nabla}_{\langle\alpha}q_{\beta\rangle}
        -\dot{u}_{\langle\alpha}q_{\beta\rangle} \\
        &+\eta_{\gamma\delta\langle\alpha}\left( \hat{\nabla}^\gamma H_{\beta\rangle}^{\phantom{\beta\rangle}\delta} 
        +2\dot{u}^{\gamma}H_{\beta\rangle}^{\phantom{\beta\rangle}\delta} - \omega^\gamma E_{\beta\rangle}^{\phantom{\beta\rangle}\delta}
        - \frac{1}{2}\omega^\gamma \pi_{\beta\rangle}^{\phantom{\beta\rangle}\delta} \right) \ .
      \end{split}
      \end{equation}
    \item
      The \emph{magnetic propagation equation} is:
      \begin{equation}
       \begin{split}
        \dot{H}_{\langle\alpha\beta\rangle}=&-\eta_{\gamma\delta\langle\alpha}\left(
        \hat{\nabla}^\gamma E_{\beta\rangle}^{\phantom{\beta\rangle}\delta}-\frac{1}{2}\hat{\nabla}^{\gamma}\pi_{\beta\rangle}
        ^{\phantom{\beta\rangle}\delta}+2\dot{u}^\gamma E_{\beta\rangle}^{\phantom{\beta\rangle}\delta}
        +\omega^{\gamma}H_{\beta\rangle}^{\phantom{\beta\rangle}\delta}+\frac{1}{2}q^\gamma\sigma_{\beta\rangle}^{\phantom{\beta\rangle}\delta}\right)
        \\ &-\Theta H_{\alpha\beta} +3 \sigma_{\langle\alpha}^{\phantom{\langle\alpha}\gamma}H_{\beta\rangle\gamma}
        -\frac{3}{2}\omega_{\langle\alpha}q_{\beta\rangle} \ .
       \end{split}
      \end{equation}
  \end{itemize}

\subsubsection{Constraint equations}

  \begin{itemize}
    \item
    The \emph{electric constraint equation} is:
      \begin{equation}
       \begin{split}
        \hat{\nabla}_{\beta}E^{\beta\gamma}=&-3\omega_{\beta}H^\beta_{\phantom{\beta}\alpha} +\eta_\alpha^{\phantom{\alpha}\beta\gamma}
        \left( \sigma_{\beta\delta} H^\delta_{\phantom{\delta}\gamma}-\frac{3}{2}\omega_\beta q_\gamma \right) \\
        &+\frac{1}{3}\hat{\nabla}_{\alpha}\mu
        -\frac{1}{2}\hat{\nabla}_\beta\pi^{\beta\gamma}-\frac{1}{3}\Theta q_{\alpha}+\frac{1}{2}\sigma_{\alpha\beta}q^{\beta} \ ,
       \end{split}
      \end{equation}
      where the spatial gradient of the energy density acts as a source term. It can be regarded as a vector analogue 
      of the Newtonian Poisson equation, enabling tidal action at a distance.
    \item
    The \emph{magnetic constraint equation} is:
      \begin{equation}
       \begin{split}
        \hat{\nabla}_\beta H^{\beta\gamma}=& 3\omega_\beta E^\beta_{\phantom{\beta}\alpha}
        -\eta_{\alpha}^{\phantom{\alpha}\beta\gamma}\sigma_{\beta}^{\phantom{\beta}\delta}
        \left( E_{\gamma\delta}+\frac{1}{2}\pi_{\gamma\delta} \right) \\
        &+\left( \mu +p \right)\omega_{\alpha}-\frac{1}{2}\pi_{\alpha\beta}\omega^{\beta}
        -\frac{1}{2}\eta_\alpha^{\phantom{\alpha}\beta\gamma}\hat{\nabla}_\beta q_\gamma \ ,
       \end{split}
      \end{equation}
      where the fluid vorticity acts as a source.
  \end{itemize}

\subsection{Twice-contracted Bianchi identities}
\label{frombianchitwicecont}

The twice contracted Bianchi identities \re{twicebianchi} arise writing the Bianchi identities \re{bianchi} in the form
  \begin{equation}
    \nabla_{\epsilon}R_{\alpha\beta\gamma\delta} + \nabla_{\gamma}R_{\alpha\beta\delta\epsilon} + \nabla_{\delta}R_{\alpha\beta\epsilon\gamma} = 0 \ ,
  \end{equation}
and contracting twice. Inserting the Einstein equation \re{efe} gives rise to equation \re{evot}, i.e. the vanishing of the covariant derivative of the energy-momentum tensor. We can insert in this expression the form \re{ttensor} of $T_{\alpha\beta}$ and use again \re{split} in order to obtain two further propagation equations.

\subsubsection{Propagation equations}

\begin{itemize}
  \item
    Projecting parallel to $u_{\alpha}$ with the projector \re{prou} we have
      \begin{equation}
        \dot{\mu}=-\Theta\left( \mu +p\right)-\hat{\nabla}_{\alpha}q^{\alpha}
        -2\dot{u}_{\alpha}q^\alpha-\sigma_{\alpha\beta}\pi^{\alpha\beta} \ ,
          \label{enconseq}
      \end{equation}
    while
  \item
    projecting orthogonal to $u_\alpha$ we obtain
      \begin{equation}
        \hat{\nabla}^{\alpha} p=-\left( \mu + p \right) \dot{u}^\alpha -\dot{q}^{\langle \alpha \rangle}
        -\hat{\nabla}_{\beta}\pi^{\alpha\beta}-\frac{4}{3}\Theta q^{\alpha} 
        -\sigma^\alpha_{\phantom{\alpha}\beta}q^\beta-\dot{u}_\beta \pi^{\alpha\beta}
        +\eta^{\alpha\beta\gamma}\omega_{\beta}q_{\gamma} \ .
          \label{momconseq}
      \end{equation}
\end{itemize}
Sometimes these equations are called the \emph{energy conservation equation} (or \emph{continuity equation}) and the \emph{momentum conservation equation}, respectively, even if they show that in the general case the aforementioned quantities are not conserved.

\chapter{Frobenius' theorem}
  \label{fbthchapter}

In this chapter we go in detail through Frobenius' theorem, which is a fundamental result in differential topology. In order to achieve maximal generality, we present the theorem in an abstract way, by using vector fields first, and employing differential forms later. Stated this way, Frobenius' theorem is a powerful tool for mathematics, and for helping to understand physical aspects of our universe.

In general relativity the existence of spacetime is given, while space and time, considered separately, are derived concepts. We usually choose a four-velocity vector field, and use it to define time. Then arises the question of whether or not a three-dimensional space exists, or if we have only spacetime. Frobenius' theorem gives the answer to this question: a family of three-dimensional spaces orthogonal to the four-velocity vector field exists \ev{if and only if} the vorticity vanishes.

This result is of fundamental importance for our discussion on average quantities, because we want to average over three-dimensional hypersurfaces. We can always consider three-dimensional sections of the four-dimensional manifold, but three-dimensional sections that are orthogonal to the four-velocity vector field (and so can be interpreted as three-dimensional physical spaces) exist if and only if the vorticity vanishes.

After some brief mathematical remarks (most of which can be found in \cite{jost} and \cite{dario}), we give the vector formulation of the theorem (which can be found for instance in \cite[pp. 434--435]{wald} and \cite[pp. 440--441]{dario}), the dual formulation (see \cite[pp. 435--436]{wald}) and the proof (see \cite[pp. 441--442]{dario}). Then we analyse the consequences from a physical point of view (some useful insights can be found in \cite{ellisbrunihwang}).

\section{Mathematical remarks}

First of all, we want to recall some known mathematical concepts. Since a deep explanation of these results, and proofs of many of them, are beyond the aim of this work, we just summarize them, explaining in detail only what we need.

\subsection{Lie brackets}
\label{liebracketssection}

Let $\m$ be a manifold and $\tm$ the \emph{tangent bundle} to this manifold. A \emph{vector field $X$} is usually defined as a map $X\in C^\infty\left(\m,\tm\right)$ (see \cite[p. 38]{jost}), i.e. a section of the tangent bundle $\tm$. Regarded in this way, a vector field is a differential operator acting on smooth functions on $\m$, $f\in C^\infty\left(\m,\mathbb{R}\right)$, and its action gives back another function $X\left(f\right)\in C^\infty\left(\m,\mathbb{R}\right)$. In local coordinates $x^\alpha$, when the basis of $\tm$ is $ \{ \frac{\partial}{\partial x^\alpha}, \alpha=1,2,\dots,dim\left(\m\right) \} $, the vector field acting on $f$ can be expressed as $X\left(f\right)=X^\alpha \frac{\partial f}{\partial x^\alpha}$. Given another vector field $Y$, we can consider the action of $Y$ on $X\left(f\right)$, i.e. $Y\left(X\left(f\right)\right)$. Now we can define a useful object (see \cite[p. 52]{jost}).
  \begin{definition}(Lie brackets of vector fields)
    Given two vector fields X, Y on $\m$, and a function $f\in C^\infty\left(\m,\mathbb{R}\right)$, we define the \emph{Lie brackets of the vector fields X and Y} (sometimes called the \emph{commutator between $X$ and $Y$}) to be:
    \begin{equation}
      \left[X,Y\right]f\doteq X\left(Y\left(f\right)\right)-Y\left(X\left(f\right)\right) \ .
      \label{liebrackets1}
    \end{equation}
  \end{definition}

If we want to express the Lie brackets in terms of the covariant derivative, we can write them in the following way:\footnote{ See\cite[p. 31]{wald}.}
  \begin{equation}
    \left[X,Y\right]f=\left( X^\alpha\nabla_\alpha Y^\gamma - Y^\beta\nabla_\beta X^\gamma \right) \nabla_\gamma f\ ,
      \label{liecov}
  \end{equation}
which gives for the components the expression
  \begin{equation}
    \left[X,Y\right]^\gamma=\left( X^\alpha\nabla_\alpha Y^\gamma - Y^\beta\nabla_\beta X^\gamma \right)\ .
      \label{liebrackcomp}
  \end{equation}

This object is independent of the choice of coordinates, but sometimes it is useful to express it in local coordinates as
  \begin{equation}
    \left[X,Y\right]f=\left( X^\alpha\frac{\partial Y^\gamma}{\partial x^\alpha}-Y^\beta\frac{\partial X^\gamma}{\partial x^\beta} \right) \frac{\partial f}{\partial x^\gamma} \ ,
  \end{equation}
and it is evident that the Lie brackets of two vector fields are a vector field.

We say that two vector fields $X$ and $Y$ \emph{commute} when their Lie brackets vanish, i.e.
  \begin{equation}
    \left[ X , Y \right] = 0 \ .
  \end{equation}

The Lie brackets $\left[.,.\right]$ are a map from $\Gamma\left(\tm\right)\times\Gamma\left(\tm\right)$ to $\Gamma\left(\tm\right)$ with the following properties:\footnote{$\Gamma\left(\tm\right)$ is the vector space of the smooth sections of the tangent bundle of $\m$.}
  \begin{subequations}
    \label{lieprop}
      \begin{align}
        &\left[X,Y\right]=-\left[Y,X\right] \ ,
         \label{antis} \quad\text{(\emph{Antisymmetry})} \\
        &\left[X+Y,Z+K\right]=\left[X,Z\right]+\left[Y,Z\right]+\left[X,K\right]+\left[Y,K\right] \, ,\, \text{(\emph{Bilinearity})}
         \label{bilin} \\
        &\left[fX,gY\right]=fg\left[X,Y\right]+fX\left(g\right)Y-gY\left(f\right)X 
         \label{liefunc} \ , \\
        &\left[X,\left[Y,Z\right]\right]+\left[Z,\left[X,Y\right]\right]
         +\left[Y,\left[Z,X\right]\right]=0 \ , \quad\text{(\emph{Jacobi identity})}
         \label{jacobi}
      \end{align}
  \end{subequations}
where $f$ and $g$ are two functions.
The Lie brackets equip the real vector space $\Gamma\left(\tm\right)$ with the structure of a \emph{Lie algebra}.
From \re{antis} it follows immediately that
  \begin{equation}
    \left[X,X\right]=0\quad\forall \, \text{vector fields X on $\m$.}
      \label{xx0}
  \end{equation}

\subsection{One-parameter group of diffeomorphisms}

Consider now a vector field $X\in C^\infty \left(\m,\tm\right)$, a point $p\in\m$, an open interval $I\subset\mathbb{R}$ containing zero, and the smooth curve 
  \begin{equation}
    \begin{split}
      c: \, I\subset\mathbb{R}&\longrightarrow\m \\
            t&\longrightarrow c\left(t\right) \qquad c\left(0\right)=p \ .
    \end{split}
  \end{equation}
\begin{definition}(Integral curve)\label{intcurvdef}
  The curve $c\left( t \right)$ is an \emph{integral curve of the vector field $X$ passing through $p$} \ev{if and only if}
  \begin{equation}
    X|_{c\left(t\right)}=\frac{d c\left(t\right)}{dt}.
      \label{intcurv}
  \end{equation}
\end{definition}

The intuitive meaning of the above definition is that, given the vector field $X$ defined over $\m$, the integral curve $c\left(t\right)$ is a curve on $\m$ with the property that the tangent vector to this curve at every point $p\in\m$ is given by $X|_p$, i.e. the corresponding vector of the vector field $X$ at the point $p$.
At every point of the manifold the vector field $X$ gives us only one vector $X|_p$, so it is intuitively clear that two integral curves cannot cross. It is possible to demonstrate (see section \ref{intcurvecase}) that one and only one such curve passes through each point $p\in\m$.

Now we can consider other points $p_1,p_2,\dots$, and the integral lines that pass through them. Changing our point of view, we can think of freezing the parameter $t$ at a fixed value, and see what happens when the point $p$ varies. This change of point of view may be formalized as follows.
Given the curve $c\left(t\right)$ on $\m$ as above, the map
      \begin{equation}
        \begin{split}
          \phi_t : \ \m &\longrightarrow \phi_t \left( \m \right) \\
                     \ p &\longrightarrow \phi_t\left(p\right)\doteq c_p\left(t\right)
          \label{flow}
        \end{split}
      \end{equation}
is a diffeomorphisms, and we can make the following definitions.
  \begin{definition}(Flow of a vector field)
    The $C^\infty$ map $\phi_t\left(p\right):\mathbb{R}\times\m\rightarrow\m$ defined above is called the \emph{flow of the vector field $X$}.
  \end{definition}
  \begin{definition}(One-parameter group of diffeomorphisms)
    A \emph{one-parameter group of diffeomorphisms} is a $C^\infty$ map $\phi_t:\mathbb{R}\times\m\rightarrow\m$ that for fixed $t$ is defined as in \re{flow}, and for all $t,s \in \mathbb{R}$ satisfies $\phi_t\circ\phi_s=\phi_{t+s}$.
     \label{onepargroupofdif}
  \end{definition}
From the way we have constructed it, it is obvious that to a one-parameter group of diffeomorphisms is associated a vector field, which is precisely the vector field X associated to $\phi_t\left(q\right)$ through equation \re{intcurv}.

\subsection{Lie derivative of vector fields}

\emph{The Lie derivative} is an object that evaluates the change of a \emph{tensor field} along the \ev{flow} of a vector field. This can be seen from its definition.
  \begin{definition}(Lie derivative)
    Consider a vector field $X$ on $\m$, a tensor field $T$ on $\m$, and a one-parameter group of diffeomorphisms $\phi_t\left(p\right)$, $t\in I\subset\mathbb{R}$, related to $X$ as in the previous section. The Lie derivative of $T$ along $X$ is defined as\footnote{See \cite[p. 54]{jost}.}
    \begin{equation}
      L_X T \doteq \frac{d}{dt}\left(\phi_t^\ast T\right)_{t=0},
    \end{equation}
   where $\phi_t^\ast T$ is the \emph{pullback of $T$ via $\phi_t$}. 
  \end{definition}

We are interested in evaluating the change of a vector field $Y$ along the flow of the vector field $X$. In this particular case it can be demonstrated (see \cite[p. 54]{jost}) that the \emph{Lie derivative of a vector field $Y$ along the vector field $X$} turns out to be
  \begin{equation}
    L_X Y = \left[ X,Y \right],
      \label{liederbrac}
  \end{equation}
i.e. precisely the Lie Brackets \re{liebrackets1} of the two vector fields.

\section{Frobenius' theorem}

Frobenius' theorem can be seen from different points of view, enlightening different aspects of it. So several equivalent formulations of the statement are possible. We focus on the formulations that are most interesting for our purposes. First we give a \emph{vector formulation}, which is easier to figure out because of its geometrical character, and gives immediately an important result (see section \ref{intcurvecase}). This formulation can be found for instance in \cite[pp. 434--435]{wald}, and in \cite[pp. 440--441]{dario}. Then we give a \emph{dual formulation}, which makes use of differential forms, and is useful in order to relate the geometrical meaning of this theorem to the physical quantities we use in general relativity, in particular the vorticity $\omega_{\alpha\beta}$. This formulation can be found for instance in \cite[pp. 435--436]{wald}.\footnote{In mathematical analysis the theorem is usually formulated in a different way, and it gives necessary and sufficient conditions for finding a maximal set of independent solutions of an overdetermined system of first order homogeneous linear p.d.e., but this formulation is beyond the scope of this work.}

  \subsection{Vector fields formulation}
  
Let $\left(\m,g\right)$ be an $n$-dimensional manifold, $\tpm\simeq\mathbb{R}^n$ the \emph{tangent space to the manifold at the point $p$},\footnote{$\tpm$ has the same dimension as $\m$; this is intuitively easy to understand, and a rigorous proof can be found in \cite[p. 15]{wald}.} and $\tm\doteq \{\cup_{p\in\mathcal{M}} T_p\mathcal{M}\}$ the  
\emph{tangent space to the manifold}.
At every point $p\in\m$ let $\pw\subset\tpm$ be a subspace of the tangent space $\tpm$, with dimension $dim\left(\pw\right)=k<n$.

Let us define some concepts that we need in order to give the statement of the theorem.
  \begin{definition}(Distribution)
     We define a \emph{$k$-dimensional distribution $W$} on the manifold $\m$ to be an assignment of a subspace $\pw$ for every $p\in\m$. So $W\doteq \{ \bigcup_{p\in\m} W_p \}$. Sometimes $\w$ is called a \emph{ $k$-dimensional specification of subspaces $\pw$}.
  \end{definition}
  \begin{definition}(Smooth distribution)
    The distribution $\w$ is said to be \emph{smooth} (or $C^\infty$) if $\forall \ p\in\m$ there is an open neighborhood $U$ of $p$ and $k$ smooth vector fields $X_1,\ldots,X_k$ on $U$ that span $\pw \ \forall \ p \in U$.
  \end{definition}
  \begin{definition}(lie in)
    A vector field $X:N\subset\m\rightarrow\tm$ on an open domain $N$ is said to \emph{lie in} the distribution W if $X|_p\in \pw \ \forall \ p \in N$.
  \end{definition}

  \begin{definition}(Integral submanifold)\footnote{The concept of \ev{integral curve} is given in definition \re{intcurvdef}, page \pageref{intcurvdef}. Integral curves are a particular case of integral submanifolds, obtained when $k=1$.}
  \label{intsubdef}
    We define an \emph{integral submanifold of $W$ on $\m$} as a $k$-dimensional submanifold $\s$  embedded in $\m$ (through the \emph{embedding $\phi$}) such that all vector fields tangent to the submanifold \ev{lie in} $\w$:
      \begin{equation}
        \phi_{*} \left( T_p \s \right)=\w \left( \phi \left( p \right)\right)
      \end{equation}
    i.e. such that the tangent space to this submanifold at each $p\in\s$ coincides with \w.
  \end{definition}
  
Once we have defined the above objects, a question may arise: \ev{what are the conditions that allow us to find integral submanifolds of $\w$?} The answer is contained in Frobenius' theorem.
    \begin{theorem}[Frobenius' theorem, vector fields formulation] \label{fbthvec}
      A smooth $k$-dimensional distribution $\w$ on a manifold $\m$ possesses integral submanifolds \ev{if and only if} it is involutive.
    \end{theorem}
    
\begin{definition}(Involutive distribution)
    A distribution $\w$ is said to be \emph{involutive} if for any pair of vector fields \mbox{$X:\m\rightarrow\tm$}, \mbox{$Y:\m\rightarrow\tm$}
\ev{lying in $\w$}, their Lie brackets $[X,Y]$ also \ev{lie in} $\w$.\footnote{Note that $[X,Y]:\m\rightarrow\tm$, i.e. they are a vector field too, see section \ref{liebracketssection}.}
  \end{definition}
  
This result will be proved in section \ref{dim}.

  \subsection{Dual formulation}
  
  In order to see what this theorem tell us about vorticity, it is useful to write it in an equivalent form.\footnote{The consequences obtained by the use of this formulation are presented in section \ref{k=n-1}.}
  Given $\pw\subset\tpm$ as above, we can consider the one-forms $\alpha\in\cotpm$ whose components satisfy the condition:\footnote{Precisely, a one-form $\alpha$ is a section of the \emph{cotangent bundle $\cotm$}, i.e. $\alpha \in C^\infty\left(\m,\cotm\right)$.}
  \begin{equation}
    \alpha_\nu X^\nu|_p = 0 \quad \forall \ X|_p \ \in \ \pw \ .
      \label{forma}
  \end{equation}
We can see that such $\alpha$ span an $\left( n-k\right)$-subspace $V_P^{\ast} \ \subset \ \cotpm$ of the tangent space at $p$. It is also possible to show that, conversely, an $\left( n-k\right)$-subspace $V_P^{\ast}$ of $\cotpm$ defines a $k$-dimensional subspace $\pw$ of $\tpm$ via equation \re{forma}.

We can construct a distribution $\w$ by collecting together all the subspaces $\pw$ that consist at each point of all vectors $X|_p$ satisfying \mbox{$X|_p^\nu\alpha_\nu=0$} \mbox{$\forall \ \alpha \ \in \ V_p^{\ast}$}.
Now it is obvious to reformulate the above question in terms of $V^{\ast}$: \ev{what are the conditions that a smooth specification $V^{\ast}$ of $\left(n-k\right)$-dimensional subspaces of one-forms at each point must satisfy in order to have the property that the associated distribution $\w$ admits integral submanifolds?}

Frobenius' theorem, as formulated in the previous section, gives the answer to this question: integral submanifolds exist \ev{if and only if} $\forall \ \alpha \ \in \ V^{\ast}$ and $\forall$ vector fields $Y,Z\in W$ (i.e. for which $\alpha_\nu Y^\nu=0=\alpha_\mu Z^\mu$) also $[Y,Z]$ \ev{lie in} $W$, i.e.
  \begin{equation}
    \alpha_\nu \left[Y,Z\right]^\nu=0.
      \label{quellasopra}
  \end{equation}
This condition determines an important property of $\alpha$. In order to see this, we can insert equation \re{liebrackcomp} into equation \re{quellasopra}, so we obtain
  \begin{equation}
    \begin{split}
      0=&\alpha_\nu \left(Y^\mu\nabla_\mu Z^\nu - Z^\mu \nabla_\mu Y^\nu \right) \\
       =&-Z^\nu Y^\mu \nabla_\mu \alpha_\nu + Y^\nu Z^\mu \nabla_\mu \alpha\nu  \\
       =&2Y^\nu Z^\mu \nabla_{[\mu}\alpha_{\nu]} \ , \\
       \label{rebirth}
    \end{split}
  \end{equation}
  where the second equation arises from the fact that $\nabla_\nu \left( \alpha_\mu Z^\mu \right)=0$ and $\nabla_\nu \left( \alpha_\mu Y^\mu \right)=0$.
  The condition that equation \re{rebirth} holds for the vector fields $Y$ and $Z$ that lie in the subspace annihilated by $V^{\ast}$ is equivalent to the condition
  \begin{equation}
    \nabla_{[\mu}\alpha_{\nu]}=\sum_{\alpha=1}^{n-k} \mu^\alpha_{[\mu}v^\alpha_{\nu]} \ ,
      \label{coso}
  \end{equation}
where each $v^\alpha$ is an arbitrary one-form and each $\mu^\alpha \ \in \ V^{\ast}$ (see \cite[p. 436]{wald}).

Finally we are able to reformulate Frobenius' theorem in terms of differential forms as follows.
  
    \begin{theorem}[Frobenius' theorem, dual formulation]
      Let $V^{\ast}$ be a smooth specification of an $\left(n-k\right)$-dimensional subspace of one-forms. Then the associated $k$-dimensional subspace $\w$ of the tangent space $\tm$ admits integral submanifolds \ev{if and only if} $\forall \, \alpha \, \in \, V^{\ast}$ we have $d\alpha=\sum_{\alpha}\mu^{\alpha}\wedge v^{\alpha}$, with each $v^\alpha$ being an arbitrary one-form, and each $\mu^\alpha \ \in \ V^{\ast}$.
    \end{theorem}

\section{Demonstration}
\label{dim}
We demonstrate Frobenius' theorem as expressed in the vector form. We have already showed that the dual form is equivalent to that one, so it will be demonstrated as well. This demonstration is split into two parts: first we show that if we can find integral submanifolds of $W$, then $W$ is involutive; then we show that if $W$ is involutive, it possesses integral submanifolds.

\subsubsection{\ev{If} $\w$ possesses integral submanifolds \ev{then} it is involutive.}

\begin{proof}
Given that we can find integral submanifolds of $W$, we can span $W$ in a neighborhood of any point by coordinate vector fields $X_1,X_2,\dots,X_k$ in $\m$ such that $\left[X_\mu,X_\nu\right]=0$.\footnote{This can be understood by thinking that $W$ is a collection of subspaces $\pw$ of the tangent spaces $\tpm$, and every $\tpm$ in local coordinates is spanned by the basis $\left\{\partial_1,\partial_2,\dots,\partial_n\right\}$, whose vectors obviously commute. } Any vector field that lies in the distribution $\w$ can be expressed as a linear combination of these coordinate vector fields, so if $Y$ and $Z$ are two vector fields we have $Y=\sum_\mu f_\mu X_\mu$ and $Z= \sum_\nu g_\nu X_\nu$. The commutator between the two vector fields can then be written as
 \begin{equation}
   \begin{split}
     \left[ Y,Z \right] =& \sum_{\mu,\nu} \left[f_\mu X_\mu ,g_\nu X_\nu \right]  \\
                        =& \sum_{\mu,\nu} \bigl(f_\mu X_\mu\left(g_\nu X_\nu\bigr) - g_\nu X_\nu\left( f_\mu X_\mu \right) \right) \\
                        =& \sum_{\mu,\nu}\bigl(f_\mu g_\nu\left[X_\mu,X_\nu\right]+f_\mu X_\mu\left(g_\nu\right)X_\nu
                           -g_\nu X_\nu\left(f_\mu\right) X_\mu \bigr) \\
                        =& \sum_{\mu,\nu} \bigl(f_\mu X_\mu\left(g_\nu\right) - g_\nu X_\mu\left( f_\mu \right) \bigr) X_\nu \ ,
     \label{yz}
   \end{split}
 \end{equation}
where in order to obtain the second line we have used the definition \ref{liebrackets1} of the Lie brackets, and the third line follows directly from the first because of the property \re{liefunc}.
Equation \re{yz} show that $\left[ Y , Z \right]$ can be expressed in the form of a vector field belonging to $W$, so $\left[Y,Z\right]\in W$, i.e. $W$ is involutive.
\end{proof}

\subsubsection{\ev{If} $\w$ is involutive \ev{then} it possesses integral submanifolds.}

\begin{proof}

We can prove the theorem by \ev{induction} on the dimension k, through three subsequent steps. First, we write the theorem in a way that is different from the vector field formulation given above, but equivalent. Second, we demonstrate that the statement holds for the case $k=1$. Third, we show that \ev{if} it holds for dimension $k$, it also holds for dimension $k+1$. Therefore the conclusion holds for every value of $k$.
\\

First, we can rewrite the theorem in the following way. A smooth $k$-dimensional distribution $W^k$ on a manifold $\m$ is \ev{involutive} \ev{if and only if} every point $p\in \m $ lies in a coordinate chart $\left(U;x^\alpha\right)$ such that the coordinate vector fields $\partial/\partial x^\alpha$ for $\alpha=1,\dots,k$ span $W^k$ at each point of $U$.
\\

Second, consider $k=1$. The statement now reads: if a smooth one-dimensional distribution $W^1$ on a manifold $\m$ is involutive,\footnote{Note that, as shown by equation \re{tutteinvolute}, \ev{every} one-dimensional distribution $W^1$ on a manifold $\m$ is involutive.} then every point $p \in \m$ lies in a coordinate chart $\left(U;x^\alpha\right)$ such that the coordinate vector field $\partial/\partial x^1$ spans $W^1$ at each point of $U$.
This follows directly from the fact that,
\ev{if $X$ is a vector field on a manifold $\m$ such that $X|_p\neq0 \ \forall \ p\in\m$, then there exists a coordinate chart $\left(U;x^\alpha\right)$ at $p$ such that}\footnote{The proof is outlined in \cite[p. 435]{dario}.}
  \begin{equation}
    X=\frac{\partial}{\partial x^{1}} \ .
      \label{usefulstatement}
  \end{equation}
\\

Third, suppose that the statement is true for all $k$-dimensional distributions. We now show that it also holds for a $\left(k+1\right)$-dimensional distribution.

Let $W^{k+1}$ be a $\left(k+1\right)$-dimensional involutive distribution spanned at all points of an open set $A$ by vector fields $\{ X_1,\dots,X_{k+1} \}$.
At any point $p\in A$ there exist coordinates $\left(V;y^i\right)$ such that $X_{k+1}=\partial_{y^{k+1}}$.

Set
  \begin{equation}
    Y_{\alpha}\doteq X_{\alpha} - \left( X_{\alpha} y^{k+1} \right) X_{k+1} \ , \quad 
    Y_{k+1} \doteq X_{k+1} \ ,
  \end{equation}
where the Greek indexes $\alpha,\beta,\dots$ run from 1 to k. Then the vector fields $Y_1, Y_2,\dots,Y_{k+1}$ clearly span $W^{k+1}$ on $V$, and
  \begin{equation}
    Y_{k+1} y^{k+1} =X_{k+1} y^{k+1} = \partial_{y^{k+1}} y^{k+1} = 1 \ , \qquad Y_\alpha y^{k+1} = 0 \ .
      \label{dim1lab}
  \end{equation}
  
Since $W^{k+1}$ is involutive we can write the commutators in the form
  \begin{align}
    \left[ Y_\alpha , Y_\beta \right] &= C_{\alpha\beta}^\gamma Y_\gamma+ a_{\alpha\beta}Y_{k+1} \ , \\
    \left[ Y_\alpha , Y_{k+1} \right] &= C_{\alpha}^\gamma Y_\gamma+ a_{\alpha}Y_{k+1} \ .
  \end{align}
Applying both sides of these equations to the coordinate function $y^{k+1}$ and using \re{dim1lab}, we find $a_{\alpha\beta}=a_{\alpha}=0$, whence
  \begin{align}
    \left[ Y_\alpha , Y_\beta \right] = C^\gamma _{\alpha\beta} Y_\gamma \ , \\
    \left[ Y_\alpha , Y_{k+1} \right] = C^\gamma _{\alpha} Y_\gamma \ .
      \label{secondarefdim}
  \end{align}

The distribution $W^{k}$ spanned by $Y_1 , Y_2 ,\dots, Y_{k}$ is therefore involutive on $V$. So, because we have assumed that the statement of the theorem holds for every $k $-dimensional distribution, there exists a coordinate chart $\left( \Omega ; z^i \right)$ such that $W^{k}$ is spanned by $\{ \partial_{z^1}, \dots, \partial_{z^{k}} \}$.

Set
  \begin{equation}
    \frac{\partial}{\partial z^\alpha} \doteq A^\beta_{\phantom{\beta}\alpha} Y_\beta \ ,
  \end{equation}
where $ A^\beta_{\phantom{\beta}\alpha}$ are the elements of a non singular matrix of functions on $\Omega$. The original distribution $W^{k+1}$ is spanned on $\Omega$ by the set of vector fields
  \begin{equation}
    \{ \partial_{z^1}, \dots , \partial_{z^{k}} , Y_{k+1} \}\ .
  \end{equation}
It follows then from \re{secondarefdim} that
  \begin{equation}
    \left[ \partial_{z^\alpha}, Y_{k+1} \right] = K_\alpha^{\phantom{\alpha}\beta} \partial_{z^\beta}
      \label{1544}
  \end{equation}
for some functions $K_\alpha^{\phantom{\alpha}\beta}$.

We can write
  \begin{equation}
    Y_{k+1} = \sum_{\alpha=1}^{k} \xi^\alpha \partial_{z^\alpha} + \sum_{a=k+1}^{n} \xi^a \partial_{z^a} \ .
      \label{329}
  \end{equation}
Applying equation \re{1544} to the coordinate functions $z^a \left(a=k+1,\dots, n \right) $ and using \re{329}, after some calculations we find that
  \begin{equation}
    \frac{\partial \xi^a}{\partial z^\alpha} = 0 \ .
  \end{equation}
Hence $\xi^a = \xi^a \left( z^{k+1} , \dots , z^n \right)$ for all $a\geq {k+1}$.

Since $Y_{k+1}$ is linearly independent of the vectors $\partial_{z^\alpha}$, the distribution $W^{k+1}$ is spanned by the set of vectors $\{ \partial_{z^1} , \partial_{z^2} , \dots , \partial_{z^{k}} , Z \}$, where
  \begin{equation}
    Z \doteq Y_{k+1} - \xi^\alpha \partial_{z^\alpha} = \xi^a \left( z^{k+1}, \dots, z^n \right) \partial_{z^a} \ ,
  \end{equation}
where in order to get the last part of the equation we have used \re{329}.

Because of the statement involving equation \re{usefulstatement}, there exists a coordinate transformation not involving the first $k$ coordinates,
  \begin{equation}
    \begin{split}
      x^{k+1}&=x^{k+1} \left( z^{k+1}, \dots,z^n \right) , \\
      x^{k+2}&=x^{k+2} \left( z^{k+1}, \dots,z^n \right), \dots, x^n=x^n \left( z^{k+1}, \dots,z^n \right)
    \end{split}
  \end{equation}
such that $Z=\partial_{x^{k+1}}$. Setting $x^1 = z^1, \dots, x^{k}=z^{k},$ we have the coordinates $\left(U;x^i\right)$ in which $W^{k+1}$ is spanned by $\{ \partial_{x^1}, \dots , \partial_{x^{k}}, \partial_{x^{k+1}} \}$ .
This proves the statement of the theorem for a $\left(k+1\right)$-dimensional distribution, given that it holds for a $k$-dimensional distribution. So by induction Frobenius' theorem is proved.

\end{proof}

\section{Results}

We have presented Frobenius' theorem in the most abstract way, operating with only geometrical objects. This leads to a result that does not involve the physics discussed in the first chapter of this work, but only concepts from differential geometry, so it represents a very powerful tool, that can be used in several different contexts. We are interested in what this theorem teaches us about the description of the universe carried out through general relativity that we presented in chapter \ref{3+1covform}.\footnote{Let us clarify how to match the different notations used in these two chapters. We indicate by $n$ the dimension of the manifold $\m$, and $k$ is the dimension of the smooth specification of a collection $\w$ of subsets $\pw\subset\tpm$, while when we write <<3+1 covariant formalism>> we are talking about dimensions to which we attribute a physical meaning (intuitively, a three-dimensional space and a one-dimensional time), so in terms of $n$ and $k$ we are using a <<$\left(k\right)+\left(n-k\right)$-covariant formalism>>.} In order to do this, we specify different choices of the dimensions $n$, $k$ of the manifold and the smooth distribution, and show which results arise. Only some of the subsequent cases contain interesting physical results, while others have a more pedagogical nature, being useful in order to understand more general (and more complicated) cases.

\subsection{Case k=1, n generic}
\label{intcurvecase}

The first interesting situation is when we have a generic $n$-dimensional manifold, and we choose the dimension of the subspaces $\pw\subset\tpm$ that constitutes the smooth distribution $\w$ to be $k=1$.

In this case one smooth vector field $X$ is enough to span $W$, and the problem of \ev{finding integral submanifolds of the smooth distribution $W$} reduces to the problem of \ev{finding integral curves of the vector field $X$}.\footnote{The concepts of integral submanifold and integral curve are given in definition \ref{intsubdef}, page \pageref{intsubdef}, and definition \ref{intcurvdef}, page \pageref{intcurvdef} respectively.} The vector field formulation of Frobenius' theorem now reads: \ev{the smooth one-dimensional distribution $\w$ spanned by the vector field $X$ possesses integral curves} if and only if \ev{it is involutive, i.e. for every two vector fields $Y$ and $Z$ lying in the distribution $W$ also their commutator $\left[Y,Z\right]$ lies in $W$}. But because $W$ is one-dimensional, the vector fields $Y$ and $Z$ must be of the form: $Y=fX$, $Z=gX$, and so
    \begin{align}
      \left[Y,Z\right]=&\left[fX,gX\right] 
      \notag \\
                      =&fg\left[X,X\right]+fX\left(g\right)X-gX\left(f\right)X 
      \notag \\
                      =&\left(fX\left(g\right)-gX\left(f\right)\right)X,
      \label{tutteinvolute}
    \end{align}
where we have used property \re{liefunc} in obtaining the second line and property \re{xx0} for the third. From the above expression it is clear that $\left[Y,Z\right]$ also lies in $\w$, because $X$ lies in $\w$.

So Frobenius' theorem tells us that \ev{for every smooth vector field a family of integral curves can be found}. It is also possible to demonstrate that this family is unique.

We have obtained this result from Frobenius' theorem, but doing this is like killing a fly with a 15-inch shell. The same problem can be formulated in terms of first order ordinary differential equations (as already stated, Frobenius' theorem in mathematical analysis is formulated in terms of differential equations) so the same result is obtained in a simpler way. As shown in \cite[p. 18]{wald}, using a coordinate system we found that the problem of finding such a family reduces to solving the system
  \begin{equation}
    \frac{dx^\mu}{dt}=X^\mu\left(x^1,\dots,x^n\right)
  \end{equation}
of ordinary differential equations in $\mathbb{R}^n$, where $X^\mu$ is the $\mu$th component of the vector field $X$ in the coordinate basis $\{\frac{\partial}{\partial x^\mu}\}$. The solution of this system (given the initial condition) exists and is unique, due to the Cauchy-Lipschitz theorem.\footnote{Usually referred to as <<teorema di esistenza e unicità per un problema di Cauchy>> in the Italian literature.} It follows that for every smooth vector field a family of integral curves exists and is unique.

In section \ref{velvecfield} we assumed the velocity of matter to be well described by a unique vector field of components $u_{\mu}$ that exists at each point of spacetime. At the time we said that as a consequence there is a family of preferred world lines representing the motion of matter (a congruence). Now we have shown that this family exists and is unique because of Frobenius' theorem.

\subsection{Case k=n-1, n generic}
\label{k=n-1}

Given a generic vector field $\xi^\nu$, we can find a condition that is equivalent to the fact that $\xi^\nu$ is hypersurface orthogonal. 
Chose $k=n-1$, so $V^{\ast}$ is a one-dimensional subspace of $\cotm$. Chose $V^{\ast}$ to be spanned by $\xi_\mu=g_{\mu\nu}\xi^\nu$. Now the condition \re{coso} reduces to:
  \begin{equation}
    \nabla_{[\mu}\xi_{\nu]}=\xi_{[\mu}v_{\nu]} \ ,
      \label{coso2}
  \end{equation}
in fact, because now $V^\ast$ is one-dimensional, there is only one $\mu^i_\nu$, and we have chosen it to be $\mu_\nu=\xi_\nu$. Equation \re{coso2} is equivalent to $\xi_{[\mu}\nabla_{\nu}\xi_{\epsilon]}=0$, so we can state that
  \begin{equation}
    \xi^\nu \text{ is hypersurface orthogonal} \quad \Leftrightarrow \quad \xi_{[\mu}\nabla_{\nu}\xi_{\epsilon]}=0\ .
      \label{hsort}
  \end{equation}
  
\subsubsection{The $\left(3+1\right)$-dimensional case.}
\label{3+1frob}

Now we are ready to show the physical content of Frobenius' theorem. In the previous chapter we have explained how to describe the universe in general relativity. The fluid flow was described by the velocity vector field $u^\nu$ (see section \ref{velvecfield}), and we were working in four dimensions (so $n=4$, $k=3$).

Because the volume element $\eta^{\alpha\mu\nu\epsilon}$ acts as a skew-symmetrization operator, equation \re{hsort} is equivalent to
  \begin{equation}
    \eta^{\alpha\mu\nu\epsilon}\xi_{\mu}\nabla_{\nu}\xi_{\epsilon}=0 \ .
      \label{amuzzo}
  \end{equation}

If now we choose as the vector field $\xi^\nu$ our velocity vector field $u^\nu$, we get:
  \begin{equation}
    \begin{split}
      0=& \eta^{\alpha\mu\nu\epsilon}u_{\mu}\nabla_{\nu}u_{\epsilon} \\
       =& \eta^{\alpha\mu\nu\epsilon}u_{\mu}\hat{\nabla}_{\nu}u_{\epsilon} \\
       =& \eta^{\alpha\mu\nu\epsilon}u_{\mu}\hat{\nabla}_{[\nu}u_{\epsilon]} \\
       =& \eta^{\alpha\mu\nu\epsilon}u_{\mu}\omega_{\epsilon\nu} \\
       =& \eta^{\alpha\nu\epsilon}\omega_{\epsilon\nu} \\
       =& -2\omega^{\alpha} \\
    \end{split}
   \label{omega=zero}
  \end{equation}
where the second line follows from \re{3volel} and the second equation of \re{antiseta}, the third line from the second equation of \re{volelemxxx}, the fourth line from the definition \re{defomega} of $\omega_{\nu\epsilon}$, the fifth line from \re{3volel} and the last line from \re{vorticityvector}.\footnote{Note that the tensor $u^{\mu}\omega_{\epsilon\nu}$ is sometimes referred to as the \emph{defect tensor}, as in \cite{ellisbrunihwang}. It expresses the fact that the commutator of two vector fields $X^\alpha,Y^\beta\in W$ in the case of $\omega\neq0$ does not live in $W$, through the fact that $U_\alpha^{\phantom{\alpha}\beta}\left[X,Y\right]^\alpha=-u_{\alpha}u^\beta \left[X,Y\right]^\alpha = \left[X,Y\right]^\beta - h_\alpha^{\phantom{\alpha}\beta}\left[X,Y\right]^\alpha = -2 u^\beta \omega_{\mu\nu}X^\mu Y^\nu$, with $U_\alpha^{\phantom{\alpha}\beta}$ defined in \re{prou}.}

So we can finally rewrite Frobenius' theorem in a way that clearly shows its connection with the physical description given in chapter \ref{3+1covform}.
  \begin{theorem}[Frobenius' theorem, physical point of view] \label{fbth}
    A vector field $u^\alpha$ (that represents the four-velocity of matter) is hypersurface orthogonal \ev{if and only if} the vorticity of this vector field vanishes, i.e. $\omega^{\alpha}=0$.
  \end{theorem}
This means that there exist three-dimensional spaces orthogonal to the velocity vector field $u^\alpha$ \ev{if and only if} the vector field has no vorticity.

We can think about it intuitively: in general relativity the pivotal concept is spacetime, while space and time are derived concepts. The spacetime is described by a four-dimensional manifold $\mathcal{M}$, equipped with a Lorentzian metric. There are no preferred directions on this manifold that are associated to the concept of time. The concept of physical time arises from defining some observers to which we associate a four-velocity $u^\alpha$ that describes their motion, and considering their proper time along the flow lines of the vector field $u^\alpha$.

The definitions of the projector $U_\mu^\nu$ and $h_\mu^\nu$, equations \re{prou} and \re{proh}, give a way to project quantities along $u^\alpha$ and orthogonal to it.

\ev{If the vorticity is zero}, then there exist three-dimensional hypersurfaces orthogonal to $u^\alpha$, and we can interpret every one of them as the familiar three-dimensional space considered at a given time. In this case we say that we have chosen a \emph{foliation} of the spacetime into three-dimensional space \ev{plus} time.
In this case $h_{\alpha\beta}$, which projects orthogonal to $u^\alpha$ in these three-dimensional spaces, gives the spatial projection.

On the other hand, \ev{if the vorticity is non-zero}, then there do not exist three-dimensional hypersurfaces orthogonal to the four-velocity, and the action of $h_{\alpha\beta}$ is just to project quantities orthogonal to $u^{\alpha}$, but these projections are not related to a three-dimensional space.

Before concluding, let us clarify the meaning of <<vorticity>>. The vorticity tensor was defined in equation \re{defomega} as $\omega_{\alpha\beta}\doteq \hat{\nabla}_{[\beta} u_{\alpha]}$. Because it is antisymmetric and orthogonal to $u^\alpha$, it has only three independent components. This means that it contains exactly the same amount of information as the vorticity vector $\omega^{\alpha}$ defined in \re{vorticityvector}, and we can use one or the other equivalently. Using the latter makes things more clear, because it corresponds precisely to the usual vorticity vector used in classical fluid dynamics, i.e. the curl of the three-velocity of the fluid. From this parallel we can easily see that the direction of the vorticity vector give us the axis of rotation (which is spatial, and in general may change with time), and the vorticity scalar is the norm of the vector.
  
\subsubsection{The $\left(2+1\right)$-dimensional case}

If we take a look at the simpler case where $n=3$ and $k=2$ we can proceed in a similar way.

The statement \re{hsort} is unchanged, and we can try to proceed as we did in \re{omega=zero}. Now, because of the requirement of being orthogonal to $u^{\alpha}$ and its skew-symmetry, $\omega_{\alpha\beta}$ possesses only one independent component; so it is equivalent to a single scalar quantity, and a vorticity vector cannot be defined. The spacetime volume element is $\eta^{\alpha\beta\gamma}$ and the space volume element is $\eta^{\alpha\beta}\doteq\eta^{\alpha\beta\gamma}u_\gamma$, so we can consider the scalar quantity $\omega\doteq\eta^{\alpha\beta}\omega_{\alpha\beta}$.

We can rewrite the theorem this way: a family of surfaces orthogonal to the vector field $u^\alpha$ (that represents the velocity of matter) exists \ev{if and only if} the vorticity vanishes, i.e. $\omega=0$.

The physical meaning of Frobenius' theorem is the same, with the difference that now the hypersurfaces orthogonal to the vector field $u^\alpha$ are two-dimensional, and in this case it is easier to visualize what is going on.

\subsubsection{An intuitive explanation}

In this section, our aim is to provide the intuitive idea of what is going on. We work in $2+1$ dimensions, because in this case it is easier to visualize what happens; anyway the subsequent explanation can easily be extended to the $\left(3+1\right)$-dimensional case. Because of these reasons, in what follows we sometimes proceed in a way that is a bit informal, addressing the intuition of the reader rather than his mathematical knowledge.

The fundamental issue is that we have a vector field $u^\alpha$ (with its flow lines), and we want to know under which conditions there exist surfaces orthogonal to this vector field (i.e. to its flow lines).

In order to do this, given $u^\alpha$ we can construct a distribution $\w$ by collecting together all the subspaces $\pw$ that consist at each point of all vectors $Z|_p$ satisfying the condition \mbox{$Z|_p^\nu u_\nu=0$}.

Frobenius' theorem, as formulated in the previous sections, tells us that surfaces orthogonal to $u^\alpha$ (which are integral surfaces of $W$), exist if and only if $\forall$ vector fields $X,Y$ orthogonal to $u^\alpha$, also $[X,Y]$ is orthogonal to $u^\alpha$.
It is useful to rewrite the previous sentence as: ``if and only if $\forall$ $X,Y$ $/$ $u_\nu X^\nu=0=u_\mu Y^\mu$ (i.e. $X,Y\in W$), we have $u_\nu \left[X,Y\right]^\nu=0$ (i.e. $\left[X,Y\right]$ lie in $W$)''.
      
In section \ref{dim} we provide a quite technical demonstration of the fact that this ``if and only if'' hold. Now we want to show that the same issue, at least in $2+1$ dimensions, can be intuitively understood.

First of all, we remind that $L_X Y = \left[ X,Y \right]$ (see equation \re{liederbrac}), i.e. the Lie brackets $[X,Y]$ are the Lie derivative of the vector field $Y$ along the vector field $X$, and the Lie derivative is an object which evaluates the change of the vector field $Y$ along the \ev{flow} of the vector field $X$.

Which is the meaning of having Lie brackets that lie in $W$? Why does this condition allow to find surfaces orthogonal to $u^\alpha$?

\begin{figure}
\includegraphics[scale=0.62]{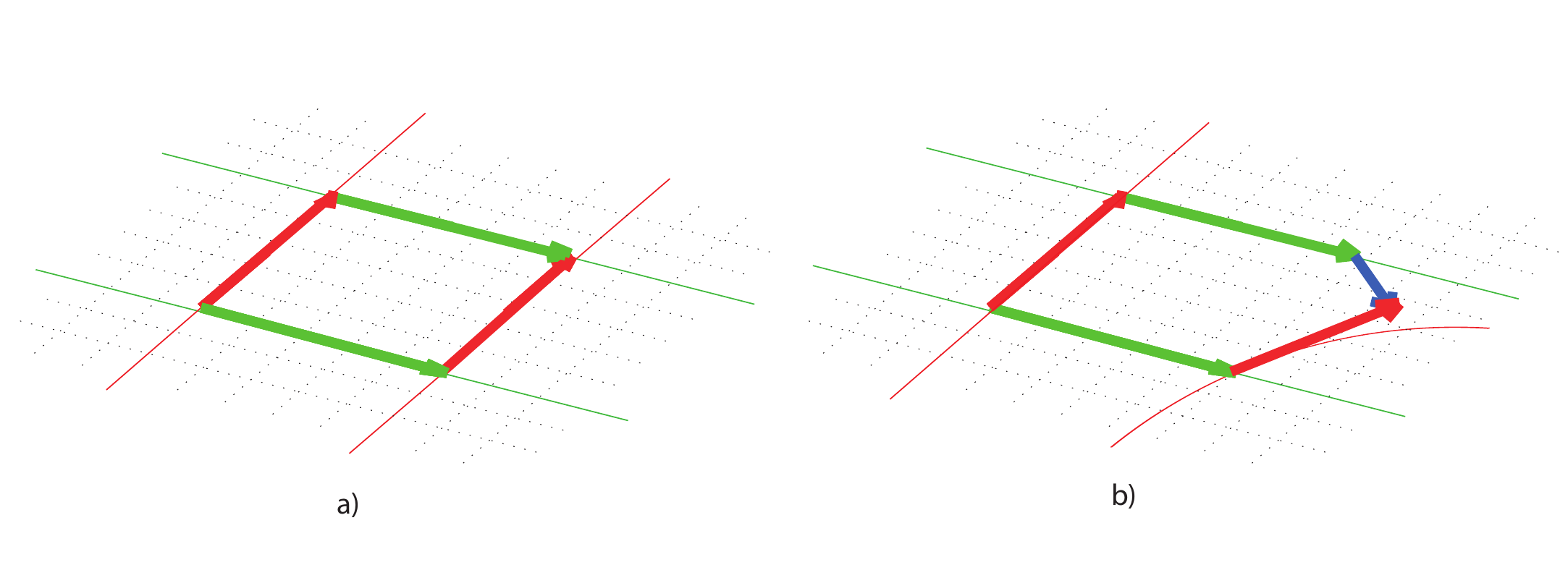}
\caption{Lie derivative inside the spam}
\label{liein}
\centering
\end{figure}

\begin{figure}
\centering
\includegraphics[scale=0.7]{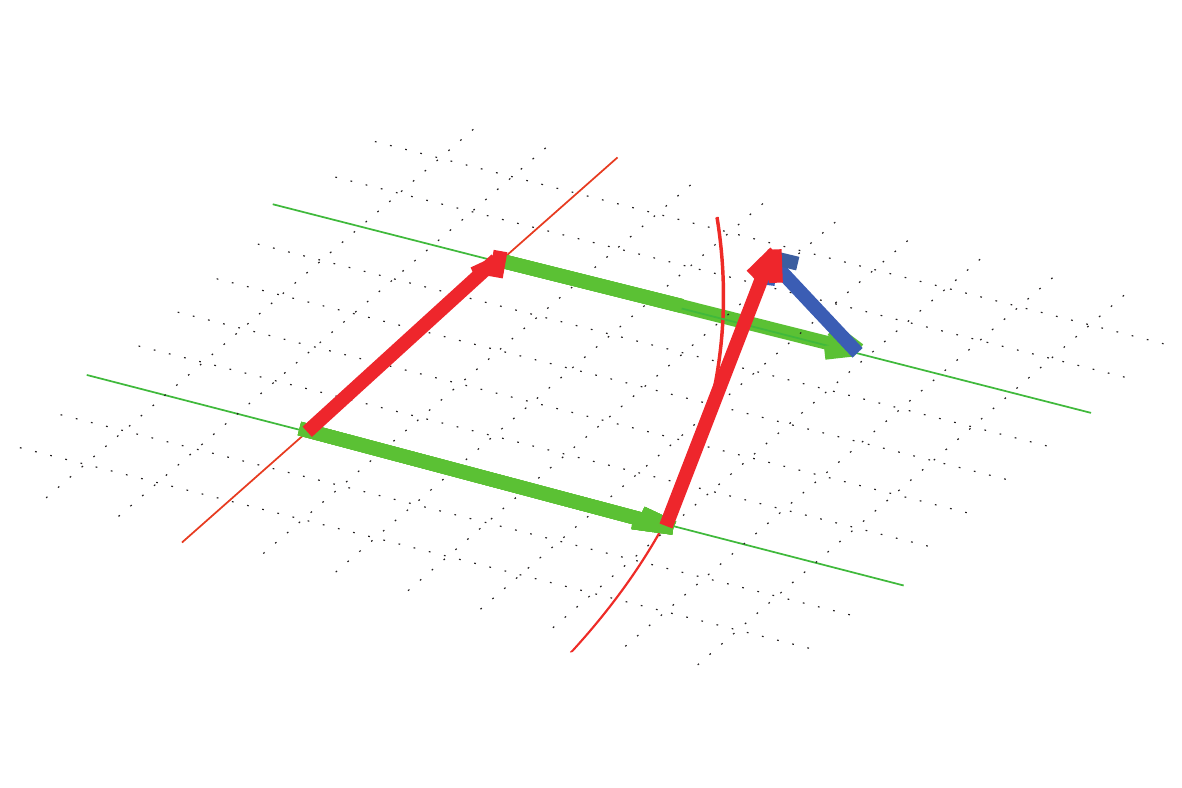}
\caption{Lie derivative outside the spam}
\label{lieout}
\end{figure}

In order to provide an answer to these questions, we need to visualize what the above condition really means.
Using the above remarks, we can obtain useful information from a couple of images. In figure \ref{liein} and figure \ref{lieout} we have plotted two vector fields (let us call $X$ the green one, and $Y$ the red one) with their flow lines. Note that in these figures, in order to avoid confusion, only two vectors for every vector field have been plotted.  Note also that the vector fields used are defined on a manifold that is three dimensional, but in our figures we represent them only on a two dimensional plane, because in this manner the figures are more clear. 

In figure \ref{liein}a we have vanishing Lie brackets. The two vector fields are both constant, so the vector field $Y$ does not change along the flow of the vector field $X$. This means that $L_X Y = \left[ X,Y \right]=0$. For instance, imagine to have in figure \ref{liein}a a vector field $u^\alpha$ which is constant everywhere, and vertically directed. In this case, the two vector fields represented in this figure (that are constant everywhere, orthogonal to each other and orthogonal to $u^\alpha$) can be used to span surfaces which are orthogonal to $u^\alpha$.\footnote{Note that the right procedure is the following. Consider the vectors $X|_p$ and $Y|_p$; at every point $p$ they span a surface $W_p$. We can collect the $W_p$ in $W$. $W$ is the tangent space of a surface, that being flat coincide with $W$ itself; this is the surface we were looking for.} These surfaces are nothing more than flat surfaces, that foliate the $\left(2+1\right)$-dimensional spacetime into spatial surfaces, as the pages of a book foliate the book itself.

Without opening it, we can fold a book (if it has no rigid cover); the pages still foliate it, but now they are no longer flat. If we imagine them to be spatial (non-flat) surfaces, we see that we can find a vector field $u^\alpha$ that is orthogonal to them (and now it is non-constant). In this way we can visualize how it appears the spacetime in 2+1 dimensions when we foliate it into spatial surfaces, parametrized by time.

Our aim is to understand why there are situations where we cannot work out such a foliation, so let us focus on the Lie brackets. 
From figure \ref{liein}b we can understand what it means for two vector fields to have Lie brackets which lie in $W$. In fact, if we imagine to have again a vector field $u^\alpha$ constant and vertically directed, $W$ is the surface spanned by the vector fields $X$ and $Y$ in the figure. The vector field $Y$ (in red) is no longer constant, and in the figure the amount by which it has changed along the flow lines of the vector field $X$ is represented by the blue vector. In this case we see that the Lie brackets $\left[X,Y\right]=L_X Y$ lie in $W$, that is the surface spanned by the two vector fields. It constitutes one of the (flat) surfaces orthogonal to $u^\alpha$ that can be found, and which foliate the spacetime. 

Now, with the help of figure \ref{lieout} we can understand what it means for two vector fields to have Lie brackets outside $W$. If, given the vector field $u^\alpha$, at every point the vector fields $X$ and $Y$ represented in figure \ref{lieout} are orthogonal to it, then a surface orthogonal to $u^\alpha$ cannot be constructed as a consequence of the fact that $\left[X,Y\right]\notin W$. This can be seen as follows.
By construction, $W$ is the collection of the $W_p$ that, at every point $p$, are the collection of the vectors which at the point $p$ are orthogonal to $u^\alpha$. If the vector fields $X$ and $Y$ are orthogonal to $u^\alpha$, this means that when we evaluate them at each point $p$ we obtain vectors $X|_p$ and $Y|_p$ which are orthogonal to $u^\alpha$, so $W$ is the collection of the $W_p$, which are spanned, point by point, by the vectors $X|_p$ and $Y|_p$. From the figure we see that now the amount (in blue) by which $Y$ has changed along the flow lines of the vector field $X$ does not lie in $W$. So in this case the Lie brackets $\left[X,Y\right]=L_X Y$ do not lie in $W$. From the figure we see also that now we cannot repeat the procedure outlined above to produce a two-dimensional surface orthogonal to $u^\alpha$. $W$, defined as a collection of $W_p$, does not define a surface.

One question may arise: is it true that in this way we do not obtain surfaces orthogonal to $u^\alpha$, or may be we obtain surfaces orthogonal to $u^\alpha$, but simply they are curved, instead of being flat?

The answer is that with the vector fields $X$ and $Y$ of figure \ref{lieout} we do not obtain surfaces orthogonal to $u^\alpha$. In order to obtain (curved) surfaces orthogonal to $u^\alpha$ we must have a vector field $X$ which changes along the flow lines of the vector field $Y$ in such a way that their Lie brackets remain in $W$.

\begin{figure}
\includegraphics[scale=1.1]{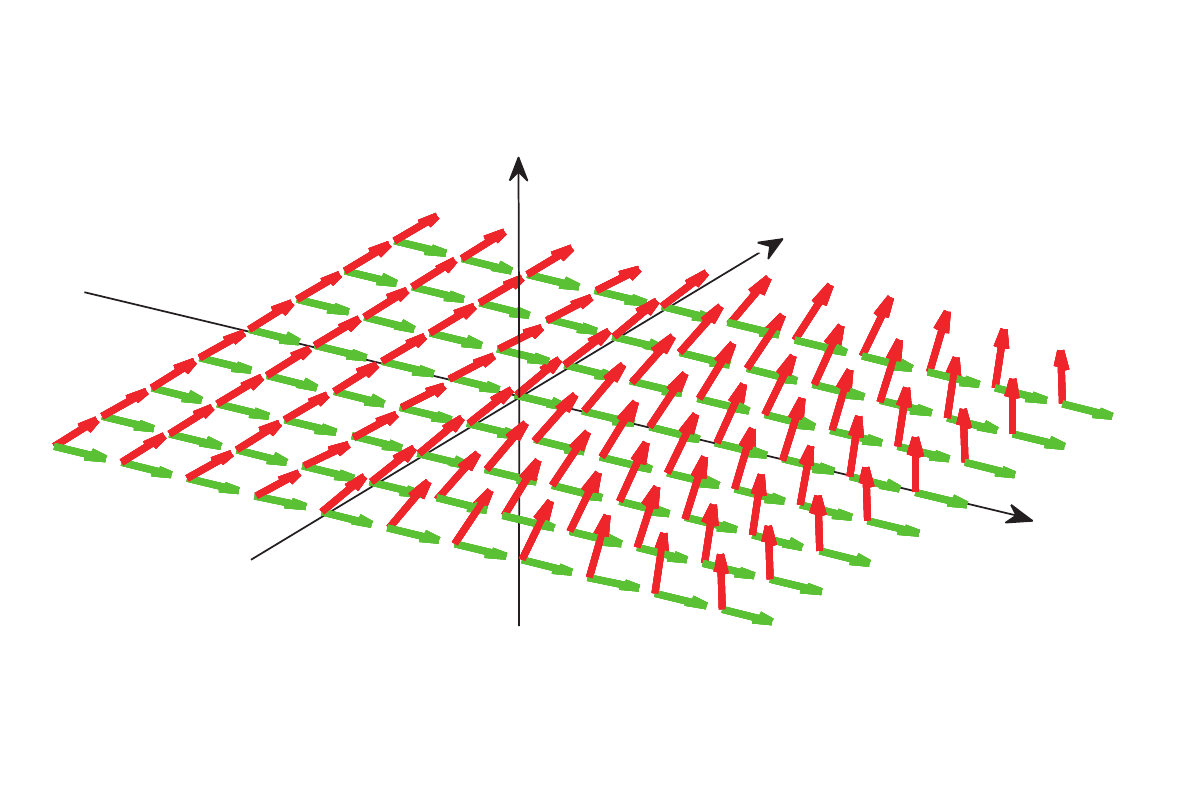}
\caption{A suggestion that inspires the understanding of Frobenius Theorem}
\label{frobeniusfigure}
\centering
\end{figure}

Finally, we can collect the above results in figure \ref{frobeniusfigure}, that can help in visualizing the content of Frobenius theorem. We consider an $u^\alpha$ which is orthogonal to two vector fields $X$ and $Y$ with the following structure. $X$ is constant everywhere. $Y$ is orthogonal to $X$, and constant in a region of the space, while in another region it turns around the direction of $X$. This means that in the first region (on the left in figure \ref{frobeniusfigure}) we have the situation of figure \ref{liein}a, while in the other region (on the right in figure \ref{frobeniusfigure}) we have the situation of figure \ref{lieout}. So, in the region on the left the vector fields $X$ and $Y$ have Lie brackets which lie in $W$, the space $W_p$ spanned at each point by $X|_p$ and $Y|_p$ are orthogonal to a surface. This is the surface orthogonal to $u^\alpha$ which we were looking for. On the other hand, in the region on the right they have Lie brackets which do not lie in $W$, and the space $W_p$ are not tangent to the same surface, as it is clear from figure \ref{frobeniusfigure}, so a surface orthogonal to $u^\alpha$ cannot be found.

The situation acquires its full physical meaning when we recall (as we shown in section \ref{3+1frob} for the $\left(3+1\right)$-dimensional case, and extended in this section to the 2+1 dimensions) that the Lie brackets lie in $W$ if and only if the vorticity vanishes.

\chapter{Backreaction}
  \label{backreaction}
  
This chapter explains what backreaction is, and shows some of the results that have been obtained up to now. We explain the physical idea that underlies the concept of backreaction. Then we present backreaction in the case of irrotational dust, introducing the Buchert equations. Finally we present the equations that hold in the case of general matter with non-zero vorticity.

\section{From statistical homogeneity and isotropy to backreaction}
  \label{statomisot}

The early universe is close to exact homogeneity and isotropy in two ways:\footnote{Exact homogeneity and isotropy are \ev{local} symmetries: they mean that all points and all directions are equivalent. These two assumptions are mathematically defined in section \ref{twoassumptions}.} the amplitude of the perturbations around homogeneity and isotropy in the density field is small, and the distribution of the perturbations is statistically homogeneous and isotropic.\footnote{Let $\rho\left(\vec{r}\right)$ be the microscopic density function and $\mathcal{P}\left[\rho\left(\vec{r}\right)\right]$ its probability density functional. This functional can be interpreted as the joint probability density function of the random variables $\rho\left(\vec{r}\right)$ at every point $\vec{r}$. Then we can represent a mass density field as a stochastic process that consists in extracting the value of the random variables $\rho\left(\vec{r}\right)$ at any point of the space. We say that the stochastic process is statistically homogeneous when the above functional is invariant under any spatial translation. If it is also invariant under spatial rotation, then the density field is statistically isotropic.}
These properties are verified by observations of the cosmic microwave background and large scale structure.
Therefore the early universe can be well described by a FLRW model with linear perturbations.

In the universe at late times structures form, and because they are non-linear exact homogeneity and isotropy are broken. Density fluctuations are of order one, so a FLRW model plus linear perturbations become inappropriate. But if we consider the universe above a certain scale, it may appear, over that scale, homogeneous and isotropic. This is encoded in the concepts of statistical homogeneity and isotropy, which still hold. This roughly means that if we take a box anywhere in the universe, larger than the homogeneity scale, the mean quantities in the box do not depend on its location, orientation or size.\footnote{The definition of homogeneity scale can be found in \cite{syloslabini}.} So we can, instead of studying the behavior of local quantities, look at average ones.

In the usual procedure, i.e. describing the universe with a homogeneous and isotropic model, we employ smooth quantities that are thought to represent some sort of an average of the corresponding quantities of the real universe, and we evolve them through the Einstein equation.\footnote{Note that in the usual approach, no mathematical operation of averaging is employed. We simply describe the universe with a model whose smooth quantities are thought to represent on average the corresponding quantities of the real universe.} In this way we are evolving the physical quantities after having smoothed out the inhomogeneities.

However, physically one should first evolve the inhomogeneous quantities using the Einstein equation, and then take the average. The Einstein equation is non-linear, so the two procedures of time evolution and averaging do not commute.
The fact that the average evolution of a clumpy space is not the same as the evolution of a smooth space (i.e. a space where the inhomogeneities have been smoothed in the sense outlined above, before evolving) is referred to as backreaction.

In order to investigate this, we must first define a procedure of averaging. The averaging of quantities in general relativity is an involved issue, because the metric is a dynamical variable which enters non-linearly in the Einstein equation, and in a generic spacetime there are no preferred time slices one could average over.

Throughout this work we follow the approach of Buchert.
The problem of averaging was studied in Newtonian cosmologies in \cite{buchert1996}, and in general relativity in \cite{buchert1999}.\footnote{Note that many different procedures of averaging have been carried out. For instance, a more geometrical approach can be found in \cite{buchertecarfora}.}
In that paper the Buchert equations, the analogues of the Friedmann equations for averaged inhomogeneous dust cosmologies, are derived. In \cite{buchert2001} the case of an ideal fluid is analysed. The generalization of the procedure to the case of general matter and non-vanishing vorticity is worked out, in the covariant formalism, in \cite{syksy2009b}.

Whether the homogeneity scale exists, and, if so, what is its value, are still open questions (see \cite{syksy2009b}, \cite{syloslabini}).\footnote{Note that in statistical physics the existence of a homogeneity scale is called \emph{spatial homogeneity}.
See for instance \cite{syloslabini}. Note also that this can lead to a certain confusion, because the expression ``spatial homogeneity'' interpreted in the framework of general relativity may refer to exact homogeneity.} Some studies suggest that an approximate value may be 100 Mpc, some others propose 300 Mpc, and there are also studies that argue that there is no evidence for the homogeneity scale at all.
Anyway, in general the existence of an homogeneity scale is expected. On the other hand, whether it exists or not it is not clear from the observations of large scale structure. In what follows we assume the existence of an homogeneity scale, for otherwise to consider mean quantities would be useless.

 \section{Irrotational dust}
Let us consider what we obtain when we describe the universe assuming that the energy density of matter dominates everywhere over the pressure, the anisotropic stress and the energy flux. In this case we can consider the matter as dust,\footnote{See section \ref{partfluids}.} i.e.
  \begin{equation}
    T_{\alpha\beta}=\mu u_\alpha u_\beta\ .
  \end{equation}

Since we have chosen the pressure to be zero, the fluid elements do not interact with each other, so the motion is geodesic and $u^\alpha$ is a tangent vector to time-like geodesics, i.e. $\dot{u}^\alpha=0$.\footnote{A proof of this fact is given in a footnote to equation \re{dotuzero}.}
  
As stated by Frobenius' theorem (formulation \re{fbth}, page \pageref{fbth}), there exists a family of spatial hypersurfaces that are orthogonal to $u^\alpha$ and which provide a foliation that fills the spacetime exactly once \ev{if and only if} the vorticity is zero. If we use comoving coordinates \re{normcoord}, the flow-orthogonal hypersurfaces coincide with the hypersurfaces of constant proper time of comoving observers.

In this section we investigate what happens in the simplest case, i.e. assuming $\omega_{\alpha\beta}=0$.\footnote{We recall that $\omega_{\alpha\beta}=0\Leftrightarrow\omega^{\alpha}=0\Leftrightarrow\omega=0$, as discussed in section \ref{vorticitysection}.}
Vorticity contributes positively to acceleration, as we will see in section \ref{nzvortsec}, so setting it to zero gives a lower bound to acceleration, which in the irrotational case is always non-positive (this is just an expression of the fact that, for matter satisfying the strong energy condition \re{strong}, gravity is always attractive).

\subsection{Defining the average}
\label{defavsec}

Given a foliation of spacetime, taking the spatial average of equations for scalar fields is a covariant operation, so we proceed without any choice of particular coordinates.

Following \cite{syksy2009}, we define the \emph{spatial average $\av{f}$ of a scalar quantity $f$} as its integral over the hypersurface of constant proper time $t$ orthogonal to $u^\alpha$, divided by the volume of the hypersurface:
  \begin{equation}
     \av{f}\left(t\right)\doteq\frac{\int_t f \epsilon}{\int_t \epsilon}\ ,
       \label{averagedef}
  \end{equation}
where $\epsilon$ is the volume form of the spatial hypersurface considered, i.e. (in analogy to section \ref{volelsec}) we have
  \begin{equation}
    \epsilon\doteq\sqrt{ \lvert ^{\left( 3 \right)}g \rvert}dx^{1}\wedge dx^{2}\wedge dx^{3} \ ,
  \end{equation}
and $ ^{\left(3\right)} g \doteq \det\left( ^{\left(3\right)} g\left(t,x^1,x^2,x^3\right) \right) $ is the determinant of the metric on the hypersurface of constant proper time.\footnote{In the case of irrotational dust the metric $ ^{\left(3\right)}g_{\alpha\beta}$ coincides with the restriction of the projector $h_{\alpha\beta}$ to the three-dimensional hypersurface of averaging.}

Let us now define the \emph{scale factor $a\left(t\right)$} as the volume of the hypersurface of constant proper time to the power $1/3$, i.e.
  \begin{equation}
    a\left(t\right)\doteq\left( \frac{\int_t \epsilon}{\int_{t_0} \epsilon} \right)^{\frac{1}{3}} \ ,
      \label{adef}
  \end{equation}
normalized to unity at time $t_0$, which (as stated in section \ref{exprate}) we take to be today.
The \emph{Hubble parameter} is now defined as\footnote{Compare this definition to the one given in \re{hubbleparam}.}
  \begin{equation}
    H\left(t\right)\doteq\frac{\dot{a}\left(t\right)}{a\left(t\right)}\ .
  \end{equation}

Using definition \re{averagedef} we can now calculate the \emph{average expansion rate $\av{\Theta}$} which turns out to be
  \begin{equation}
    \av{\Theta}\left(t\right)\doteq\frac{\int_t \Theta \epsilon}{\int_t \epsilon}
                             =\frac{\partial_t\int_t \epsilon}{\int_t \epsilon}
                             =3\frac{\dot{a}}{a} \ .
                     \label{avtheta}
  \end{equation}
We later prove this in the most general case, leading to equation \re{Navtheta}.

The time evolution and the average procedure do not commute, and this fact is expressed by the \emph{commutation rule}:
  \begin{equation}
    \partial_t\av{f}=\av{\dot{f}}+\av{f \Theta} - \av{f}\av{\Theta}\ ,
      \label{commrule}
  \end{equation}
explicitly calculated in the general case in section \ref{gencommrulesec}.

\subsection{The scalar equations}

In chapter \ref{3+1covform} we have shown how to turn the Einstein equation \re{efe} into a set of evolution and constraint equations (see sections \ref{fromricci}, \ref{frombianchi} and \ref{frombianchitwicecont}). Now we want to obtain the corresponding equations for averaged quantities.

Because only scalar quantities can be straightforwardly averaged, we turn our attention to the scalar part of the Einstein equation. We are interested only in the Raychaudhuri equation \re{raycheq} and the continuity equation \re{enconseq}.\footnote{A third scalar equation already mentioned, the geometrical identity \re{vortdiveq}, is not considered here because it deals with vorticity alone.}
We consider also another scalar equation, the Hamiltonian constraint, so the equations that we want to average (written for irrotational dust) are:\footnote{Note that in this chapter, according to the motivations presented in the introduction, we assume the cosmological constant to be zero.}
  \begin{align}
    \dot{\Theta} + \frac{1}{3}\Theta^2 = -\frac{1}{2} \mu - 2 \sigma^2  \qquad &\text{(Raychaudhuri equation)}
    \label{scalar1} \\
    \frac{1}{3} \Theta^2 = \mu - \frac{1}{2} {\tr} + \sigma^2  \qquad &\text{(Hamiltonian constraint)} 
     \label{hamconst} \\
    \dot{\mu} + \Theta \mu = 0  \qquad &\text{(continuity equation)}
    \label{scalar3} \ ,
  \end{align}
where $\tr$ is the Ricci scalar on the three-dimensional space orthogonal to $u^\alpha$.
The first and second equations are local generalizations of the Friedmann equations,\footnote{See section \ref{friedmannsection}, where the corresponding equations for an ideal fluid are used in order to obtain the Friedmann equations.} while the continuity equation shows that mass is conserved.\footnote{Note that this is true in the case of dust, but not in the case of general matter, as stated after equation \re{enconseq}.} We derive the Hamiltonian constraint in section \ref{obtainhcsection}.
These are the equations that relate the local quantities of interest; in the next section we obtain their averaged counterparts.

\subsection{Buchert equations}

Now we want to obtain the averaged equations that arise from application of the outlined procedure to the above set of scalar equations.
Considering the scalar equations \re{scalar1}, \re{hamconst} and \re{scalar3}, taking the average (as defined in \re{averagedef}), using the commutation rule \re{commrule} and the relation \re{avtheta}, we have the \emph{Buchert equations}:
   \begin{align}
    3\frac{\ddot{a}}{a}=-\frac{1}{2} \av{\mu} +  \q  \qquad &\text{(averaged Rayc. eq.)}
    \label{avraycheq} \\
    3\frac{\dot{a}^2}{a^2}=\av{\mu}-\frac{1}{2}\av{ \tr}-\frac{1}{2} \q  \qquad &\text{(averaged Ham. const.)}
    \label{avhamcon} \\
    \partial_t\av{\mu}+3\frac{\dot{a}}{a}\av{\mu}=0  \qquad &\text{(averaged cont. eq.)} \ ,
    \label{avconteq}
  \end{align}
  where the \emph{backreaction variable $ \q$} is defined as
    \begin{equation}
       \q \doteq \frac{2}{3}\left( \av{\Theta^2} - \av{\Theta}^2\right) - 2 \av{\sigma^2} \ ,
        \label{backreactiondef}
    \end{equation}
  and it contains the effects of inhomogeneities and anisotropies.

These equations are different from the Friedmann equations \re{friedmann1} and \re{friedmann2} for two reasons.

The first difference lies in the slightly different meaning that the scale factor $a\left(t\right)$ has in the two different sets of equations. In the FLRW case the scale factor $a\left(t\right)$ is a component of the metric (see appendix A) and indicates how the space is evolving locally, while this is not the case in the present context, where it gives the total volume of a region, as follows from its definition \re{adef}.

The second and more important difference is that the Buchert equations contain the backreaction term, which is zero in the FLRW case. It follows that we can have acceleration, and the average spatial curvature can have a non-trivial evolution.

The Buchert equations are the generalization of the Friedmann equations to an inhomogeneous dust universe. If the backreaction $ \q$ is small, they reduce to the Friedmann equations written for the averaged quantities. Since $ \q$ is quadratic in small perturbations, it is obvious that for small $ \q$ the linear perturbation theory is applicable.\footnote{Note that it may be that inhomogeneities are large, but that $ \q$ is small. So, also in a universe with significant inhomogeneities it is possible that the Buchert equations reduce to the Friedmann equations written for the averaged quantities.}

\subsubsection{Acceleration from backreaction}
 
As we can see in equation \re{backreactiondef}, the backreaction variable has two parts; one originates from $\Theta$, the other from $\sigma$.
The second term is the average of the squared shear scalar, which is also present in the local equations, and vanishes in the homogeneous and isotropic case. When present, its contribution is always negative, so it decelerates the expansion. On the other hand the first term, the variance of the expansion rate, also vanishes if the expansion is homogeneous, but it has no local counterpart. If the expansion is not homogeneous, this term (being a variance) is always positive, and so it acts to accelerate the expansion. So from the averaged Raychaudhuri equation we can see that, if the variance is large enough w.r.t. shear and energy density, the expansion rate accelerates, even if the local expansion (governed by the Raychaudhuri equation) decelerates everywhere.

This is precisely the way backreaction provides acceleration, which in the FLRW case requires the presence of either dark energy or the cosmological constant, as the Friedmann equations \re{friedmann1} and \re{friedmann2} show. From the mathematical point of view, this is evident from the equations; however, from the physical point of view it may appear difficult to imagine how it is possible that the expansion rate accelerates while the local expansion rate decelerates everywhere. The explanation of how this happens is given in \cite{syksy2006a} and \cite{syksy2006b}, where a toy model made of two regions, one overdense and the other underdense, is developed for this purpose.

A semi-realistic model which generalizes the previous one and uses as a starting point a spatially flat FLRW model with a linear Gaussian field of density fluctuations is contained in \cite{syksy2008a}, \cite{syksy2008b}. In this case it turns out that, due to backreaction, there is less deceleration.

These are approximate models, and the question of whether or not the inhomogeneities present in the real universe give the right amount of acceleration needed to explain observations is still open.

\subsubsection{The relation between backreaction and average spatial curvature}
  \label{intconsec}

The averaged Raychaudhuri equation \re{avraycheq} and the averaged Hamiltonian constraint \re{avhamcon} form a system of two equations for the three unknown variables $a\fdt$, $\avtr\fdt$ and $ \q\fdt$, so we cannot obtain these three quantities from this set of equations alone. We can take the time-derivative of equation \re{avhamcon}, and insert both equations \re{avraycheq} and \re{avhamcon} into the result. This procedure yields a relation between the averaged Ricci scalar $\avtr$ and the backreaction $ \q$ which is of fundamental importance:
    \begin{equation}
      \partial_t \avtr + 2 \frac{\dot{a}}{a} \avtr = - \partial_t  \q - 6 \frac{\dot{a}}{a}  \q \ ,
        \label{intcond}
    \end{equation}
a necessary \emph{integrability condition} between equations \re{avraycheq} and \re{avhamcon}.
The meaning of this equation is that, in general, the evolution of backreaction and averaged scalar curvature influence each other.

Immediately we see that there are some particular solutions for the variables $ \q$ and $\avtr$ expressed as functions of $a\fdt$ that are particularly interesting.

\paragraph{First: [$\avtr=0 \Rightarrow \q= \q\left(t_0\right)a^{-6}$].}
If we consider a portion of the universe that is \emph{spatially flat on average}, i.e. $\avtr=0$, then solving equation \re{intcond} gives us
  \begin{equation}
     \q= \q\left(t_0\right)a^{-6}.
  \end{equation}
  
\paragraph{Second: [$ \q=0 \Rightarrow\avtr=\avtr\left(t_0\right)a^{-2}$].}
For \emph{vanishing backreaction} we obtain that the average scalar curvature is given by
  \begin{equation}
    \avtr=\avtr\left(t_0\right)a^{-2}.
  \end{equation}
Here $a\fdt$ is given by the Friedmann equations, and we are describing a system that is FLRW \ev{on average}.
This does not mean that there are no inhomogeneities, but just that their effects compensate each other, so backreaction vanishes.

\paragraph{Third: [$ \q= \q\left(t_0\right)a^{-6}$, $\avtr=\avtr\left(t_0\right)a^{-2}$].}
An obvious solution of \re{intcond} in the case of non-vanishing backreaction and non-vanishing average Ricci scalar, given the previous two solutions, is given by:
  \begin{equation}
     \q= \q\left(t_0\right)a^{-6}  \quad \text{and}  \quad \avtr=\avtr\left(t_0\right)a^{-2} \ .
  \end{equation}
In this case both sides of \re{intcond} are zero, so we have \ev{independent evolution} of the backreaction variable and the average Ricci scalar.

\section{General matter with non-zero vorticity}
  \label{nzvortsec}
   
   In the previous section we have described the universe as filled with irrotational dust, i.e. we have approximated the vorticity as being zero, and the matter density to be by far the biggest component of the energy-momentum tensor. However, the treatment of matter as irrotational dust is just an approximation. On one hand, rotation and velocity dispersion are important for stabilizing structures on small scales; on the other, including matter other than dust is necessary in order to treat backreaction in the early universe (as during inflation or preheating). Even if vorticity and the non-dust nature of matter may be expected to be unimportant for the overall cosmological evolution of the real universe at late times, it is good to include such effects in order to establish under which conditions they can be neglected.
In practice, we want now to show how to generalize the Buchert equations, first given in \cite{buchert1999}. A generalization to the case of an ideal fluid is given in \cite{buchert2001} using the ADM formalism, while in \cite{gaspmarven} the covariant formalism is used, with an arbitrary averaging hypersurface. The generalization to the case of general matter content, with an arbitrary hypersurface of averaging, and using the covariant formalism, is developed in \cite{syksy2009b}; this is the treatment we shall follow in this thesis.

  \subsection{Spacetime geometry}
    \subsubsection{Statistical homogeneity and isotropy}

As explained in section \ref{statomisot}, we assume that a homogeneity scale exists.
Even if we want to include vorticity in the description, it is still possible to foliate the spacetime into three-dimensional hypersurfaces.\footnote{In fact, Frobenius' theorem tells us that in this case no foliation of the spacetime into hypersurfaces that are orthogonal to the four-velocity vector field $u^\alpha$ (relative to the vorticity that does not vanish) exists, but we can \ev{always} choose a foliation of the spacetime, that in this case will not be orthogonal to $u^\alpha$, so it will not be interpreted as the three-dimensional space orthogonal to the observer's velocity.} So we assume that there exists a foliation of the spacetime into spatial hypersurfaces of statistical homogeneity and isotropy (i.e. three-dimensional spaces, each possessing statistical homogeneity and isotropy), which we denote by $\mathcal{N}$. We denote as $t$ the time that is constant on one of these hypersurfaces, and we use the notation $\mathcal{N}\fdt$ when referring to a particular hypersurface. The assumptions of statistical homogeneity and isotropy concern only average quantities evaluated over large scales, so the local dynamics can be complex. But over scales larger than the homogeneity scale there are no preferred locations or directions, and no correlations. Moreover, integrated flux of any quantity through the boundary of a volume larger than the homogeneity scale vanishes.

   \subsubsection{The two frames}
   We denote the normal vector to the hypersurfaces of statistical homogeneity and isotropy $\mathcal{N}$ as $n^\alpha$. We denote the velocity of the observers, which is completely general (i.e. not assumed to be irrotational or geodesic), by $u^\alpha$. Both are normalized to unity,
   \begin{equation}
     n_\alpha n^\alpha = -1 = u_\alpha u^\alpha \ .
   \end{equation}
   
   Now we have two different frames, so it is useful, in order to avoid confusion, to give again some of the definitions involving $u^\alpha$ and $n^\alpha$.\footnote{In this section we present only the equations we need, while a wider discussion about the transformations under a four-velocity boost, and also some of the relations we use, can be found in \cite[appendix A.2]{ilcerchiosichiude}.}
  The spacetime metric is $g_{\alpha \beta}$, while the tensors that project onto the hypersurface $\mathcal{N}$ orthogonal to $n^\alpha$ and the rest space orthogonal to $u^\alpha$ are defined, respectively, and in analogy to \re{proh}, as
  \begin{equation}
    h_{\alpha\beta}=g_{\alpha\beta} + n_\alpha n_\beta  \qquad \text{and}  \qquad h^{\left(u\right)}_{\alpha\beta}=g_{\alpha\beta} + u_\alpha u_\beta \ .
  \end{equation}
Moreover, restricting the projection tensor $h_{\alpha\beta}$ to $\mathcal{N}$ gives the metric on $\mathcal{N}$.
In analogy to \re{spatderiv} the spatial derivative of a tensor quantity is defined as
\begin{equation}
\hat{\nabla}_{\mu}T^{\alpha \ldots \beta} _{\phantom{\alpha \ldots \beta} \gamma \ldots \delta}=h^{\alpha}_{\phantom{\alpha} \lambda }  h^{\rho}_{\phantom{\rho} \gamma } \ldots h^{\beta}_{\phantom{\beta} \nu } h^{\sigma}_{\phantom{\sigma} \delta } h^{\phi}_{\phantom{\phi} \mu}\nabla_{\phi} T^{\lambda \ldots \nu} _{\phantom{\lambda \ldots \nu} \rho \ldots \sigma} \ ,
\end{equation}
with projection over every free index, and similarly for scalars and vectors.
The volume element on $\mathcal{N}$ is $\eta^{\alpha\beta\gamma}\doteq\eta^{\alpha\beta\gamma\delta}n_\delta$, where $\eta^{\alpha\beta\gamma\delta}$ is the spacetime volume element.

While $t$ is the \ev{time that is constant on $\mathcal{N}$}, we denote by $s$ the \ev{proper time of the frame of statistical homogeneity and isotropy}. They do not coincide, unless $\dot{n}_\alpha =0$. The derivatives w.r.t. the two different times are, respectively,
  \begin{equation}
    \partial_s = n^\alpha \nabla_\alpha  \qquad \text{and}  \qquad \partial_t=m^\alpha\nabla_\alpha \ ,
      \label{partialst}
  \end{equation}
and we denote the first one also with an overdot. Here $m^\alpha\doteq\dot{t}^{-1}n^\alpha$, and $n_\alpha$ can be written as $n_\alpha=-\dot{t}^{-1} \partial_{\alpha} t$.

We define
  \begin{equation}
    \Gamma\doteq-n_\alpha m^\alpha \ ,
  \end{equation}
so we have
  \begin{equation}
    \Gamma = \dot{t}^{-1} = \partial_t s  \qquad \text{and}  \qquad m^\alpha= \Gamma n^\alpha \ ,
      \label{dsdt}
  \end{equation}
where physically $\Gamma$ describes the time dilation due to the non geodesic motion of the $n^\alpha$ frame.
When $\dot{n}^\alpha=0$, i.e. when $\mathcal{N}$ is a hypersurface of constant proper time, then we can set $\Gamma=1$, obtaining $t=s$. Inserting the second equation of \re{dsdt} in the second equation of \re{partialst}, and using the first equation of \re{partialst}, we obtain
  \begin{equation}
    \partial_t=\Gamma\partial_s \ .
  \end{equation}
Note that in addition to $s$ and $t$ we have also the proper time $\tau$ measured by the observers, defined by $u^\alpha$. 

\subsubsection{Fluid kinematics}

We can split the covariant derivatives of $n^\alpha$ and $u^\alpha$ in a way analogous to the splitting \re{split}:
  \begin{align}
    \nabla_{\beta} n_{\alpha}&=- \dot{n}_\alpha n_\beta+\hat{\nabla}_{\beta}n_{\alpha}=- \dot{n}_\alpha n_\beta+\frac{1}{3}\Theta h_{\alpha\beta} +           \sigma_{\alpha\beta} \ , \label{splitn} \\
    \nabla_{\beta} u_{\alpha}&=
    -u_\beta u^\gamma\nabla_\gamma{u}_\alpha+\frac{1}{3}\Theta^{\left(u\right)} h^{\left(u\right)}_{\alpha\beta} 
    + \omega^{\left(u\right)}_{\alpha \beta} +\sigma^{\left(u\right)}_{\alpha\beta}\ ,
  \end{align}
where the various quantities are defined in analogy to those in section \ref{veldec}.

The vorticity $\omega_{\alpha\beta}\doteq\hat{\nabla}_{[\beta}n_{\alpha]}$ is absent in equation \re{splitn} because it is zero, a result that follows through Frobenius' theorem (statement \ref{fbth}) from the fact that $n^\alpha$ is hypersurface orthogonal.

\subsubsection{The energy-momentum tensor}

In analogy to equation \re{ttensor} we can also split the energy-momentum tensor with respect to $n^\alpha$ or $u^\alpha$, respectively, as:
\begin{align}
  T_{\alpha\beta} &= \mu^{\left(n\right)} n_\alpha n_\beta + q^{\left(n\right)}_\alpha n_\beta + n_\alpha q^{\left(n\right)}_\beta + p^{\left(n\right)} h_{\alpha\beta} + \pi^{\left(n\right)}_{\alpha\beta} \ , 
  \label{tn} \\
  T_{\alpha\beta} &= \mu^{\left(u\right)} u_\alpha u_\beta + q^{\left(u\right)}_\alpha u_\beta + u_\alpha q^{\left(u\right)}_\beta + p^{\left(u\right)} h_{\alpha\beta} + \pi^{\left(u\right)}_{\alpha\beta} \ ,
\end{align}
where the quantities are defined as in section \ref{tdec}.
The second decomposition give us the quantities measured by the observers, which are moving with velocity $u^\alpha$.

Note that the non-dust terms present in the two decompositions can have different origins. From one side, it is possible that the matter we are describing cannot be approximated with dust in any frame. On the other, they can arise from the fact that an ideal fluid looks non-ideal to a non-comoving observer.

 \subsection{The averages}
 \label{genavproc}
 
Turning our attention to the procedure of averaging, before calculating the generalization of the Buchert equations we need to give some definitions.

First of all, we define the \emph{average of a scalar $f$ on the hypersurface $\mathcal{N}$} as
  \begin{equation}
    \av{f}\fdt \doteq \frac{\int f \epsilon}{\int \epsilon} \ .
      \label{avgendef}
  \end{equation}
As already pointed out, the time $t$ is the one that is constant on $\mathcal{N}$, and it is not a proper time, neither the proper time associated to $n^\alpha$ (that is, $s$), nor the proper time measured by the observers (that is, $\tau$).
This definition is analogous to \re{averagedef}. While in that case the averages are taken over three-dimensional surfaces which could be interpreted as the usual three-dimensional surfaces, now we are averaging over the three-dimensional hypersurfaces $\mathcal{N}\left(t\right)$ of statistical homogeneity and isotropy.

  \subsubsection{Commutation rule}
    \label{gencommrulesec}
    
The commutation rule between averaging and taking the derivative with respect to the time $t$ is now given by
  \begin{equation}
    \partial_t\av{f}=\av{\partial_t f} + \av{\Gamma \Theta f} - \av{f} \av{\Gamma \Theta} \ .
      \label{gencommrule}
  \end{equation}

To prove this, consider the derivative with respect to $t$ of the average value of $f$, which, using the definition \re{avgendef}, turns out to be
  \begin{equation}
    \partial_t \av{f} = \partial_t \frac{\int f \epsilon}{\int \epsilon}
    = \frac{\partial_t \int f \epsilon}{\int \epsilon} - \av{f} \frac{\partial_t \int \epsilon}{\int \epsilon} \ .
  \end{equation}
Now, inserting the equations
  \begin{equation}
    \partial_t \int \epsilon = \int \Gamma \Theta \epsilon
      \label{primointegrale}
  \end{equation}
and
  \begin{equation}
    \partial_t \int f \epsilon = \int \Gamma f \Theta \epsilon + \int \partial_t f \epsilon
      \label{secondointegrale}
  \end{equation}
and using again the definition \re{avgendef}, we get the desired result \re{gencommrule}.

The equations \re{primointegrale} and \re{secondointegrale} can be proven for instance by the use of comoving coordinates.

  \subsubsection{The scale factor}
  
In analogy to equation \re{adef}, we define the \emph{scale factor $a\left(t\right)$} as the volume of the hypersurface $\mathcal{N}\left(t\right)$ to power $1/3$, i.e.
  \begin{equation}
    a\left(t\right)\doteq\left( \frac{\int_t \epsilon}{\int_{t_0} \epsilon} \right)^{\frac{1}{3}} \ ,
      \label{adefgenmatcase}
  \end{equation}
normalized to unity at time $t_0$.
  
Because of the way we have defined it, $\Theta$ alone gives the rate of change of the local volume element with respect to time $s$; so $\Gamma \Theta $ gives the rate of change with respect to time $t$. 

Using definition \re{adefgenmatcase} and equation \re{primointegrale} the \emph{average expansion rate} turns out to be
  \begin{equation}
    3\frac{\partial_t a}{a}=\av{\Gamma\Theta} \ .
      \label{avgt}
  \end{equation}

\subsection{The scalar equations}

In order to generalize the Buchert equations \re{avraycheq}, \re{avhamcon} and \re{avconteq} to the case of general matter content with vorticity, we want to average the scalar part of the evolution equations that arise from the Einstein equation. So we consider the most general version of the Raychaudhuri equation \re{raycheq}, the Hamiltonian constraint \re{3+1Hamconst} and the continuity equation \re{enconseq}. Now we want to express these equations in terms of $n^\alpha$, so we have to take into account the decompositions \re{splitn} and \re{tn}, and remember that, by definition, in this frame the vorticity vanishes.

The three scalar equations that we are going to average now read:
  \begin{align}
    \dot{\Theta} + \frac{1}{3}\Theta^2 = -\frac{1}{2} \left( \mu^{\left(n\right)}+ 3 p^{\left(n\right)}\right) 
    -2\sigma^2 + \dot{n}_\alpha \dot{n}^\alpha + \hat{\nabla}_\alpha \dot{n}^\alpha  \quad &\text{Raych. eq.} 
        \label{genraych} \\
    \frac{1}{3} \Theta^2 = \mu^{\left(n\right)} - \frac{1}{2} \tr + \sigma^2 \quad &\text{Ham. const.} 
        \label{genhc} \\
    \dot{\mu}^{\left(n\right)} + \Theta \left(\mu^{\left(n\right)}+p^{\left(n\right)}\right)= -\hat{\nabla}_{\alpha}q^{\left(n\right)\alpha}
    -2\dot{n}_{\alpha}q^{\left(n\right)\alpha}-\sigma_{\alpha\beta}\pi^{\left(n\right)\alpha\beta} \quad &\text{Cont. eq.} 
        \label{genc} \ ,
  \end{align}
where $\tr$ is the Ricci scalar of the hypersurface $\mathcal{N}$.

 \subsection{The averaged equations}

In order to take the average of the above equations, we are going to use the procedure outlined in section \ref{defavsec} and generalized in section \ref{genavproc}.

The resulting equations, which represent a generalization of the Buchert equations to the case of general matter with non-zero vorticity, are:
\begin{align}
    3\frac{\partial_t^2 a}{a}=&-\frac{1}{2} \av{\mu^{\left(n\right)}+3p^{\left(n\right)}}+\av{\dot{n}_\alpha\dot{n}^\alpha}
    +\av{\hat{\nabla}_\alpha n^\alpha} +  \q 
    \notag \\
    &+ \av{\frac{1}{3}\left(\Gamma^2-1\right)\Theta^2 + \left(1-\Gamma^{-2}\right)\Gamma\partial_t\Theta
    +\Theta\partial_t\Gamma} 
        \label{genavraych} \\
    3\frac{\left(\partial_t a\right)^2}{a^2}=&\av{\mu^{\left(n\right)}}-\frac{1}{2}\avtr-\frac{1}{2} \q
    + \frac{1}{3}\av{\left(\Gamma^2-1\right)\Theta^2}   
        \label{genavhc} \\
   \partial_t\av{\mu^{\left(n\right)}}&+3\frac{\partial_t a}{a}\av{\mu^{\left(n\right)}+p^{\left(n\right)}}=
   -\av{\Gamma\Theta p^{\left(n\right)}}+\av{\Gamma\Theta}\av{p^{\left(n\right)}}
   \notag \\
   &-\av{\Gamma\dot{n}_\alpha q^{\left(n\right)\alpha} 
   + \Gamma \sigma_{\alpha\beta}\pi^{\left(n\right)\alpha\beta}}
   -\av{\hat{\nabla}_\alpha\left(\Gamma q^{\left(n\right)\alpha}\right)}  
        \label{genavc} \ ,
  \end{align}
where the backreaction variable $ \q$ is now defined as
  \begin{equation}
     \q \doteq \frac{2}{3} \left( \av{\Gamma \Theta^2} - \av{\Gamma \Theta}^2 \right) - 2 \av{\sigma^2} \ .
  \end{equation}

Let us explain how they have been derived, showing the calculation that leads, for instance, to the third one; the others can be obtained in a similar manner.

Start from the continuity equation \re{genc}. First of all note that, because it is expressed in terms of $n^\alpha$, the derivatives involved are derivatives w.r.t. time $s$. So \ev{changing to derivatives w.r.t. time $t$}, i.e. using $\partial_t f = \Gamma \partial_s f$, we get
  \begin{equation}
    \frac{1}{\Gamma} \partial_t \mu^{\left(n\right)} + \Theta \left( \mu^{\left(n\right)} + p^{\left(n\right)} \right) 
    = - \hat{\nabla}_\alpha q^{\left(n\right) \alpha}-2\dot{n}_\alpha q^{\left(n\right)\alpha}-\sigma_{\alpha\beta}\pi^{\left(n\right)\alpha\beta} \ .
  \end{equation}
Now multiplying by $\Gamma$ and \ev{taking the average} we obtain
  \begin{equation}
   \begin{split}
    \av{\partial_t \mu^{\left(n\right)}} + \av{\Theta\Gamma\left(\mu^{\left(n\right)}+p^{\left(n\right)}\right)} =
    &-\av{\Gamma \hat{\nabla}_\alpha q^{\left(n\right)\alpha}} -2\av{\Gamma\dot{n}_\alpha q^{\left(n\right)\alpha}} \\
    &-\av{\Gamma \sigma_{\alpha\beta} \pi^{\left(n\right)\alpha\beta}} \ .
   \end{split}
  \end{equation}
The \ev{use of the commutation rule \re{gencommrule}}, written for the scalar function $\mu^{\left(n\right)}$, leads to
  \begin{equation}
   \begin{split}
    \partial_t\av{\mu^{\left(n\right)}} + \av{\Gamma\Theta} \av{\mu^{\left(n\right)} + p^{\left( n \right)}} =
    & -\av{\Theta\Gamma p^{\left(n\right)}} + \av{p^{\left(n\right)}\av{\Gamma\Theta}}  \\
    & -\av{\Gamma\hat{\nabla}_{\alpha}q^{\left(n\right)\alpha}} -2\av{\Gamma \dot{n}_\alpha q^{\left(n\right)\alpha}} \\
    & -\av{\Gamma\sigma_{\alpha\beta}\pi^{\left(n\right)\alpha\beta}} \ .
   \end{split}
  \end{equation}
Finally, \ev{inserting \re{avgt}} and \ev{using $\dot{n}_\alpha=\Gamma^{-1}\hat{\nabla}_{\alpha}\Gamma$} gives, after some calculation, the desired result \re{genavc}.

\chapter{Backreaction in N+1 and 2+1 dimensions}

In this chapter we calculate the $\left(N+1\right)$-dimensional analogues of the Raychaudhuri equation and the Hamiltonian constraint, in the case of general matter and non-vanishing vorticity and shear. After generalizing the averaging procedure to the case of higher dimensions, we apply it to the above scalar equations (in the irrotational dust case), and we derive their averaged counterparts.
We obtain the $\left(N+1\right)$-dimensional integrability condition, and we point out some particular solutions.

We show in a detailed manner how to calculate these relations.\footnote{Note that in this chapter the indexes run over the values $0,1,\dots,N$.} By setting $N$ equal to $3$, we obtain the same equations used in the previous chapter.

We obtain also the $\left(2+1\right)$-dimensional version, and analyse it. For an even-dimensional hypersurface the geometry and the topology are related, a result encoded in the \emph{Gauss-Bonnet theorem}. We state the theorem, clarify some concepts needed in order to understand its meaning, and analyse what it tells us about backreaction in 2+1 dimensions.

In this chapter, for the sake of generality, we restore the cosmological constant $\Lambda$.

\section{The (N+1)-dimensional equations}
In this section we obtain the $\left(N+1\right)$-dimensional version of the Raychaudhuri equation and the Hamiltonian constraint. Setting $N=2$, we get the relations for the $\left(2+1\right)$-dimensional case.
For the sake of completeness we also obtain, with the choice $N=3$, the equations for the $\left(3+1\right)$-dimensional case.

In working out the calculations we proceed as in the $\left(3+1\right)$-dimensional case, but taking into account the following differences.
From the property $g_{\alpha\beta}g^{\beta\gamma}=\delta_\alpha^{\phantom{\alpha}\gamma}$ it follows that
  \begin{equation}
    g_{\alpha\beta}g^{\alpha\beta}=N+1\ ,
      \label{Npropmet}
  \end{equation}
and so, instead of the second equation \re{proph}, we have
  \begin{equation}
    h_\alpha^{\phantom{\alpha}\alpha}=N\ .
      \label{Ntraceh}
  \end{equation}
Then the covariant derivative of the velocity vector field (which is now an $\left(N+1\right)$-velocity) can be decomposed in analogy to equation \re{split}, page \pageref{split}, as
  \begin{equation}
     \nabla_{\beta} u_{\alpha}=-\dot{u}_\alpha u_\beta +\frac{1}{N}\Theta h_{\alpha\beta} + \omega_{\alpha \beta} + \sigma_{\alpha\beta} \ ,
      \label{Nvdec}
  \end{equation}
where $\dot{u}^\alpha$, $\Theta$, $\omega_{\alpha\beta}$ and $\sigma_{\alpha\beta}$ are defined as before.

While the Einstein equation \re{efe} remains unchanged, as does the splitting of the energy-momentum tensor \re{ttensor}, inserting the latter into the former and taking the trace we obtain the trace of the Ricci tensor, i.e. the Ricci scalar:
  \begin{equation}
    R=\left(\frac{2}{1-N}\right)\left[-\mu-\left(N+1\right)\Lambda+Np\right] \ .
      \label{Nricciscalar}
  \end{equation}
Inserting this expression into the Einstein equation we get for the Ricci tensor
  \begin{equation}
   R_{\alpha\beta}=T_{\alpha\beta}+\left(\frac{g_{\alpha\beta}}{1-N}\right)\left[-\mu-2\Lambda+Np\right] \ ,
      \label{Nriccitensor}
  \end{equation}
which we use in what follows.
  
\subsection{The (N+1)-dimensional Raychaudhuri equation}

In order to obtain the $\left(N+1\right)$-dimensional analogue of the Raychaudhuri equation \re{raycheq} we start from the Ricci identities \re{ricciid}. We project them along $u^\alpha$ with the projector \re{prou}, and we take the trace with the metric tensor. Then we can work on the various terms we obtain using the properties and the definitions given in sections \ref{kinvarsec}, the splitting of the covariant derivative of the velocity vector field \re{Nvdec}, and the expression \re{Nriccitensor} for the Ricci tensor. After some calculations we obtain the \emph{$\left(N+1\right)$-dimensional Raychaudhuri equation}:
  \begin{align}
	  \dot{\Theta}+\frac{\Theta^2}{N} =& \hat{\nabla}_\alpha \dot{u}^\alpha + \dot{u}_\alpha \dot{u}^\alpha + 2\left( \omega^2 - \sigma^2 \right)
	     \notag \\
	  &+\frac{1}{N-1}\left[\left(2-N\right)\mu-Np+2\Lambda\right]\ .
	     \label{Nraycheq}
	\end{align}
That with the choice $N=3$ gives the usual $\left(3+1\right)$-dimensional form \re{raycheq}, i.e.
	\begin{equation}
	  \dot{\Theta}+\frac{1}{3}\Theta^2 
	  = \hat{\nabla}_\alpha \dot{u}^\alpha + \dot{u}_\alpha \dot{u}^\alpha 
	  + 2\left( \omega^2 - \sigma^2 \right)
	  - \frac{1}{2}\left(\mu + 3p\right) + \Lambda\ ,
	     \label{3+1raycheq}
	\end{equation}
while with $N=2$ we obtain the \emph{$\left(2+1\right)$-dimensional Raychaudhuri equation}:
	\begin{equation}
	    \quad \dot{\Theta}+\frac{1}{2}\Theta^2 
	  = \hat{\nabla}_\alpha \dot{u}^\alpha + \dot{u}_\alpha \dot{u}^\alpha 
	  + 2\left( \omega^2 - \sigma^2 \right)
	  - 2p + 2\Lambda\ .
	    \label{2+1raycheq}
	\end{equation}

Apart from the numerical factors, the three equations have the same form, except for one important difference: in the $\left(2+1\right)$-dimensional case the term containing the energy density $\mu$ is absent.

\subsubsection{Irrotational dust}
Now we want to put ourselves in the particular case of irrotational dust, so $\omega=0$, $p=q^\alpha=\pi_{\alpha\beta}=0=\dot{u}^\alpha$. We get
  \begin{align}
     \dot{\Theta} + \frac{\Theta^2}{N} =\frac{N-2}{1-N} \mu - 2 \sigma^2 +\frac{2}{N-1}\Lambda 
     \qquad &\text{(N+1)}
      \label{Nraychdust}   \\
     \dot{\Theta}+\frac{1}{3}\Theta^2 = -\frac{1}{2} \mu - 2 \sigma^2  + \Lambda
     \qquad &\text{(3+1)}
      \label{3+1raychdust}  \\
     \dot{\Theta} + \frac{1}{2}\Theta^2 = -2 \sigma^2   + 2 \Lambda
     \qquad &\text{(2+1)} \ ,
      \label{2+1raychdust}
  \end{align}
where again the main difference arises from the fact that in the $\left(2+1\right)$-dimensional case the term containing $\mu$ is absent.

\subsection{The (N+1)-dimensional Hamiltonian constraint}
\label{obtainhcsection}

In order to obtain an $\left(N+1\right)$-dimensional version of the Hamiltonian constraint \re{hamconst}, we generalize the procedure used in \cite{ellisbrunihwang} to the case of $N+1$ dimensions.\footnote{The Hamiltonian constraint is reported at the beginning of that paper, while its derivation, plus some interesting discussions, can be found in the appendix.}

The \emph{fully orthogonally projected covariant derivative $\hat{\nabla}_\alpha$} has already been defined in equation \re{spatderiv}, and we recall that, if $u_{\mu}$ has non-zero vorticity, then $\hat{\nabla}$ is not an $N$-dimensional covariant derivative. This is due to Frobenius' theorem (statement \ref{fbth}, page \pageref{fbth}), from which it follows that in the case of non-vanishing vorticity the projector $h_{\alpha\beta}$ does not project on hypersurfaces orthogonal to $u^\alpha$, because these hypersurfaces do not exist.

While the elements of the Riemann tensor are given as usual, for a generic vector field of components $X^\alpha$, by
  \begin{equation}
    R^\alpha_{\phantom{\alpha}\beta\gamma\delta}X^\beta\doteq\nabla_\gamma\nabla_\delta X^\alpha-\nabla_\delta\nabla_\gamma X^\alpha \ ,
      \label{usualriemann}
  \end{equation}
we define
  \begin{equation}
    \nmur_{\alpha\beta\gamma\delta}\doteq\left(R_{\alpha\beta\gamma\delta}\right)_\bot + k_{\alpha\delta}k_{\beta\gamma}-k_{\alpha\gamma}k_{\beta\delta} \ .
      \label{n-1rabcd}
  \end{equation}
Here we have used\footnote{Defined in this way, $k_{\alpha\beta}$ is minus the \emph{extrinsic curvature tensor}.}
  \begin{equation}
    k_{\alpha\beta}\doteq \hat{\nabla}_{\beta}u_{\alpha}=\frac{1}{N}\Theta h_{\alpha\beta}+\omega_{\alpha\beta}+\sigma_{\alpha\beta} \ ,
      \label{kdef}
  \end{equation}
and
  \begin{equation}
    \left(R_{\alpha\beta\gamma\delta}\right)_\bot\doteq h_\alpha^{\phantom{\alpha}\mu}h_\beta^{\phantom{\beta}\nu}h_\gamma^{\phantom{\gamma}\sigma}h_\delta^{\phantom{\delta}\rho}R_{\mu\nu\sigma\rho} \ .
  \end{equation}
So $\nmur_{\alpha\beta\gamma\delta}$ has been defined not by the derivative $\hat{\nabla}_\alpha$ alone, but also by its embedding.
This tensor is the \emph{Riemann tensor of the hypersurfaces orthogonal to $u^\alpha$} \ev{if and only if} vorticity is zero.
We can rewrite it as
  \begin{equation}
    \nmur^{\alpha\beta}_{\phantom{\alpha\beta}\gamma\delta}=\left(R^{\alpha\beta}_{\phantom{\alpha\beta}\gamma\delta}\right)_\bot - 2 k^{[\alpha}_{\phantom{[\alpha}[\gamma} k^{\beta]}_{\phantom{\beta]}\delta]} \ .
  \end{equation}
  
Taking the trace we obtain
  \begin{equation}
    \nmur_{\alpha\gamma}\doteq\nmur^{\phantom{\alpha}\beta}_{\alpha\phantom{\beta}\gamma\beta}=h^{\beta\delta}\left(R_{\alpha\beta\gamma\delta}\right)_\bot + k_{\alpha\beta}k^\beta_{\phantom{\beta}\gamma}-k_{\alpha\gamma}\Theta \ ,
  \end{equation}
which is the Ricci tensor of $N$-dimensional hypersurfaces \ev{if and only if} vorticity vanishes.
Taking the trace of this expression by using $h^{\alpha\gamma}$, and inserting equation \re{kdef} we obtain, after some algebra,
  \begin{equation}
    \nmur\doteq \nmur^\alpha_{\phantom{\alpha}\alpha}=R+2u^\mu u^\nu R_{\mu\nu} + \frac{1-N}{N} \Theta^2 + 2 \left( \sigma^2 - \omega^2 \right) \ ,
  \end{equation}
which we can rewrite as
  \begin{equation}
    u^\mu u^\nu R_{\mu\nu} = \frac{1}{2} \left[ \nmur - R + \frac{N-1}{N} \Theta^2 -2 \left(\sigma^2-\omega^2\right) \right] \ .
      \label{uur}
  \end{equation}
  
Now we can project the Einstein equation \re{efe} in such a way that we obtain
  \begin{equation}
    u^\mu u^\nu R_{\mu\nu} - \frac{1}{2} u^\mu u^\nu g_{\mu\nu} R + u^\mu u^\nu g_{\mu\nu} \Lambda = u^\mu u^\nu T_{\mu\nu} \ ,
  \end{equation}
and inserting equations \re{uur}, \re{Nricciscalar}, \re{ttensor} we get the \emph{$\left(N+1\right)$-dimensional Hamiltonian constraint}
  \begin{equation}
    \frac{1}{2} \nmur + \frac{N-1}{2N} \Theta^2 - \sigma^2 = - \omega^2 + \mu + \Lambda \ .
      \label{Nhamconst}
  \end{equation}
In the case of $N=3$ we obtain again the expression
  \begin{equation}
    \frac{1}{2} \tr + \frac{1}{3} \Theta^2 - \sigma^2 = - \omega^2 + \mu + \Lambda \ ,
      \label{3+1Hamconst}
  \end{equation}
while with the choice $N=2$ we have the \emph{$\left(2+1\right)$-dimensional Hamiltonian constraint}
  \begin{equation}
    \frac{1}{2} \dr + \frac{1}{4} \Theta^2 - \sigma^2 = - \omega^2 + \mu + \Lambda \ .
      \label{2+1hamconst}
  \end{equation}

\subsubsection{Irrotational dust}
Turning our attention to the particular case of irrotational dust, we have to take into account that $\omega=0$, and $p=q^\alpha=\pi_{\alpha\beta}=0=\dot{u}^\alpha$. We find
  \begin{align}
     \frac{1}{2} \nmur + \frac{N-1}{2N} \Theta^2 - \sigma^2 = \mu +\Lambda \qquad &\text{(N+1)}
      \label{Nhcdust} \\
     \frac{1}{2} \tr + \frac{1}{3} \Theta^2 - \sigma^2 = \mu +\Lambda \qquad &\text{(3+1)}
      \label{3+1hcdust} \\
     \frac{1}{2} \dr + \frac{1}{4} \Theta^2 - \sigma^2 = \mu +\Lambda \qquad &\text{(2+1)} \ .
      \label{2+1hcdust}
  \end{align}

\section{The averaged (N+1)-dimensional equations}

We want to obtain average equations that hold for the general $\left(N+1\right)$-dimensional case, so we need to generalize to higher dimensions the procedure of averaging previously defined. Note that we average the equations obtained for the case of irrotational dust.

\subsubsection{Defining the averaging procedure.} Generalizing equation \re{averagedef}, we define the average value of the scalar quantity $f$ over the $N$-dimensional hypersurface of averaging as
  \begin{equation}
    \av{f}\fdt\doteq \frac{\int_t f \epsilon}{\int_t \epsilon}\ ,
      \label{Nav}
  \end{equation}
where $\epsilon$ is the volume form of the $N$-dimensional hypersurface of averaging, i.e.
  \begin{equation}
    \epsilon=\sqrt{ \lvert ^{\left(N\right)}g \rvert}dx^{1}\wedge\dots\wedge dx^{N}\ ,
      \label{Nvolumeform}
  \end{equation}
with $^{\left(N\right)}g$ being defined as the determinant $^{\left(N\right)}g\doteq det \left( ^{\left(N\right)}g_{\alpha\beta} \right)$.\footnote{See section \re{volelsec} for a detailed discussion about the volume element.}

Let us point out that now the averages are taken over $N$-dimensional hypersurfaces rather than over three-dimensional ones, so the volume element is different, because it is the $N$-dimensional, and not the three-dimensional one.

\subsubsection{The commutation rule.} If we go step by step through the calculation that gives us the commutation rule \re{commrule}, contained in section \ref{gencommrulesec}, we see that no difference arises between the case of $3+1$ dimensions and the case of $N+1$ dimensions, apart from the fact that the averaged quantities are defined differently. So in the case of irrotational dust the commutation rule is again:
  \begin{equation}
    \partial_t \av{f} = \av{\partial_t f} + \av{\Gamma \Theta f } - \av{f}\av{\Gamma\Theta} \ .
      \label{Ncommrule}
  \end{equation}

\subsubsection{The average expansion rate.}
\label{navtheta}

 In the $\left(N+1\right)$-dimensional case, generalizing \re{adef} we define the scale factor as
   \begin{equation}
     a\fdt\doteq\left(\frac{\int_t \epsilon}{\int_{t_0}\epsilon}\right)^{\frac{1}{N}} \ ,
       \label{ndimadef}
   \end{equation}
 so the average expansion rate $\av{\Theta}$ turns out to be:
   \begin{equation}
     \av{\Gamma \Theta}=N\frac {\dot{a}}{a} \ .
       \label{Navtheta}
   \end{equation}
 
This can be proved as follows.
  Considering the time derivative of equation \re{ndimadef} we get
  \begin{align}
      \dot{a}\fdt=&\partial_t \left(\frac{\int_t \epsilon}{\int_{t_0}\epsilon}\right)^\frac{1}{N} 
                  \notag \\
                 =&\left(\frac{\int_t\epsilon}{\int_{t_0}\epsilon}\right)^{\frac{1}{N}}\frac{1}{N}\frac{\partial_t\int_t\epsilon}{\int_t\epsilon} 
                  \notag \\
                 =&a\fdt\frac{1}{N}\frac{\partial_t\int_{t}\epsilon}{\int_{t}\epsilon} 
                  \notag \\
                 =&a\fdt\frac{1}{N}\frac{\int_{t}\Gamma\Theta\epsilon}{\int_{t}\epsilon} \ ,
          \label{contone}
  \end{align}
where we have used equation \re{ndimadef} in obtaining the third line, and the equation
  \begin{equation}
    \int_t \Gamma \Theta \epsilon = \partial_t \int_t \epsilon \ ,
      \label{partialtheta}
  \end{equation}
which is the same as \re{primointegrale}, which does not change in the $\left(N+1\right)$-dimensional case, for the last one.
  Rewriting equation \re{contone} as
   \begin{equation}
     N\frac{\dot{a}\fdt}{a\fdt}=\frac{\int_{t}\Gamma \Theta\epsilon}{\int_{t}\epsilon}
   \end{equation}
  and using \re{averagedef} we get the result \re{Navtheta}.

Let us note that the Hubble parameter is still defined as
  \begin{equation}
    H\fdt\doteq\left(\frac{\dot{a}\fdt}{a\fdt}\right) \ ,
      \label{nhubbledef}
  \end{equation}
but now $a\fdt$ is given by equation \re{ndimadef} rather than \re{adef}.

\subsection[The (N+1)-dim. averaged Raychaudhuri equation]{The (N+1)-dim. averaged \mbox{Raychaudhuri equation}}

Let us now apply the averaging procedure outlined above to equation \re{Nraychdust}. Note that now $\Gamma=0$, because the equations we average hold in the case of vanishing vorticity. Taking the average as defined by \re{Nav} and using the commutation rule \re{Ncommrule} where we insert \re{Navtheta}, with $a$ defined by \re{ndimadef}, after some calculations we obtain the \emph{averaged $N$-dimensional Raychaudhuri equation}
  \begin{equation}
    N\frac{\ddot{a}}{a}= \frac{N-2}{1-N} \av{\mu} + \nq + \frac{2}{N-1} \Lambda \ ,
      \label{avnraycheq}
  \end{equation}
where we have defined the \emph{$\left(N+1\right)$-dimensional backreaction variable $\nq$} as
  \begin{equation}
    \nq\doteq\frac{N-1}{N}\left(\av{\Theta^2}-\av{\Theta}^2\right)-2\av{\sigma^2} \ .
      \label{nqdef}
  \end{equation}
  
Putting $N=3$ we obtain again equation \re{avraycheq}:
  \begin{equation}
    3\frac{\ddot{a}}{a}=-\frac{1}{2} \av{\mu} +  \q +\Lambda \ ,
  \end{equation}
while for the value $N=2$ we get the \emph{averaged $\left(2+1\right)$-dimensional Raychaudhuri equation}
  \begin{equation}
    2\frac{\ddot{a}}{a} = \tq +2\Lambda \ ,
  \end{equation}
where the \emph{$\left(2+1\right)$-dimensional backreaction variable $\tq$} is
  \begin{equation}
    \tq\doteq\frac{1}{2}\left(\av{\Theta^2}-\av{\Theta}^2\right)-2\av{\sigma^2} \ .
      \label{tqdef}
  \end{equation}
These equations have been obtained in the case of irrotational dust.
  
Once again we notice that the most important difference between the $\left(2+1\right)$-dimensional and the $\left(3+1\right)$-dimensional case is that in the former the term containing $\av{\mu}$ is not present.

\subsection{The (N+1)-dim. averaged Hamiltonian constraint}

With the same averaging procedure applied to \re{Nhcdust} we get the \emph{averaged $\left(N+1\right)$-dimensional Hamiltonian constraint}
  \begin{equation}
    \frac{\left(N-1\right)N}{2}\frac{\dot{a}^2}{a^2}= \av{\mu} - \frac{1}{2} \av{ \nmur}-\frac{1}{2} \nq +\Lambda \ ,
      \label{avnhamconst}
  \end{equation}
with $\nq$ given by equation \re{nqdef}. 
In the particular case of $N=3$ we obtain again \re{avhamcon}:
  \begin{equation}
    3 \frac{\dot{a}^2}{a^2}= \av{\mu} - \frac{1}{2} \av{ \tr} - \frac{1}{2}  \q +\Lambda \ ,
  \end{equation}
and with the choice $N=2$ we have the \emph{averaged $\left(2+1\right)$-dimensional Hamiltonian constraint}
  \begin{equation}
    \frac{\dot{a}^2}{a^2}= \av{\mu} - \frac{1}{2} \av{ \dr} - \frac{1}{2} \tq +\Lambda\ ,
  \end{equation}
with $\tq$ defined by \re{tqdef}.

\subsection{The (N+1)-dimensional integrability condition}

Even in the $\left(N+1\right)$-dimensional case we have two equations, the averaged Raychaudhuri equation \re{avnraycheq} and the averaged Hamiltonian constraint \re{avnhamconst}, which form a system for the three unknown variables $a\fdt$, $\avnmur\fdt$ and $\nq\fdt$. We cannot solve this system for the three  variables, but we can use the two equations in order to obtain a single formula that relates $\avnmur\fdt$ and $\nq\fdt$. This generalizes what has been done in section \ref{intconsec}, and the interpretation of the results is similar.

Consider the averaged $\left(N+1\right)$-dimensional Hamiltonian constraint \re{avnhamconst}, take the time derivative and insert the two equations \re{avnraycheq} and \re{avnhamconst}. If we insert the Hubble parameter $H$ defined in \re{nhubbledef}, and the continuity equation, which in the $\left(N+1\right)$-dimensional case reads again $\partial_t\av{\mu}+\av{\Theta} \av{\mu}=0$ (with $\Theta$ now being defined by \re{Navtheta}), we obtain the \emph{integrability condition} between the averaged $\left(N+1\right)$-dimensional Hamiltonian constraint and the averaged $\left(N+1\right)$-dimensional Raychaudhuri equation, which reads
  \begin{equation}
    2NH\nq+\partial_t\nq= -2H \avnmur - \partial_t\avnmur \ .
  \end{equation}
In the particular case of $N=3$ we have
  \begin{equation}
    6H \q + \partial_t \q = -2H \avtr - \partial_t \avtr \ .
  \end{equation}
In the case $N=2$ we get
  \begin{equation}
    4H\tq + \partial_t \tq = -2H\avdr - \partial_t \avdr \ .
  \end{equation}
  
The meaning of these equations is that the evolution of backreaction and averaged scalar curvature influence each other.
Note that the cosmological constant cancels out in the calculations sketched above.

\subsection{Particular solutions}

In analogy to what happens in section \re{intconsec}, here also are given some particular solutions for the variables $ \q$ and $\avtr$ expressed as functions of $a\fdt$.

In the $\left(N+1\right)$-dimensional case they are:
  \begin{itemize}
    \item
      \begin{equation}
        \nq =0 \Rightarrow \avnmur \propto a^{-2} \ ,
          \label{nsol1}
      \end{equation}
    \item
      \begin{equation}
        \avnmur=0 \Rightarrow  \nq\propto a^{-2N} \ ,
          \label{nsol2}
      \end{equation}
    \item
      \begin{equation}
        \nq\propto a^{-2N}  \quad \text{and}  \quad \avnmur \propto a^{-2} \ .
          \label{nsol3}
      \end{equation}
  \end{itemize}
  
So in the $\left(2+1\right)$-dimensional case they read:
  \begin{itemize}
    \item
      \begin{equation}
        \tq=0 \Rightarrow \avdr \propto a^{-2} \ ,
          \label{2sol1}
      \end{equation}
    \item
      \begin{equation}
        \avdr=0 \Rightarrow  \tq\propto a^{-4} \ ,
          \label{2sol2}
      \end{equation}
    \item
      \begin{equation}
        \tq\propto a^{-4}  \quad \text{and}  \quad \avdr \propto a^{-2} \ ,
          \label{2sol3}
      \end{equation}
  \end{itemize}
while in the $\left(3+1\right)$-dimensional case we obtain what has already been discussed in section \ref{intconsec}.

\section{Geometry and topology}

In the case of even-dimensional manifolds, their geometry can be related to their topology. In $2$ dimensions this is encoded in the Gauss-Bonnet theorem (see for instance \cite[p. 235]{jost}), which relates the curvature of a surface to its Euler characteristic (this object is presented for instance in \cite[section 6.1]{jost} and in \cite[section 17.4]{dario}; anyway here we proceed in a simpler way). A \emph{generalized Gauss-Bonnet theorem}, that relates powers of the Riemann tensor to the Euler characteristic, is valid in $2m$ dimensions, with $m$ integer. For odd dimensional manifolds, and in particular for three-dimensional manifolds, there is no analogue of these theorems.\footnote{A discrete analogue is the \emph{Descartes theorem}. This theorem can be found for instance in \cite[chapter 42]{gravitation}, as the discrete part of equation $(42.1)$, which contains also a particular case of the Gauss-Bonnet theorem. The whole chapter 42, which is about Regge Calculus, may provide useful insights on the topic of this section.}

For a two-dimensional manifold, i.e. a surface, the Ricci scalar assumes a particularly interesting form. Using the Gauss-Bonnet theorem we can write the average Ricci scalar in a way which shows that it is proportional to $a^{-2}$; the constant of proportionality is related to the topology of the surface. We obtain that $\avdr = C a^{-2}$, where the constant $C$ can also be zero. Because of this relation and the integrability condition, the backreaction variable must be either $ \q=0$ or $ \q\propto a^{-4}$, i.e. $ \q=Aa^{-4}$, where $A$ is a constant that can also be zero.

\subsection{The Gauss-Bonnet theorem}
  \label{gaussbonnetsec}

First of all, we need to recall some concepts from geometry.

The \emph{Gaussian curvature} of a surface at a point on it is the product of the principal curvatures, $k_1$ and $k_2$, of the surface at the given point:\footnote{Note that it is possible to give an alternative definition of the Gaussian curvature through the use of the covariant derivative and the metric tensor.}
  \begin{equation}
    K\doteq k_1 \cdot k_2 \ .
  \end{equation}

The meaning of principal curvatures can be understood as follows. At each point $p$ on a surface embedded in three-dimensional Euclidean space one may choose a unit normal vector. A plane passing through $p$ that contains the normal is called a \emph{normal plane}, and it will also contain a unique direction tangent to the surface and cut the surface in a plane curve. The principal curvatures at $p$, denoted $k_1$ and $k_2$, are the maximum and minimum values of this curvature.

From this it is clear that the principal curvatures represent an \emph{extrinsic} measure of curvature, i.e. they describe how the surface is curved due to its embedding in a higher-dimensional space.

Conversely, the Gaussian curvature is an \ev{intrinsic} measure of curvature, i.e. its value does not depend on the way the surface may be isometrically embedded in space.\footnote{This is the content of Gauss' theorema egregium, which reads: the Gaussian curvature of a smooth surface embedded in $\mathbb{R}^3$ is invariant under local isometries. This is, for instance, the reason it is possible to fold a piece of paper, or it is possible to eat pizza folding a slice in half. Being a flat surface, its Gaussian curvature is zero. If we embed it in space through folding a slice in half, one of the principal curvatures at a point along the fold is non-zero, so the other must remain zero in order to keep the value of the Gaussian curvature constant, according to Gauss' theorema egregium.}

The \emph{Geodesic curvature} of a curve lying on a submanifold of the ambient space measures, roughly speaking, how far the curve is from being a geodesic.

Now we can state the theorem.
\begin{theorem}[Gauss-Bonnet]
Consider $\m$ to be a \ev{compact}, \ev{two-dimensional} Riemannian manifold with boundary $\partial\m$. Let $K$ be the \emph{Gaussian curvature of $\m$}, and $k_g$ the geodesic curvature of the boundary $\partial\m$. Then the Gauss-Bonnet theorem states that
  \begin{equation}
    \int_{\m} K d\Sigma + \int_{\partial\m} k_g ds = 2 \pi \ecm \ ,
      \label{gaussbonnet}
  \end{equation}
where $d\Sigma$ is the area element of the surface,\footnote{A surface is a two-dimensional manifold, or more precisely a two-dimensional topological manifold, so here with "surface" we mean the manifold $\m$.} $ds$ the line element along the boundary of $\m$ and $\ecm$ the \emph{Euler characteristic of $\m$}.
\end{theorem}
The first integral in the above equation represents the \emph{total Gaussian curvature of $\m$}, which is the integral of the Gaussian curvature over the whole manifold.

The Euler characteristic is a \emph{topological invariant}, so it describes the shape of a topological space regardless of the way it is bent.
It can be defined for a connected planar graph, for polyhedra, for surfaces, and for other objects.\footnote{For instance, the Euler characteristic is $\chi=1$ for an interval in $\mathbb{R}$, $\chi=0$ for a circle, $\chi=1+\left(-1\right)^n$ for a $n$-sphere, $\chi=1$ for the solid unit ball in any Euclidean space, $\chi=1$ for $\mathbb{R}^n$.}

A rigorous and general definition is the following: the Euler characteristic of a $N$-dimensional differentiable manifold $\m$ is the alternating sum
    \begin{equation}
      \chi\left(\m\right)=\sum^{N}_{i=0}\left(-1\right)^i b_i \ ,
    \end{equation}
where $b_i$ denotes the \emph{i-th Betti number}.\footnote{This definition and the meaning of the objects involved in it are widely explained in \cite[section 17.4]{dario}. For our purposes it is enough to illustrate the meaning of this definition in a particular case. For a polyhedron, The Euler characteristic $\chi$ is given by \emph{Euler's formula} $\chi=V-E+F$ (see \cite[p. 1174]{gravitation}), where $V$ is the number of vertices (zero-dimensional), $E$ the number of edges (one-dimensional) and $F$ the number of faces (two-dimensional).}

Now it is clear how the Gauss-Bonnet theorem connects the geometry of a surface, in the sense of its scalar curvature, with its topology, in the sense of the Euler characteristic.

When applied to compact surfaces with no boundaries, the second integral in \re{gaussbonnet} vanishes, so we obtain
  \begin{equation}
    \int_{\m} K d\Sigma = 2 \pi \ecm \ ,
      \label{gbbello}
  \end{equation}
which means that the total Gaussian curvature of the surface considered is equal to $2\pi$ times the Euler characteristic of the surface itself.\footnote{Think of the Earth. On the surface there are peaks and valleys, which give the Gaussian curvature a wide range of values on it. One may think that they average out when you consider the whole surface, giving a total Gaussian curvature equal to zero. But the surface of the Earth is topologically a $two$-sphere, and its Euler characteristic is $2$. The Gauss-Bonnet theorem then tells us that the total Gaussian curvature of its surface is $4\pi$.}

For \ev{orientable} \ev{compact} surfaces without boundary the Euler characteristic is given by
  \begin{equation}
    \ecm=2-2 \g \ ,
      \label{ecg}
  \end{equation}
where $\g$ is the \emph{genus} of the surface.

The genus $\g$ of a surface is defined as the number of tori in a connected sum decomposition of the surface.

Intuitively, any orientable compact surface without boundary is \emph{topologically equivalent} to a sphere with some handles attached, and $\g$ counts the number of handles.\footnote{For instance, a torus $T$ has $\g=1$, i.e. one handle, and in fact $\chi\left(T\right)=0.$}

Then for this kind of a surface we have:
  \begin{equation}
    \int_{\m} K d\Sigma = 2 \pi \left( 2-2 \g \right) \ .
      \label{gbbellobello}
  \end{equation}

\subsection{Curvature tensors for surfaces}

In the case of a surface, the Riemann tensor is given by (see \cite[p. 144]{jost}):
  \begin{equation}
    R_{\alpha\beta\gamma\delta}=
    K\left(g_{\alpha\gamma}g_{\beta\delta}-g_{\alpha\beta}g_{\gamma\delta}\right) \ ,
  \end{equation}
where the function $K$ is the Gaussian curvature of the surface.

If we take the trace we get the Ricci tensor, which in this case turns out to be
  \begin{equation}
    R_{\alpha\beta}\doteq R_{\alpha\beta\gamma\delta}g^{\beta\delta} = g_{\alpha\gamma}K \ ,
  \end{equation}
i.e. proportional to the metric tensor. This means that every surface is an Einstein manifold.\footnote{See footnote 25 on the page \pageref{vacuumricci}.}

Taking the trace again we obtain the Ricci scalar of the surface:
  \begin{equation}
   R\doteq g^{\alpha\gamma}R_{\alpha\gamma} = g^{\alpha\gamma} g_{\alpha\gamma} K = 2K \ .
      \label{ricciconst}
  \end{equation}

\subsection[The average Ricci scalar and the Euler characteristic]{Relation between the average Ricci scalar and the Euler characteristic}
\label{avricciandeuler}

In the case of $2+1$ dimensions we have 
 \begin{equation}
  \begin{split}
    \avdr\doteq&\frac{\int \dr \epsilon }{\int \epsilon}
    =\frac{\int \dr \sqrt{| ^{\left(2\right)}g|}dxdy}{\int \epsilon} \\
    =&\frac{\int 2K \sqrt{| ^{\left(2\right)}g|}dxdy}{\int \epsilon} \\
    =&\frac{4\pi}{\int \epsilon}\ecm \ ,
    \label{rchi}
  \end{split}
 \end{equation}
where we have in the second line used the fact that for a two-dimensional manifold the Ricci scalar is given by equation \re{ricciconst}, and in the third line the Gauss-Bonnet theorem in the form \re{gbbello}.

Using the equation \re{ndimadef} with $N=2$, we obtain\footnote{Note that equation \re{ndimadef} normalizes the scale factor $a\left( t \right)$ at the present day volume of the universe $V_0\doteq \int_{t_0} \epsilon$, i.e. $a_0\doteq a\left(t_0\right) = 1$.}
  \begin{equation}
    \avdr=\frac{4\pi \ecm a_0^2}{a^2 V_0} = \frac{4\pi \ecm}{a^2 V_0}\ .
      \label{eccoci}
  \end{equation}

So, we can state that in the $\left(2+1\right)$-dimensional case the average value of the Ricci scalar of a surface is given by
  \begin{equation}
    \avdr\fdt=Ca^{-2}\fdt \propto a^{-2}\fdt \ ,
      \label{cameno2}
  \end{equation}
where $C$ is a constant that may be zero, in which case we have a surface that on average has vanishing Ricci scalar curvature.
It is important to note that, as a consequence of the fact that the Euler characteristic is a topological invariant, the average Ricci scalar does not change for deformations of the manifold that leave its topology unchanged.

Looking for particular solutions of the integrability condition that holds between the Raychaudhuri equation and the Hamiltonian constraint, we already found this behavior. In fact we found that there are both solutions with vanishing average Ricci scalar, i.e. \re{nsol2}, and solutions with $\avdr\propto a^{-2}$, i.e. \re{nsol1} and \re{nsol3}. However, this does not mean that there is a manifold that, once averaged, presents these properties. This was the same in $N+1$ dimensions, in $3+1$ dimensions, and in $2+1$ dimensions. But while in the general $\left(N+1\right)$-dimensional (or in the $\left(3+1\right)$-dimensional) case those were just \ev{some} solutions, in the $\left(2+1\right)$-dimensional case \re{cameno2} is \ev{the only} possibility.

So, due to equations \re{2sol1}, \re{2sol2} and \re{2sol3}, we have that in the $\left(2+1\right)$-dimensional case the only possibilities for the $\left(2+1\right)$-dimensional backreaction are to vanish, or to be proportional to $a^{-4}$, so we have
  \begin{equation}
    \tq\fdt=Aa^{-4}\fdt \ ,
      \label{ambarabaccicciccocco}
  \end{equation}
where $A$ is a constant that can also vanish.

Given this, it is interesting to ask if the constant $A$ can be different from zero, or if it always vanishes. It can be demonstrated that the acceleration is zero in a $\left(2+1\right)$-dimensional FLRW universe. 
If the answer of the above question is that we always have $A=0$, then there is no backreaction in $2+1$ dimensions. On the other hand, if it turns out that there are cases in $2+1$ dimensions in which $A\neq0$, this means that we could find cases where there is backreaction, but it does not provide acceleration.

\subsection{Comments}

In section \ref{avricciandeuler} we have found that, in 2+1 dimensions, the average Ricci scalar of a surface is constrained by the condition \re{eccoci}. As a consequence, also the backreaction variable is constrained, as equation \re{ambarabaccicciccocco} shows. Being more precise, from that equation we can see that the backreaction variable is determined up to a constant. 

In the 3+1 Newtonian gravity, the situation is similar. The total energy of the system is conserved, and the backreaction variable turns out to be a boundary term, that vanishes for spatially compact cosmologies.\footnote{These issues are discussed in \cite{buchert1996}.}

So, in both cases we have a global constraint that restricts the evolution. It is interesting to note that this does not happen in the $\left(3+1\right)$-dimensional general relativistic case.

\appendix

\chapter{Appendix: FLRW models}

This work is about more general cosmological models, but let us briefly outline the homogeneous and isotropic case. The family of cosmological models that are homogeneous and isotropic is called Friedmann-Lema\^itre-Robertson-Walker (FLRW elsewhere in the text). In this appendix we describe its main characteristics and consequences, which are interesting per se, and are sometimes recalled in the text. We give also some useful astrophysical relations. We summarize the results and just sketch how to obtain them, because many of the proofs are beyond the scope of this work. Detailed discussions can be found in \cite[chapter 4]{covform3}, \cite[chapter 5]{wald}, \cite[section 2.6.2]{roncadelli} and \cite[chapter 1]{ilcerchiosichiude}. The validity of idealizing the universe as homogeneous and isotropic is discussed in the introduction to this work.

\section{Homogeneity and isotropy}

  \subsection{Two assumptions}
    \label{twoassumptions}
  
The defining assumption of Friedmann-Lema\^itre-Robertson-Walker is that of homogeneity and isotropy of the spacetime.\footnote{As specified in chapter \ref{backreaction}, in the FLRW models by homogeneity and isotropy we mean \ev{exact} homogeneity and isotropy. Here we drop the word \ev{exact} to avoid confusion, because in the context of FLRW it is usually not used.} Let us specify the mathematical meaning of these assumptions (see \cite[pp. 92--93]{wald}).
   
A spacetime is said to be \emph{(spatially) homogeneous} if there exist a one-parameter family of spacelike hypersurfaces $\Sigma_t$ foliating the spacetime, such that $\forall t, \forall p,q \in\Sigma_t$ there exists an isometry of the spacetime metric $g_{\alpha\beta}$ which maps $p$ into $q$.
   
A spacetime is said to be \emph{(spatially) isotropic} at each point if there exists a congruence of timelike curves (i.e., observers), with tangents denoted by $u^\alpha$, filling the spacetime and satisfying the following property. Given any point p and any two unit <<spatial>> tangent vectors $s_1^\alpha$, $s_2^\alpha$ (i.e. vectors at $p$ orthogonal to $u^\alpha$), there exists an isometry of the metric which leaves $p$ and $u^\alpha$ at $p$ fixed, but rotates $s_1^\alpha$ into $s_2^\alpha$. Thus in an isotropic universe it is impossible to construct a geometrically preferred tangent vector orthogonal to $u^\alpha$.

We point out that, with the assumptions of homogeneity and isotropy, the surfaces $\Sigma_t$ must be orthogonal to the tangent vectors $u^\alpha$ to the world lines of isotropic observers. This is expressed in our formalism by the vanishing of the vorticity tensor (see \mbox{chapter 3}).

An important statement to be pointed out is that isotropy at three or more points, i.e. the property that at each of these points all directions are geometrically indistinguishable (a pointwise property), implies homogeneity, i.e. that all points are geometrically indistinguishable (a global property), see \cite[p. 145]{jost}.\footnote{Note that isotropy around two points imply homogeneity in any geometry other than the spherical one, where we need three points. In this case two different centers of spherical symmetry are  possible without requiring homogeneity.}

We use comoving coordinates (see section \ref{velvecfield}).\footnote{In general only self gravitating systems, like clusters of galaxies, or isolated galaxies, can be considered comoving observers, i.e. the observers associated to comoving coordinates. This is due to the definition of comoving coordinates, which we recall from chapter 2. Choose arbitrarily a space section of the spacetime and label the fluid particles by coordinates $x^{a}$; at all later times label the same particles by the same coordinate values, so that the fluid flow lines in spacetime are the curves $x^{a}=const$. The time coordinate is then determined by measuring proper time, from the initial space section, along the flow lines. Our galaxy is part of the Local Group, so it is locally gravitationally related to other objects. This results in a local motion of the Milky Way, but we are able to take this into account, so we can approximately consider ourselves comoving observers.}

  \subsection{The FLRW metric}
  
From the two assumptions made above the form of the metric can be determined up to one free function ($a\left(t\right)$, see below) and one constant ($k$), without using the Einstein equation. This equation allows us to determine the evolution of the free function, as explained in the next section.

Spatial homogeneity and isotropy imply that the Riemann tensor $^{\left( 3 \right)}R_{\alpha\beta\gamma\delta}$ constructed from $h_{\alpha\beta}$ on $\Sigma_t$ is of the form (see \cite[p. 94]{wald}):
  \begin{equation}
    ^{\left( 3 \right)} R_{\alpha\beta\gamma\delta} = K h_{\gamma [ \alpha} h _{\beta ] \delta} \ ,
  \end{equation}
where K can be a function only of time.
This means that the curvature of every $\Sigma_t$ is described by a single scalar K, that is the Gaussian curvature, and in this case it is constant over space, so every $\Sigma_t$ represents a three-dimensional \emph{space of constant curvature}.

For the Gaussian curvature we have
  \begin{equation}
    K\doteq k_1 \cdot k_2 = \frac{k}{a^2} \ ,
      \label{gaussiancurvatureFLRW}
  \end{equation}
and so we obtain
  \begin{equation}
    ^{\left( 3 \right)} R = 6K = 6\frac{k}{a^2} \ .
      \label{3erre}
  \end{equation}

The Gaussian curvature of a space of constant curvature completely determine its geometry. In particular, there are only three possible geometries, \emph{spherical}, \emph{flat} or \emph{hyperbolic}, depending on whether the value of the Gaussian curvature is positive, zero, or negative, respectively. 

Then the metric turns out to be of the form (see \cite[p. 95]{wald}, \cite[p. 21]{covform3}):
  \begin{equation}
    ds^2=-dt^2+a^2 \left( t \right) \left[ dr^2+f^2 \left( r \right) d\Omega^2 \right] \ ,
    \label{flrw1}
  \end{equation}
where\footnote{It is common to refer to the case of positive (negative) Gaussian curvature as \emph{spatially closed} (\emph{spatially open}). These should not be confused with \emph{open} and \emph{closed manifolds}. A closed manifold is a compact manifold without boundaries; an open manifold is one that is not compact and without boundary. With these definitions, both closed and open manifolds are possible in the spatially open case, while in the spatially closed case we need a closed manifold. From the topological point of view, in each of the three cases of equation \re{geometries} several different manifolds can be used to describe the universe. The most common choices are the Euclidean space $\mathbb{E}^3$ in the case of flat geometry, and the 3-sphere $\mathbb{S}^3$ for the spherical one, while for the hyperbolic geometry a lot of different choices are considered.}
  \begin{equation}
    f \left( r \right) =
      \begin{cases}
        \sin r & \text{$k=+1$ (spherical geometry)}
        \\
        r & \text{$k=0$   \quad (flat geometry)}
        \\
        \sinh r & \text{$k=-1$ (hyperbolic geometry)}
    \end{cases}
    \label{geometries}
  \end{equation}
and
  \begin{equation}
    d\Omega^2 = d\Theta^2 + \sin^2\Theta d\phi^2
  \end{equation}
is the angular part.
The function $a\fdt$, which depends only on time, is called the scale factor.

Following \cite[p. 191]{roncadelli} we can also express the metric \re{flrw1} in the form:
  \begin{equation}
    ds^2=-dt^2+a^2 \left( t \right) \left[ \frac{dr^2}{1-kr^2}+r^2 d\Omega^2 \right] \ ,
  \end{equation}
where the geometry of space is determined by the value of $k$ as specified in the definition \re{geometries}.

  \section{Dynamics}
 
    \subsection{Friedmann equations}
    \label{friedmannsection}
The symmetries of the Friedmann models imply that all the dynamical and kinematical variables are functions of time only, and any quantity that represents inhomogeneity or anisotropy vanishes identically.

Now our aim is to determine the dynamical evolution of the universe. This can be obtained by the use of the Einstein equation \re{efe}. In order to solve it, we have to specify what kind of matter fills the universe, i.e. to give the form of the energy-momentum tensor.
    
The most general form consistent with homogeneity and isotropy that can be taken into account is the ideal fluid form \re{perfectfluid}. It allows us to describe both matter and radiation, once we specify the equation of state (see below). The energy-momentum tensor has the form
      \begin{equation}
        T_{\alpha\beta}=\mu u_\alpha u_\beta + p h_{\alpha\beta} \ .
        \label{perfectfluid2}
      \end{equation}

Moreover, from the symmetries of the FLRW it follows that
  \begin{equation}
      E_{\alpha\beta}=0 \ , \qquad
      H_{\alpha\beta}=0 \ ,
        \label{vanishweyl}
  \end{equation}
and
  \begin{equation}
      \omega_{\alpha\beta}=0 \ ,  \qquad
      \sigma_{\alpha\beta}=0 \ ,  \qquad
      \dot{u}^\alpha=0 \ ,
        \label{vanishrobevarie}
  \end{equation}
the last due to the fact that fundamental observers are free falling.\footnote{See \cite[p. 190]{roncadelli}, and section \ref{accefreefallsec}.}
In addition, due to the spatial homogeneity, all orthogonally projected covariant derivatives (e.g. $\tilde{\nabla}_{\alpha}p$,\dots) are by definition zero.

Now we have to compute the scale factor $a\left(t\right)$ from the Einstein equation.\footnote{Note that it is possible to solve the Einstein equation using the metric \re{flrw1}, as done in \cite[pp. 96--97]{wald}, but we proceed in a different way.} In principle it constitutes a set of ten independent equations, but is possible to show that the conditions assumed about the symmetry of spacetime reduce them to only two. This follows from inserting equations \re{vanishweyl} and \re{vanishrobevarie} into the set of evolution and constraint equations derived in section \ref{evoeconsteqsect}: the only non-vanishing equations are the Raychaudhuri equation \re{raycheq} and the continuity equation \re{enconseq}.

Instead of the continuity equation we can consider, together with the Raychaudhuri equation \re{raycheq}, the Hamiltonian constraint \re{3+1Hamconst}. The continuity equation is then redundant, because it can be obtained (see equation \re{consflrw}) by combining the two Friedmann equations, that we derive in the following lines from Raychaudhuri equation and the Hamiltonian constraint.
Inserting conditions \re{vanishrobevarie} and the form \re{perfectfluid2} of the energy-momentum tensor into \re{3+1Hamconst} and \re{raycheq}, we obtain for the Hamiltonian constraint the form
  \begin{equation}
    \frac{ ^{\left( 3 \right)}R }{2}= -\frac{1}{3}\Theta^2+\mu+\Lambda \ ,
  \end{equation}
and for the Raychaudhuri equation the form
  \begin{equation}
    \dot{\Theta}+\frac{\Theta^2}{3}=-\frac{1}{2} \left( \mu +3p \right) +\Lambda \ ,
  \end{equation}
where the cosmological constant $\Lambda$ is now necessary in order to describe the observed universe with the FLRW model, and it is subject to subsequent considerations.

By using again the relation \re{3erre} and remembering that $\Theta/3=\dot{a}/a = H$ (see equation \re{hubbleparam}) we obtain the two evolution equations for the scale factor $a \left( t \right)$:
  \begin{align}
   H\left(t\right)^2 = \left( \frac{ \dot{a} \left( t \right) }{a \left( t \right) } \right) ^2 &= \frac {1 }{3} \mu
    -\frac{k}{a^2 \left( t \right)} + \frac{1}{3} \Lambda \ ,
      \label{friedmann1} \\
    \dot{H}\left(t\right)+H\left(t\right)^2 = \frac{\ddot{a}\left(t\right)}{a\left(t\right)}&=
    -\frac{1}{6} \left(\mu + 3p \right) + \frac{1}{3}\Lambda \ ,
      \label{friedmann2}
  \end{align}
known as the \emph{Friedmann equations}. Once again, $k$ determines the geometry of the space: the choice $k=1$ gives the spherical geometry, $k=0$ the flat geometry, and $k=-1$ the hyperbolic one.
  
Because an ideal fluid is described by the energy-momentum tensor \re{perfectfluid2}, it is characterized by the quantities $\mu$ and $p$, which are functions only of time as a consequence of homogeneity. As explained in section \re{eqstate}, the dynamical description of a fluid requires also the specification of an equation of state that relates $\mu \left( t \right)$ and $p \left(t\right)$.

In current cosmological models the three main components that fill the universe, in terms of an FLRW model, are \emph{non-relativistic particles} (\emph{matter}, both barionic matter and cold dark matter), \emph{relativistic particles} (\emph{radiation}) and \emph{vacuum energy} (described by the cosmological constant $\Lambda$).\footnote{Actually the cosmological constant represents a modification of the Einstein equation that allows gravity to be repulsive. However, the term containing $\Lambda$ can be written in the form of an energy-momentum tensor, so it can be considered as arising from a form of energy with negative pressure, vacuum energy, which is the main candidate for dark energy. The cosmological constant, dark energy and modified gravity are treated in depth in chapter 1.} When they are in thermodynamical equilibrium, we can describe them by the equation of state (see \cite[p. 192]{roncadelli})
  \begin{equation}
    p\left(t\right)=wc^2\mu\left(t\right) \ ,
    \label{eqstateflrw}
  \end{equation}
where $w$ is constant, and assumes the values\footnote{In equation \re{wcases} we report the value of $w$ for vacuum energy. More generally, for dark energy we have $w<-1/3$, which gives $p\left(t\right)<-\frac{1}{3}\mu\left(t\right)$. This equation shows that dark energy violates the strong energy condition \re{strong}.}
  \begin{equation}
    w=
      \begin{cases}
        0 & \text{matter,} \\
        1/3 & \text{radiation,} \\
        -1 & \text{vacuum.}
      \end{cases}
      \label{wcases}
  \end{equation}

Differentiating (w.r.t. time) the first Friedmann equation \re{friedmann1}, and inserting it into the second Friedmann equation \re{friedmann2}, we obtain the conservation equation\footnote{Some algebra shows that this equation is precisely the continuity equation \re{enconseq} written in the FLRW case where equations \re{vanishrobevarie} hold, i.e. $\dot{\mu}=-3 \frac{\dot{a}}{a} \left( p + \mu \right)$.}
  \begin{equation}
    \frac{d}{dt}\left(\mu a^3 \right) = -\frac{3p}{c^2}a^2\dot{a} \ ,
      \label{consflrw}
  \end{equation}
and substituting in it the equations of state \re{eqstateflrw} we get
  \begin{equation}
    \mu \propto a^{-3\left( 1+w \right)} \ .
  \end{equation}
Inserting the definition of $w$ gives the behavior of $\mu \left( t \right)$:
  \[
    \mu\left(t\right)\propto
      \begin{cases}
        a\left(t\right)^{-3} & \text{for matter,} \\
        a\left(t\right)^{-4} & \text{for radiation,} \\
        constant  & \text{for the vacuum.}
      \end{cases}
  \]
  
  Now it is useful to define the quantities
  \begin{equation}
    \mu_c\left(t\right)\doteq 3H\left(t\right)^2 \ ,  \qquad
    \Omega\fdt\doteq\frac{\mu\left(t\right)}{\mu_c\left(t\right)} \ ,  \qquad
    H\left(t\right)\doteq\frac{\dot{a}\left(t\right)}{a\left(t\right)} \ ,
  \end{equation}
  i.e. the \emph{critical density}, the \emph{cosmic density parameter}, and the \emph{Hubble parameter} (already defined in equation \re{hubbleparam}).
Using them it is possible to rewrite equation \re{consflrw} as
    \begin{equation}
      \Omega\left(t\right)=1+\frac{kc^2}{H^2\left(t\right)a^2\left(t\right)} \ ,
    \end{equation}
  from which it is clear that the geometry of the space is determined once $\Omega$ is determined, in such a way that the universe has
    \begin{itemize}
      \item hyperbolic geometry if $\Omega<1$ \ ,
      \item flat geometry if $\Omega=1$ \ ,
      \item spherical geometry if $\Omega>1$ \ .
    \end{itemize}
    
Some other parameters are useful in the description of the model, the most important of them being
  \begin{equation}
    \Omega_M\doteq\frac{\mu_M}{\mu_c},  \qquad \text{and}  \qquad \Omega_\Lambda\doteq\frac{\mu_\Lambda}{\mu_c} \ ,
  \end{equation}
i.e. the \emph{contribution of ordinary matter and radiation}\footnote{By ordinary matter and radiation we mean the particles of the standard model, and we include dark matter of whatever origin.} to the cosmic density parameter, and the \emph{contribution of the cosmological constant}.
They allow us to write $\Omega$ as
  \begin{equation}
    \Omega=\Omega_M+\Omega_\Lambda \ ,
  \end{equation}
and they are of particular importance from the observational point of view, because they can be determined from several different kinds of observations (see for instance \cite{roncadelli}). Note that analogous relations are defined for the \emph{baryonic contribution}, the \emph{contribution due to luminous matter}, that due to \emph{dark matter}, and so on.

  \subsection{The evolution of the universe}
    \label{evouniv}
  
  Once we have obtained the FLRW metric and the Friedmann equations, we can see that different kinds of evolution are possible for a universe described by this model.\footnote{We must point out that by <<evolution of the universe>> in this section we mean only the evolution of the scale factor of the universe $a\left( t \right)$.} For simplicity in this section we assume the matter to be represented by dust, so the universe is assumed to be filled by radiation plus dust, plus possibly vacuum energy. The behavior is considerably different in the cases of vanishing and non-vanishing cosmological constant.
    \begin{itemize}
   
      \item If we put $\Lambda=0$, we are describing the universe as if it contained only ordinary matter and radiation. It can be shown (see \cite[p. 194]{roncadelli}) that in this case the spatial geometry of the universe and its evolution are uniquely related. This means that once we know the value of the parameter $\Omega$ we know the qualitative evolution of the universe. In particular:
      \begin{itemize}
        \item if $\Omega\leq1$ the geometry of the universe is hyperbolic or flat, and it expands indefinitely; while
        \item if $\Omega>1$ the geometry of the universe is spherical and from a certain instant of time it will contract.
      \end{itemize}
      It is clear that, in this case, a static universe is not allowed.\footnote{In fact Einstein generalized the original form $G_{\mu\nu}=T_{\mu\nu}$ of his equation by inserting the term $g_{\mu\nu}\Lambda$ (which represents the maximal generalization that preserves all the properties of the original) expressly in order to allow static solutions, which however turn out to be unstable.}
      Moreover, as we can see from equation \re{friedmann2} the expansion rate decelerates. This is due to the fact that gravity is always attractive for ordinary matter and radiation.
      
      Given (by one century of observations) that our universe is expanding, i.e. $\dot{a}\left(t_0\right)>0$, it follows from equation \re{friedmann2} that $\ddot{a}<0$, so the universe must have been expanding at a faster and faster rate as one goes backward in time. This means that in the past we would have had $a=0$, a singular state usually called the big bang.
      
      \item If we put $\Lambda\neq0$ we are describing a universe where, besides ordinary matter and radiation, there is another form of energy, vacuum energy.
      Now expansion and contraction are possible for each spatial geometry, depending on the values of $\mu_\Lambda$ and $\mu_{M}$.
      But the most important difference is that now the evolution can be \emph{accelerated}, which happens when we have $\mu_{\Lambda}>0$ and $\mu_M<\mu_{\Lambda}$, as seen from equation \re{friedmann2}. This shows that gravity becomes repulsive when the negative pressure of dark energy is dominant.
      \end{itemize}

Let us now turn our attention to the present day universe. The \emph{age of the universe}, i.e. the amount of time between the big bang and today, is indicated by $t_0$. The Hubble parameter is $H=H\left(t\right)$, and the Hubble constant is $H_0\doteq H\left(t_0\right).$\footnote{See equation \re{hubbleparam} and the related footnote for more details.}
For small redshift (defined below), i.e. $z\ll1$, the \emph{Hubble law} is (\cite[p. 196]{roncadelli}):
\begin{equation}
  v_r\approx H_0d \ ,
  \label{hubblelaw}
\end{equation}
where $d$ is the proper distance from the object observed (galaxies in the original work of Hubble) to the observer in the three-dimensional space defined by given cosmological time, $v_r$ is the velocity along the corresponding sight line, and the value of the Hubble constant is estimated to be $\hubbleconstant$.\footnote{There are many determinations of the Hubble constant, based on different techniques. The different results are often compatible with each other, once the uncertainties are considered. We use the value $H_0 = 73.8 \pm2.4\,km\,s^{-1}\,Mpc^{-1}$, affected by a $3.3\%$ uncertainty. This determination of the Hubble constant has been obtained in \cite{hubble}, from optical and infrared observations of Cepheid variables, using the \emph{Wide Field Camera 3} (\emph{WFC3}) on the \emph{Hubble Space Telescope} (\emph{HST}).}
A rough estimate of the value of $t_0$ is the \emph{Hubble time}:
  \begin{equation}
    \hubbletime \ .
  \end{equation}

It is useful to emphasize the interpretation of the redshift related to the Hubble law. The overall expansion of our universe, as represented by the Hubble law, forces galaxies or clusters of galaxies not gravitationally interacting to move apart. However this is not an expansion of distances between objects \ev{in} the space, but an expansion \ev{of} the space itself. This means that the redshift of cosmological origin is not due to the relative motion of source and receiver as the one due to the Doppler effect is.\footnote{In this section we deal only with the \ev{redshift of cosmological origin}, i.e. the one due to the expansion of the universe. Anyway, objects that are not self gravitating, like for instance stars inside galaxies, are also subject to a local motion, so there is also a \ev{redshift due to the Doppler effect}, which affects the light they emit. The \ev{observed redshift} is the sum of both contributions.} But we can still demonstrate that we have a redshift, which can be intuitively understood by thinking that the wavelength of a light ray traveling through the expanding space experiences the same stretch of the space itself. The wavelength of a photon changes from the value $\lambda_E$ at time $t_E$, to the value $\lambda_O$ at time $t_O > t_E$. We define the redshift as
  \begin{equation}
    z\doteq\frac{\lambda_O - \lambda_E}{\lambda_E} \ ,
  \end{equation}
which is equivalent to
  \begin{equation}
    1+z\doteq\frac{\lambda_O}{\lambda_E} \ .
  \end{equation}
It can be demonstrated that, in the framework of an FLRW model, the redshift $z$ and the scale factor $a\fdt$ are related as
  \begin{equation}
    \frac{1}{1+z}=\frac{a\left(t\right)}{a\left(t_0\right)} \ .
      \label{zdef}
  \end{equation}

\backmatter

\end{document}